\documentclass[prx, reprint, 9pt, superscriptaddress,
notitlepage, nofootinbib, longbibliography,
floatfix]{revtex4-2}
\usepackage{gpkmac}

\usepackage{xcolor}

\definecolor{gr}{rgb}{0,0,0}

\begin{document}

\title{Conformal invariance and quantum non-locality in critical hybrid circuits}
\author{Yaodong Li}
\affiliation{Department of Physics, University of California, Santa Barbara, CA 93106}
\author{Xiao Chen}
\affiliation{Kavli Institute for Theoretical Physics, University of California, Santa Barbara, CA 93106}
\affiliation{Department of Physics and Center for Theory of Quantum Matter, University of Colorado, Boulder, CO 80309}
\affiliation{Department of Physics, Boston College, Chestnut Hill, MA 02467}
\author{Andreas W. W. Ludwig}
\affiliation{Department of Physics, University of California, Santa Barbara, CA 93106}
\author{Matthew P. A. Fisher}
\affiliation{Department of Physics, University of California, Santa Barbara, CA 93106}
\date{September 14, 2021}

\begin{abstract}
We establish the emergence of {\cyan a conformal field theory} 
(CFT) in {\cyan a} (1+1)-dimensional hybrid quantum circuit right at the measurement-driven entanglement transition, by revealing space-time conformal covariance of entanglement entropies and mutual information for various subregions at different circuit depths.
While the evolution takes place in real time, 
{the spacetime manifold of the circuit 
appears to host a Euclidean field theory with imaginary time.}
Throughout the paper we investigate Clifford circuits with several different boundary conditions by injecting physical qubits at the spatial and/or temporal boundaries, all giving consistent characterizations of the underlying ``Clifford CFT''.
We emphasize (super)universal results that are consequences solely of the conformal invariance and do not depend crucially on the precise nature of the CFT.
Among these are the infinite entangling speed as a consequence of measurement-induced quantum non-locality, and the critical purification dynamics of a mixed initial state.
\end{abstract}

\maketitle




\tableofcontents


\section{Introduction}

Entanglement is a central concept in quantum physics.
It violates classical laws of physics in dramatic ways, and makes quantum communication and quantum computation fundamentally more powerful than their classical counterparts~\cite{nielsen2010qiqc}.
In recent years, the entropy of entanglement has proven useful in condensed matter physics, providing new insights and tools for understanding quantum states of matter, either in or out of equilibrium, at zero or finite temperature~\cite{Deutsch1991, Srednicki1994, Calabrese2004, Calabrese_Cardy_2009, Levin2006, Kitaev2006,Ryu2006, Nandkishore2015, abanin2018rmp}.

One of the most bizzare aspects of entanglement, namely quantum non-locality, has always involved wavefunctions subject to measurements.
The measurements, albeit local, have non-local influences on the states and their entanglement structure. 
In the famous EPR thought experiment~\cite{einstein1935can, bell1964on}, one destroys entanglement between a pair of distant qubits by making local measurements in exchange for perfectly correlated measurement outcomes.
Conversely, one can entangle a pair of distant qubits with local measurements via a mechanism similar to quantum teleportation~\cite{bennett1993teleporting,Bouwmeester1997teleportation}, without the two ever needing to talk to one another -- a phenomenon known as ``entanglement swapping'' {that has found wide applications in quantum information 
science~\cite{bennett1993teleporting, zukowski1993event, pan1998experimental, jennewein2001experimental} (see  Fig.~\ref{fig:EE_swap}(a) for an illustration).
These are examples of ``measurement-induced quantum non-locality'' in systems of a few qubits, and the experiments usually require carefully following specific protocols (that is, making the right unitary gates and right measurements at the right place and right time).
One is therefore led to the following question: can quantum non-locality show up in many-body quantum dynamics under measurements 
\textit{without} fine tuning?
Notice that this is never possible in unitary systems, since information as well as quantum entanglement must evolve in a strictly local fashion, as required by the Lieb-Robinson bound~\cite{lieb1972finite, hastings1008locality}.

\env{figure}{[t]
\includegraphics[width=.45\textwidth]{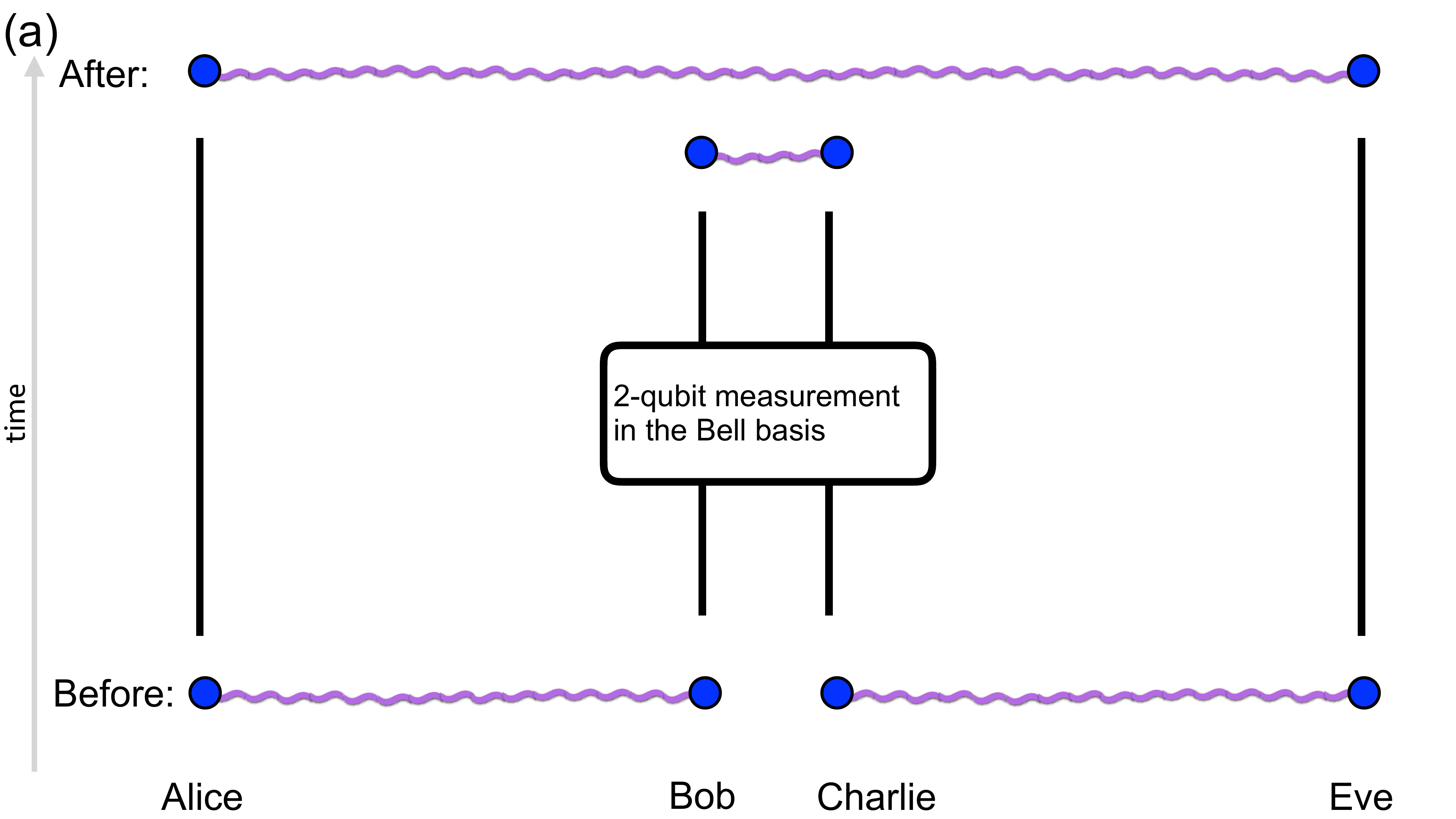}
\includegraphics[width=.45\textwidth]{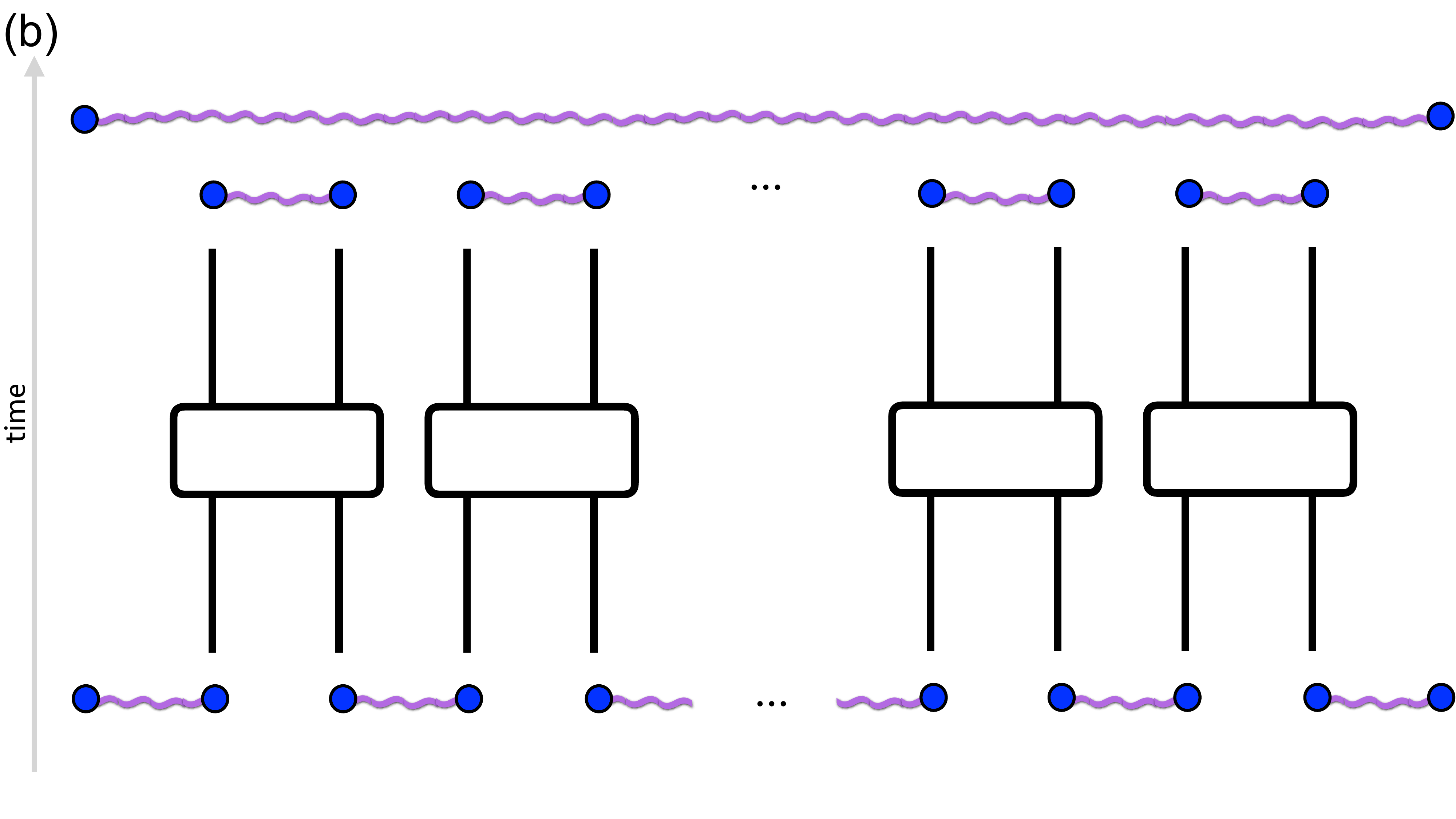}
\caption{Illustration of the entanglement swapping protocol.
(a)
An example with four parties.
The initial state consists of two Bell pairs, (Alice, Bob) and (Charlie, Eve), with Alice and Bob far apart, as well as Charlie and Eve.
Suppose Bob and Charlie are spatially proximate and they make a collective 2-qubit measurement in the Bell basis ( $\{ \frac{1}{\sqrt{2}}\(\ket{00} \pm \ket{11}\), \frac{1}{\sqrt{2}}\(\ket{01} \pm \ket{10}\)\} )$~\cite{nielsen2010qiqc}.
The measurement ``swaps'' the entangled pairs, and now we have entangled pairs (Alice, Eve) and (Bob, Charlie).
Notice that Alice and Eve never directly talked to one another, but nevertheless become entangled due to the measurement: for each one of the four possible measurement outcomes, Alice and Eve share a different state in the Bell basis.
Since the distance between Alice and Eve is arbitrary, the speed of entanglement is arbitrarily large.
However, no information is transmitted: in order for Alice and Eve to know what their wavefunction is, classical information about the mesurement outcome must be obtained from Bob and Charlie {via classical communication}.
(b)
A many-qubit example of entanglement swapping.
In this circuit with nearest neighbor gates, only a finite circuit depth is required to generate a long-range entangled Bell pair, given that the initial state is properly set-up and each two-qubit Bell measurement is perfect.
This fine-tuned example merely serves the purpose of illustrating the possibility of infinite entangling speed in many-body systems.
Notice the similarity with the actual circuit in Fig.~\ref{fig:upc_rect}.
}
\label{fig:EE_swap}
}

Recently there has been some interest in a novel phase transition in the dynamics of entanglement, driven by 
repetitive local measurements in the time evolution of an otherwise unitary quantum system~\cite{cao2018monitoring, nandkishore2018hybrid, nahum2018hybrid, li1808hybrid, li1901hybrid, choi2019qec, szyniszewski1903measurement,  gullans1905purification, choi2019spin, andreas2019hybrid,Tang2019, gullans1910scalable,huse1911tripartite,zhang2020nonuniversal,fan2020selforganized}. 
Such systems can be conveniently modelled as a ``random \textit{hybrid} 
quantum circuit''~\cite{halpern1711qiqcog}, composed of both random unitary gates that increase entanglement and (on their own) ``thermalize'' the system~\cite{nahum2017KPZ, nahum2018operator, keyserlingk2018operator},
and local measurements made at random positions of the circuit that act against this tendency;
the only tuning parameter of this model is thus the measurement strength/frequency.
The phase transition is a consequence of this competition, and separates a ``volume-law entangled phase'' and an ``area-law entangled phase'' at low/high strengths of measurements, respectively.
Various aspects of the transition have been explored, including
numerical characterizations of the critical entanglement dynamics (and ``statics'' of the steady-state)~\cite{nahum2018hybrid, li1808hybrid, li1901hybrid, gullans1905purification,gullans1910scalable, huse1911tripartite};
analytic mappings to effective statistical mechanical models~\cite{nahum2018hybrid, choi2019spin, andreas2019hybrid},
interpreting the entanglement transition as a conventional Landau ordering transition; {two \textit{different} mappings to
properties of 2D critical percolation for the transition in hybrid circuits with Haar random unitary gates (i.e. random matrices sampled from the Haar measure on the unitary group $U(4)$ on two qubits), one for the $0$th R\'enyi
(Hartley)
entropy~\cite{nahum2018hybrid}, and another for the $n$th R\'enyi entropies
with $n\geq 1$ in the limit of infinite onsite Hilbert space dimension~\cite{andreas2019hybrid};}
the ``genericity'' of the transition, e.g. discussions of conditions for the presence/absence of the transition~\cite{cao2018monitoring, nandkishore2018hybrid, choi2019qec} and evidences for the transition in more realistic contexts~\cite{szyniszewski1903measurement, Tang2019};
connections to quantum channel capacity and quantum error correction~\cite{choi2019qec, gullans1905purification, choi2019spin, gullans1910scalable};
and implications for classical simulability of unitary circuits in (2+1)-dimensions~\cite{harrow2001efficient}.

Curiosities of the transition aside, the numerical accessibility of the model alone makes it a convenient theoretical platform for investigating 
non-unitary quantum dynamics.
{For example, one can ask if the aforementioned measurement-induced quantum non-localtity shows up in such circuits.}
In Ref.~\cite{li1901hybrid}, 
{it was suggested}
that the disentangling capabilities of local measurements are indeed non-local, as evidenced by the powerlaw distribution of the ``disentanglement length'' throughout the volume-law phase {and at the critical point}.
However, this was an indirect probe {lacking an explicit information-theoretic meaning}.

On a seemingly separate note, it was found that the steady state wavefunction right at the critical point exhibits long-range correlations and {conformal
invariance~\cite{nahum2018hybrid, li1901hybrid,andreas2019hybrid}.}
{While in 
{random unitary circuits the time-evolution is well understood~\cite{nahum2017KPZ, nahum2018operator, keyserlingk2018operator},} 
exactly how the 
{long}-time critical entanglement structures of the
hybrid quantum circuits 
emerge under the real-time evolution, has not been
explicitly described (see relevant discussions in Refs.~\cite{nahum2018hybrid, andreas2019hybrid}).}

In the present work, we establish the 
{emergence of conformal symmetry}
in the {spacetime} circuit right at the critical point -- by illustrating its role in {describing} the critical entanglement dynamics -- and discuss the {physical mechanism} underpinning its emergence, namely the aforementioned non-locality induced by quantum measurements. 
Our starting point is a simple postulate that 
{
at the critical point, the spacetime manifold of the hybrid circuit 
hosts a Euclidean field theory, with the real-time direction of the circuit playing the role of imaginary/Euclidean time of the field theory (this naturally accounts for the absence of a Lieb-Robinson bound, as we briefly explain below).}
{This idea was already implicit in 
{
the mappings to 
{effective spin models~\cite{choi2019spin, andreas2019hybrid}}
relating quantum entanglement entropy 
to the \textit{boundary free energy} 
of a classical statistical mechanics model~\cite{vasseur2018rtn}.
Once time is interpreted as another spatial dimension, and entanglement entropies 
{as boundary free energies}, it is immediate that the conformal invariance -- therefore also ``criticality'' and long-range correlations -- 
{makes already detailed predictions for {entanglement dynamics}}
at the very early times. 
Long-range correlations at arbitrarily early times imply an infinite entangling speed ({as detailed in Eqs.~(\ref{eq:I_eta_aaaa}, \ref{eq:LR_bound})}), giving a positive answer to the question raised above -- 
that there is indeed a many-body version of entanglement swapping induced by measurements in the circuit,
despite
{the fact}
that the circuit is composed of completely random unitaries and measurements (as opposed to carefully designed protocols as in Fig.~\ref{fig:EE_swap}).\footnote{\label{footnote:ee_swap}There are important subtleties in this statement, which we clarify immediately below.
\env{itemize}{
\item
Entanglement itself does not contain information, and absence of lightcone in entanglement dynamics does not imply the ability to send information faster than light.
{As emphasized in the caption of Fig.~\ref{fig:EE_swap},
to verify the entanglement that has been generated by entanglement swapping, classical communication of the measurement outcomes is necessary (for specifying the pure state wavefunction after the measurement, much like in a quantum teleportation experiment).
This type of communication between ``people'' that perform and monitor the experiment is of course not included in the simple circuit model.
}
\item
The entanglement dynamics is only accessible in the pure state quantum trajectories, and is not accessible in the mixed state density matrix.
In fact, in the density matrix everything remains local, since we are only applying local operations.
To experimentally access the nonlocal entanglement one needs to prepare several copies of the same wavefunction, which requires {heavy} post-selection on the measurement outcomes.
{The need of introducing an ``experimenter'' doing all the work of recording measurement outcomes and post-selecting them, is in some sense similar to the aforementioned need of classical communication between ``parties'' in order to verify entanglement swapping.}
}
}
{This suggests}
that ``measurement-induced quantum non-locality'' is a 
consequence of broken unitarity, rather than of specific protocols/algorithms.

We establish the main results by studying the random Clifford circuit model {\cyan for a 1d chain of Qubits}, introduced in Ref.~\cite{li1901hybrid}, taking a trivial product initial state and open spatial boundary {conditions}, tuned to the transition.
{The space-time region of the circuit is thus a rectangle.}
The Gottesman-Knill theorem~\cite{gottesman9604hamming, gottesman9807heisenberg, aaronson0406chp} enables efficient simulation of Clifford circuits of up to thousands of qubits on a laptop, allowing us to perform detailed scaling analyses.
We numerically compute the entanglement entropies and mutual information for various 
{subregions}
at all time steps of the evolution, and verify that their dynamics 
are completely characterized by
{boundary}
3- and 4-point correlation functions, respectively, 
{of a CFT} 
in the finite rectangular geometry.
From the data we also extract several 
{critical exponents} 
characterizing the {underlying} Clifford CFT.

We further explore several different {sets of} boundary conditions of the Clifford circuit, by ``inserting''  physical qubits initialized in a trivial product state at the spatial and/or temporal boundaries of the finite circuit.
Remarkably, the boundary qubits become critically entangled 
through the bulk as an intermediary, despite 
{the fact}
that they never talked directly {to each other} -- another manifestation of entanglement swapping.
Numerical computations of entanglement entropies and mutual information further 
{confirm the presence of conformal symmetry,
and give}
consistent 
{estimates}
of 
corresponding 
{boundary} operator scaling dimensions
{appearing in {various \textit{different}} observables}.

Among 
{various different setups,}
 of particular interest is the one in which the initial state 
{consists of $L$ Bell pairs (i.e.
of $L$ maximally entangled pairs of qubits).}
By taking one qubit from each pair, we form a length-$L$ qubit chain which is subsequently subject to the hybrid circuit dynamics (the {``system qubits''}); the 
{remaining qubit chain (the ``environment qubits'')}
is left unevolved.
The two qubit chains appear to be on the same footing and have identical entanglement structures at all times.
In particular, while the system qubits experience the entanglement transition, the environment {qubits also know} 
about the transition.
{After tracing out the environment qubits,}
this setup is equivalent to the one in Ref.~\cite{gullans1905purification}, where 
{a mixed-state density matrix was time-evolved}.
The entanglement entropy between the system 
{qubits}
and the environment
{qubits}
is correspondingly interpreted as the 
{``purity of the system", }
and the entanglement transition is now a ``purification transition'', between a ``mixed phase'' and a ``pure phase'' characterized by, among other things, slow and fast purification dynamics, respectively.
In our CFT language, this setup maps to the same bulk theory but with a different boundary condition, so the purification transition
is indeed the same bulk transition as the transition in entanglement entropy with a pure initial state.
We show that the $(T/L)^{-1}$ decay 
{($T$ is the circuit depth)}
of the
{entanglement entropy between ``system'' and ``environment''}
at early times{, observed in the numerics of Ref.~\cite{gullans1905purification}},
follows directly from conformal symmetry,
{which in turn identifies the amplitude of that decay
as a universal (boundary) scaling dimension of the CFT (up to a factor of $\pi$).}
We also show that the universal exponential decay of {the same quantity} at late times is a consequence of crossover to a quasi-one-dimensional system{, the rate of decay being given by yet another
universal (boundary) scaling dimension of the CFT, which we identify here.}
These results are consequences solely of the conformal 
{invariance; they hold in all CFTs, and thus hold, in particular equally in other critical hybrid quantum circuits
described by CFTs, {\cyan presumably} including those with Haar unitaries.}

{We apply the same reasoning to
{the analysis of} the problem of the $0$th R\'enyi (Hartley)
entropy in random Haar circuits, which is believed~\cite{nahum2018hybrid} to be described
by two-dimensional critical { first-passage} percolation.}
{Comparison between the 
{critical properties of the von Neumann entropy in Clifford
CFT and those of the so-obtained zeroth R\'enyi (Hartley) entropy in the Haar circuits is made, and their relationship}
is discussed.}

The rest of this paper is organized as follows.
In Sec.~\ref{sec:bcft}, we introduce the random hybrid circuit model in rectangular geometry with several sets of boundary conditions.
We then give a statement of the conjecture regarding the 
{presence of conformal symmetry, as well
as}
a concrete prescription for computing the entanglement entropy {of an arbitrary segment at an arbitrary time step}.
In Sec.~\ref{sec:res_rect}, we present the main results of this paper, namely the numerical data on entanglement entropy and mutual information dynamics in the rectangular circuit, {and compare them with CFT calculations}.
In Sec.~\ref{sec:pbc}, we present results for circuits with periodic boundary condition, {that are used for fixing tuning parameters of our fitting scheme}.
In Sec.~\ref{sec:discussion}, we discuss the universality of our results, relations to other works, and possible future directions.
In Appendix~\ref{sec:cft_calc}, we 
{provide, for reference, a list of elementary facts from conformal field theory}
used in this paper.
In Appendix~\ref{sec:refQ}, we discuss purification dynamics of ``reference qubits'' recently introduced in Ref.~\cite{gullans1910scalable}, which reveals a {boundary} operator scaling dimension taking different values in the Clifford CFT and in critical percolation.
This result is further confirmed by a separate calculation in Appendix~\ref{sec:localizable_entanglement}.
In Appendix~\ref{sec:perc}, we present parallel numerics and analysis of two-dimensional critical first-passage percolation.

\section{The hybrid circuit model and the conjecture \label{sec:bcft}}

\env{figure*}{[t]
\includegraphics[width=.45\textwidth]{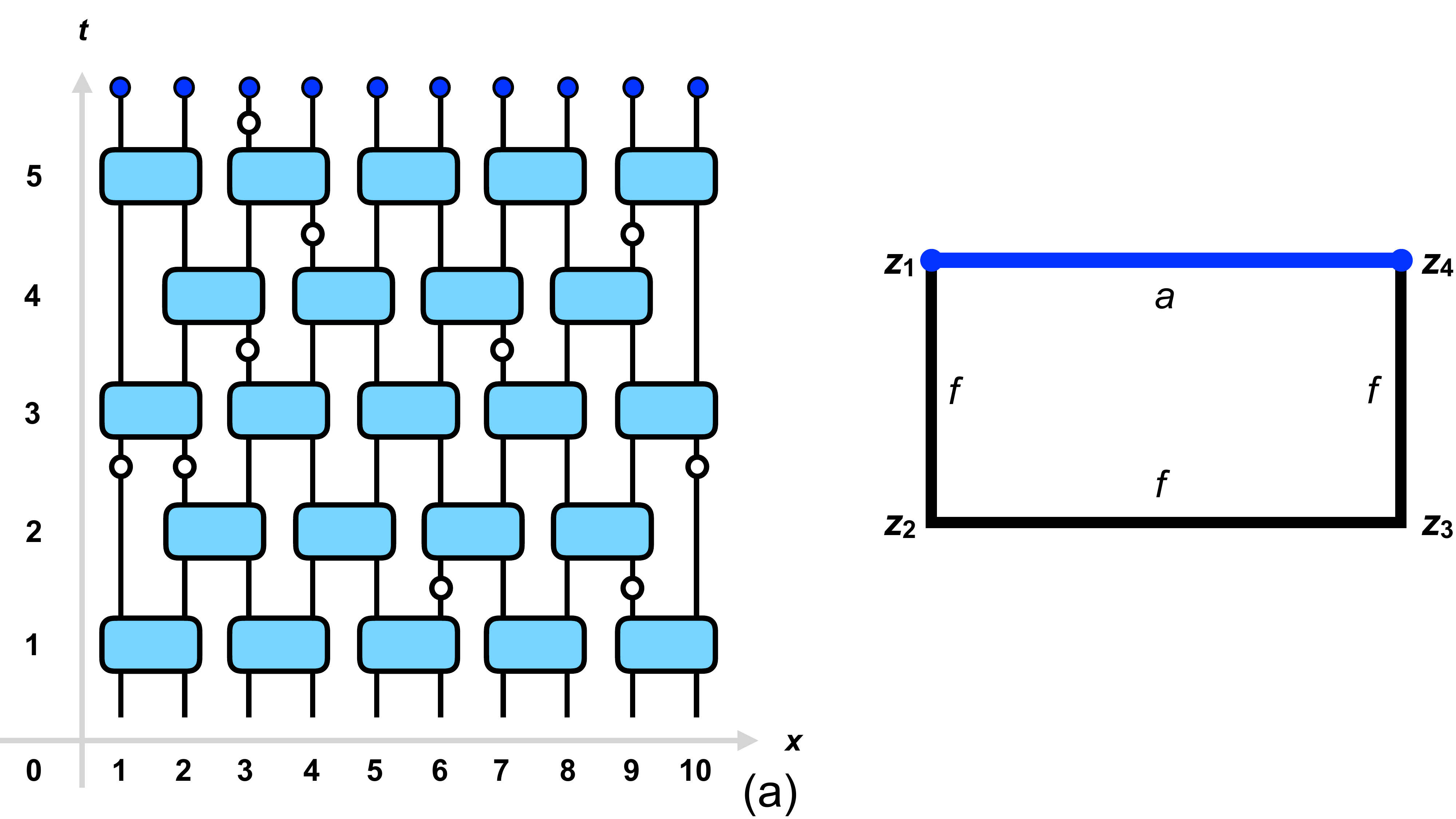}
\hfill
\includegraphics[width=.45\textwidth]{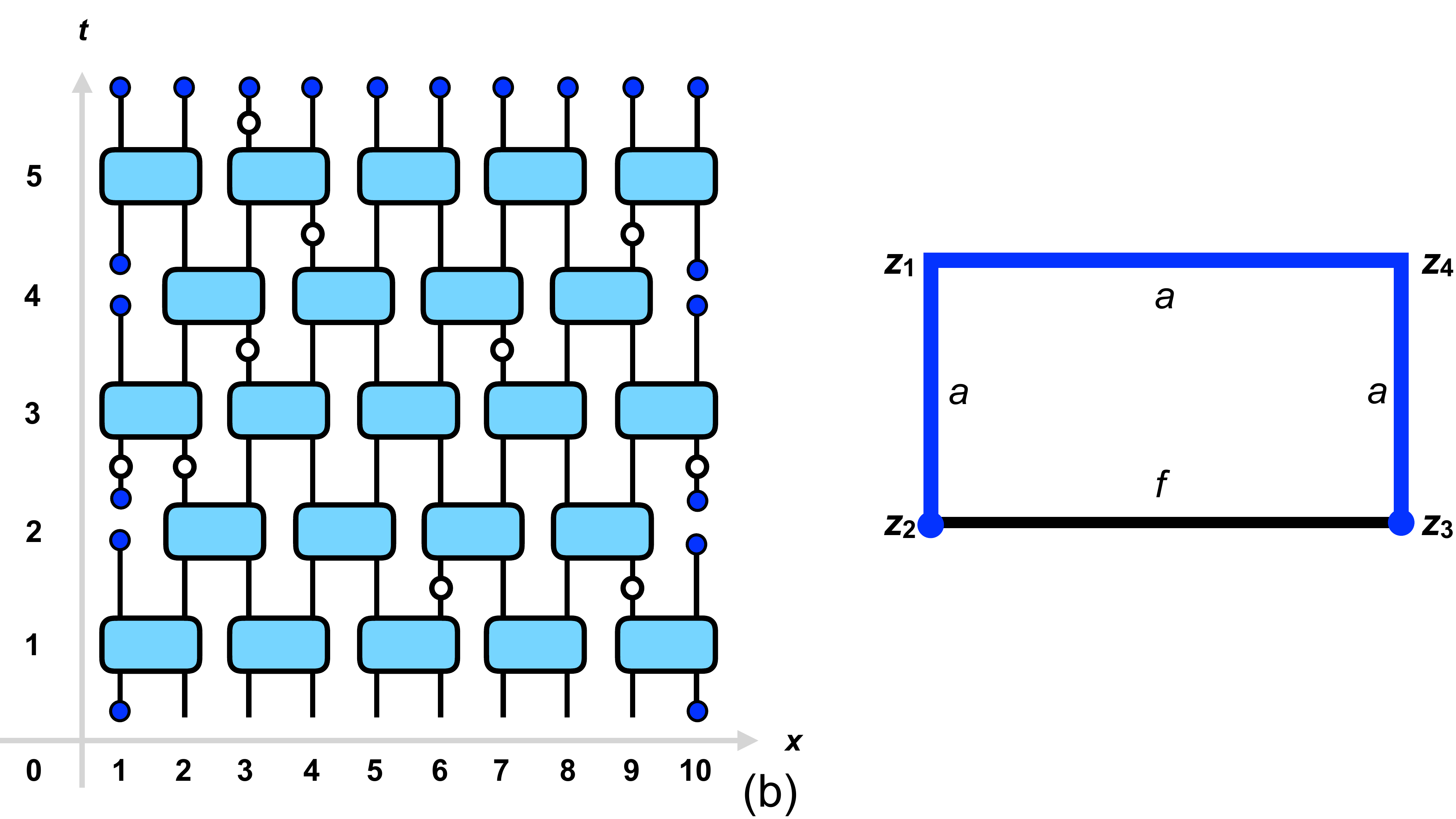}
\\
\includegraphics[width=.45\textwidth]{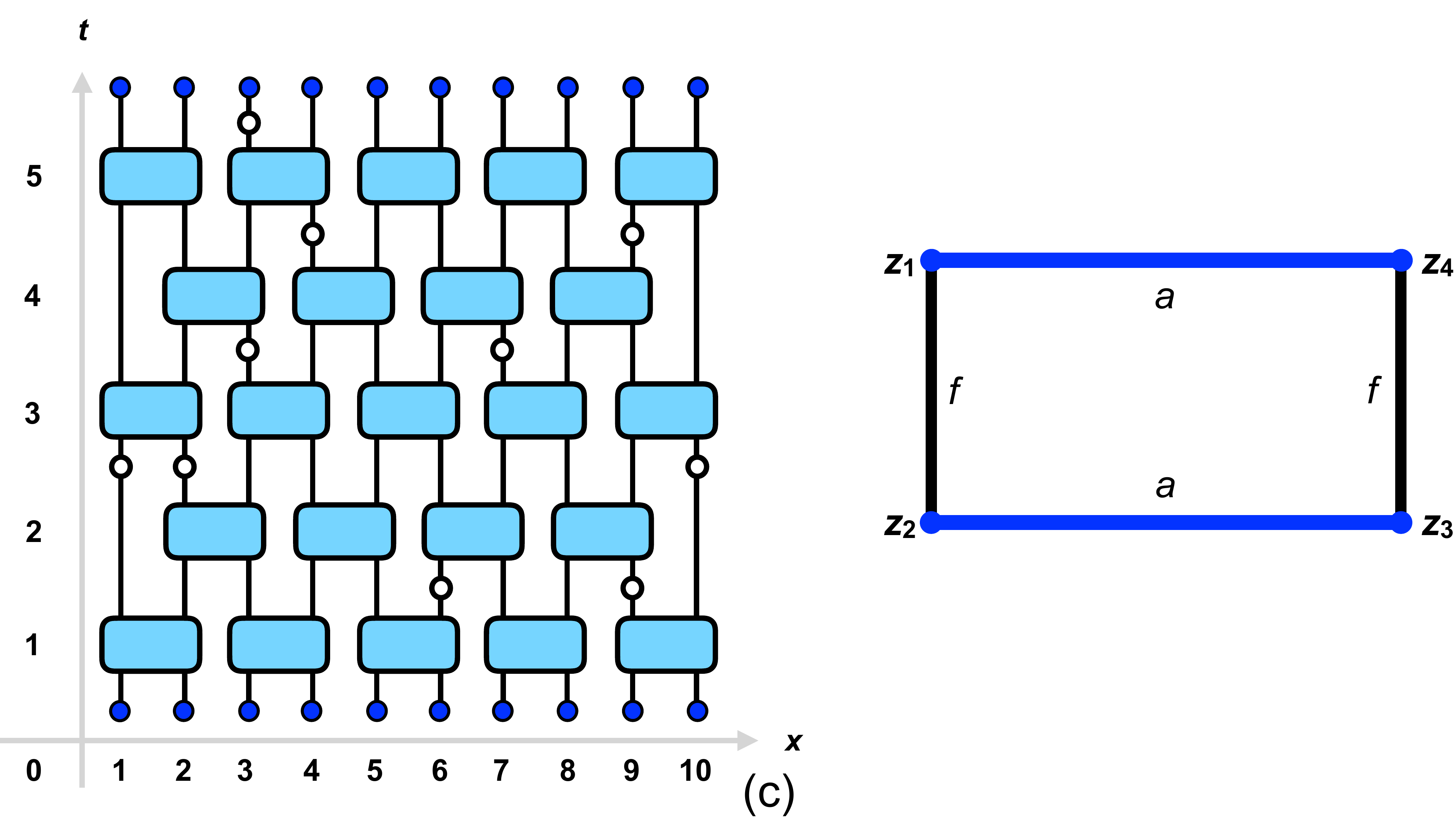}
\hfill
\includegraphics[width=.45\textwidth]{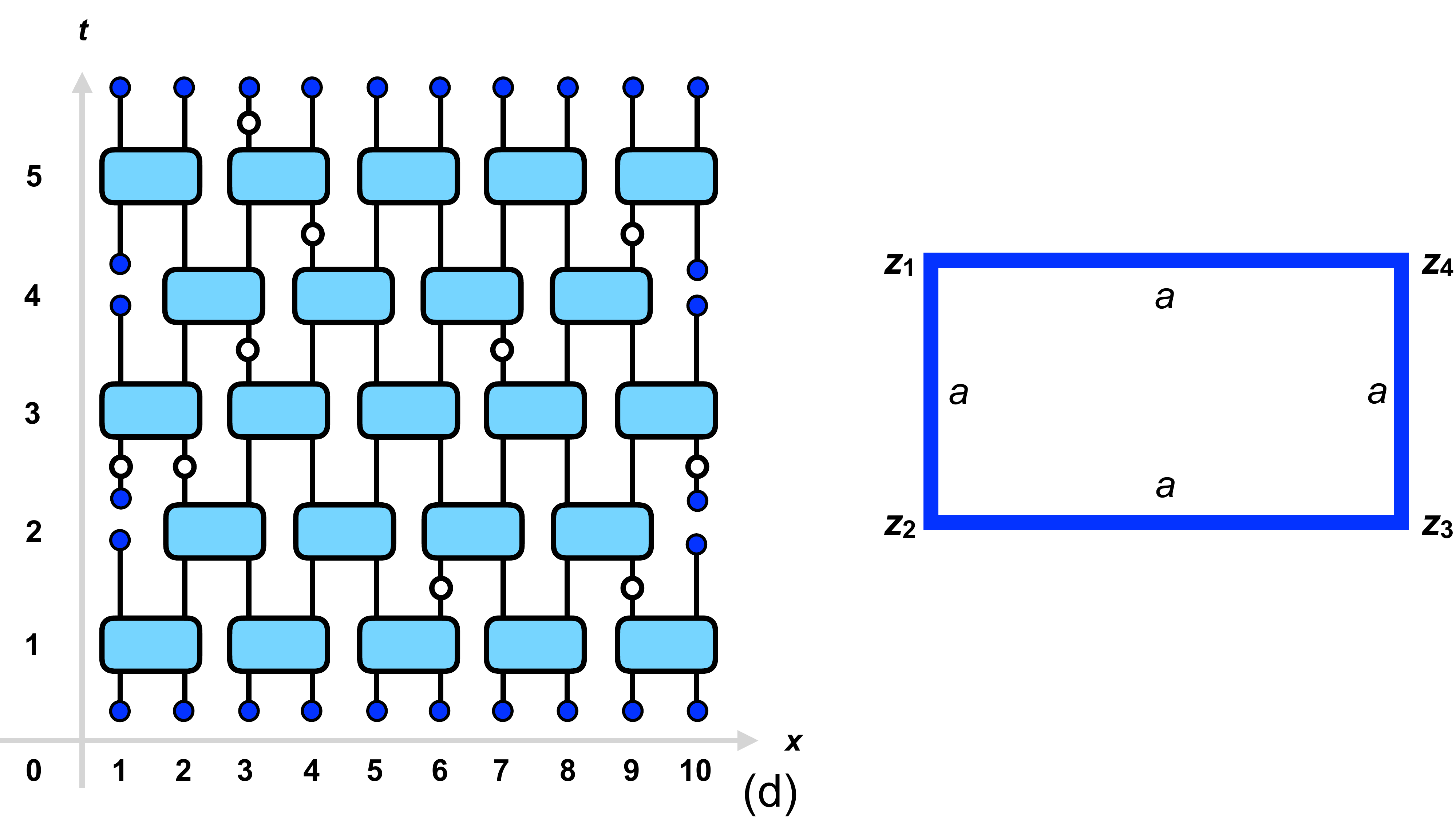}
\caption{
Random hybrid Clifford circuit model with different boundary conditions.
The rectangles in cyan represent random Clifford unitary gates, arranged in a brickwork fashion.
Between the unitary layers are projective measurements of single-site Pauli operators made at random sites at probability $p = p_c$, represented by hollowed circles.
The blue solid circles at the upper boundary represent physical qubits after evolution of circuit depth $T$.
Notice that time runs ``upwards''.
In (a), we illustrate the simplist b.c. of all, with a trivial product state and open spatial b.c.
These two are assumed to correspond to the same ``free b.c.'', denoted $\bc{f}$ and represented with black color.
The blue edge represents a ``physical qubit b.c.'', denoted $\bc{a}$.
The two boundary conditions are separated by boundary condition changing (bcc) operators at the corners, denoted $z_1$ and $z_4$.
In (b), we ``insert'' initially unentangled physical qubits at the left and right edges of the circuit at every time period, so that we have the $\bc{a}$ b.c. on three edges of the rectangle, with the other one still in $\bc{f}$.
In (c), we take the initial state of $L$ Bell pairs, and take one qubit from each pair to form a qubit chain (the system) which undergoes the circuit dynamics, leaving the other qubit chain untouched (the environment).
We put the environment {\cyan and the system} on the $t=0$ and $t=T$ boundaries, respectively, and both in the $\bc{a}$ b.c.
In (d) we combine the initial state in (c) and the ``temporal insertion'' setup, to obtain a circuit with $\bc{a}$ on all four sides.
We shall refer to the four sets of boundary conditions as (a)$\bc{fffa}$, (b)$\bc{afaa}$, (c)$\bc{fafa}$, and (d)$\bc{aaaa}$, respectively.
}
\label{fig:upc_rect}
}

\env{table*}{[t]
\centering
\begin{tabular}{c|c|c|c}
\hline
 bcc operator & Definition & Scaling dimension & Reference \\
\hline \hline
 $\phi_{\bc{f|a}},\ \phi_{\bc{f|b}}$
 &
    \includegraphics[height=.36in]{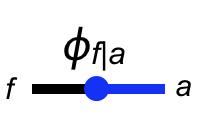}
    $\quad , \quad$
    \includegraphics[height=.36in]{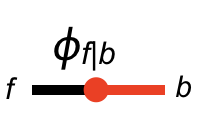}
 &
 $h_\bc{f|a} = h_\bc{f|b}$ is unknown
 &
 Eqs.~(\ref{eq:Z_background}, \ref{eq:corr_fab}, \ref{eq:corr_fb}, \ref{eq:OPE_phi_psi_sub}).
 \\ \hline
 $\phi_{\bc{a|b}}$
 &
    \hspace{-.1in}
    \includegraphics[height=.30in]{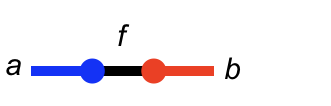}
    $\sim \quad$
    \includegraphics[height=.36in]{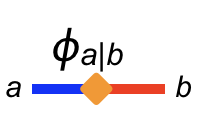}
 &
 $h_{\bc{a|b}} = 0.76 \ln(2) = 0.53$
 &
 Eqs.~(\ref{eq:corr_fab}, \ref{eq:EE_3pt_fun}, \ref{eq:OPE_phi_phi_ab}); Figs.~(\ref{fig:res_fffa}, \ref{fig:res_afaa}, \ref{fig:res_fafb}, \ref{fig:res_aaaa}, \ref{fig:res_pbc}, \ref{fig:refQ}); Refs.~\cite{li1901hybrid, gullans1905purification}
 \\ \hline
 $\phi^{(1)}_{\bc{f|a}}$
 &
    \includegraphics[height=.30in]{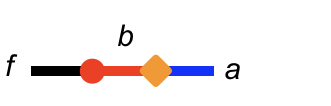} $\sim \quad \phi_{\bc{f|a}} + \phi^{(1)}_{\bc{f|a}}$
 &
 $h_\bc{f|a}^{(1)} = h_\bc{f|a} + 0.9$
 &
 Eq.~(\ref{eq:OPE_phi_psi_sub}); Figs.~(\ref{fig:res_fffa}, \ref{fig:res_afaa}).
 \\ \hline
 $\phi_\bc{f|f}^{(1)}$
 &
    \includegraphics[height=.30in]{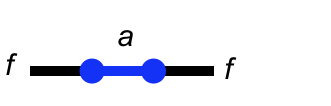} $\sim \quad 1_{\bc{f|f}} + \phi_{\bc{f|f}}^{(1)}$
 &
 $h_\bc{f|f}^{(1)} = 0.41$
 &
 Eq.~(\ref{eq:ope_phi_phi_sub}); 
 Figs.~(\ref{fig:res_fafb}, \ref{fig:refQ}, \ref{fig:refQ_power_law}, \ref{fig:localizable_MI}).
 \\ \hline
 $\phi_\bc{a|a}^{(1)}$
 &
    \includegraphics[height=.30in]{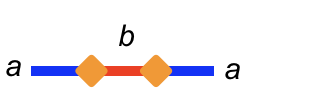} $\sim \quad 1_{\bc{a|a}} + \phi_{\bc{a|a}}^{(1)}$
 &
 $h_{\bc{a|a}}^{(1)} = 2.0$
 &
 Eq.~(\ref{eq:OPE_psi_psi}, \ref{eq:OPE_phi_phi_aa}); Fig.~(\ref{fig:res_aaaa}); Refs.~\cite{li1901hybrid}.
 \\ \hline
 -
 &
 -
 &
 $\mathit{x}_{\rm p.b.c} = 0.125$
 &
 Eq.~\eqref{eq:def_x_pbc}; Fig.~(\ref{fig:res_pbc}).
 \\ \hline
\end{tabular}
\caption{
A summary of boundary conditions (b.c.), boundary condition changing (bcc) operators, and their operator product expansions (OPE) that will appear later in this paper. 
There are three types of b.c., namely (1) $\bc{f}$, corresponding to product initial state and open spatial b.c. of the circuit; (2) $\bc{a}$, corresponding to physical qubits; and (3) $\bc{b}$, corresponding to qubits for which the entanglement entropy is computed.
Exchange symmetry between $\bc{a}$ and $\bc{b}$ is assumed.
The fundamental bcc operator is the one separating $\bc{f}$ and $\bc{a}$, which we denote as $\phi_{\bc{f|a}}$ (or its symmetric counterpart $\phi_{\bc{f|b}}$).
The OPE between $\phi_{\bc{a|f}}$ and $\phi_{\bc{f|b}}$ gives rise to a bcc operator separating $\bc{a}$ and $\bc{b}$, and we define $\phi_{\bc{a|b}}$ to be the leading term with smallest scaling dimension.
{These two operators $\phi_{\bc{f|a}}$ and $\phi_{\bc{a|b}}$ are assumed to 
{transform}
as primary fields under conformal transformations~\cite{BPZ1984}.}
We further define $\phi_{\bc{f|a}}^{(1)}$, $\phi_{\bc{f|f}}^{(1)}$, and $\phi_{\bc{a|a}}^{(1)}$ as the \textit{subleading} operators in the corresponding OPE channels.
{In these OPEs}
we have suppressed prefactors and only kept the operator content; the full form will be provided when they are encountered (see also  Appendix~\ref{sec:cft_calc}).
We summarize scaling dimensions extracted for these operators and their appearance in this paper, which we refer to for more detailed explanations.
Notice that we are unable to extract the scaling dimension for $\phi_{\bc{f|a}}$ since it does not explicitly appear in the entanglement entropy calculation.
{$x_{\rm p.b.c.}$ appearing at the bottom is not associated to scaling dimensions of bcc operators; rather, it is a universal scaling exponent of the bulk CFT (see Sec.~\ref{sec:pbc}).}
}
\label{table:bcc}
}

\subsection{The hybrid circuit models with different boundary conditions \label{model}}

Amongst various versions of the hybrid quantum circuit model~\cite{nahum2018hybrid, li1808hybrid, li1901hybrid, choi2019qec, gullans1905purification, choi2019spin, andreas2019hybrid}, we take the one with random Clifford unitaries on {\cyan pairs of} qubits (with local Hilbert space dimension $q = 2$) and projective measurements of single-site Pauli operators made in a Poissonian fashion with probability $p$, which was introduced in Ref.~\cite{li1901hybrid} and referred to as ``the random Clifford circuit''.
We focus on the critical point of the entanglement transition, taking $p = p_c \approx 0.1600$ in this particular model~\cite{li1901hybrid, gullans1905purification} (see Sec.~\ref{sec:pbc} for the location of the transition).

The circuit model is always defined together with its boundary conditions (b.c.){, which we take, for the most part of this paper, to
be open spatial boundary conditions}.
In Fig.~\ref{fig:upc_rect}, we illustrate the circuit model with the
{corresponding space-time}
geometry of a finite rectangle, with length $L$ (measured in
{terms of}
the number of qubits) and depth $T$ (measured in 
{terms of the}
number
of unitary layers), 
{where we define}
4 sets of different b.c. on its edges.
In order to introduce the circuit models, we have to make several postulates in assigning the
{boundary conditions; in this section we neither explain
the physical meanings of these boundary conditions, nor provide 
justifications of our assumptions. We}
postpone these issues to later sections: Sec.~\ref{sec:conjecture}, Sec.~\ref{sec:res_rect} and Appendix~\ref{sec:perc}.
{We proceed by listing the four sets of boundary 
conditions that we consider:}
\env{enumerate}{[(a)]
\item
The simplest of all is the one with a product initial state and open spatial b.c.
{at the right and left boundaries of the rectangle}
(Fig.~\ref{fig:upc_rect}(a) 
{-- time goes ``upwards''}).
We posit that these two map to the same b.c. (in the sense described in Sec.~\ref{sec:conjecture}), which we refer to as the ``free b.c.'', denoted $\bc{f}$.
We further posit that the physical qubits at the 
{boundary representing the quantum state at final time $t=T$}
map to a different b.c., which we refer to as the ``physical qubit'' b.c., denoted $\bc{a}$.

Since the b.c. change from $\bc{f}$ to $\bc{a}$ at
{the corners denoted by}
$z_1$ and $z_4$
{in Fig.~\ref{fig:upc_rect}(a)}, we say that there are
(analogous to Ref.~\cite{vasseur2018rtn, andreas2019hybrid})
boundary condition changing (bcc) operators $\phi_{\bc{f|a}}(z_1)$ and $\phi_{\bc{a|f}}(z_4)$ {located at these corners}.
The meaning of the bcc operators will be specified in Sec.~\ref{sec:conjecture}.

As a result, we have a circuit with 
{boundaries labeled by the sequence of boundary conditions
$\bc{fffa}$ in counter-clockwise order (starting from the left boundary of the
rectangle).}

\item
In the 2nd case, we introduce physical qubits at the left and right edges of the rectangle in the following manner (see Fig.~\ref{fig:upc_rect}(b)).
We retain $L-2$ qubits sitting at positions $x = 2, \ldots, L-1$ of the chain, and at each time step, we introduce two ``fresh'' qubits, each initially in a disentangled 1-qubit pure state (the specific state is unimportant), and ``inject'' them into the system as the 1st and the $L$-th qubit of the circuit.
The $L$-qubit chain is then evolved under the circuit dynamics for one time period (notice that one time step corresponds to two consecutive unitary layers).
After that period, we take out the 1st and the $L$-th qubit, keep them somewhere else without further actions on them, and fill their positions in the chain with two new fresh qubits in the next time period.
For a circuit 
{of depth $T$ (with $T$ even),}
by the end of its evolution, the left and right edges will each have $T/2$ qubits, 
{namely those ``fresh'' qubits that have been ``injected''
on the right and left edges,}
in addition to the $L-2$
qubits at the {final time $t = T$ (the upper edge of the rectangle), taking the same position as qubits in the previous setup (a)}.
We posit that they map to the same b.c. $\bc{a}$,
{as that discussed in the previous setup (a)}.
{As compared with Fig.~\ref{fig:upc_rect}(a),
we now have
eliminated the bcc operators at the corners denoted by
$z_1$ and $z_4$, at the cost of introducing new bcc operators
$\phi_{\bc{f|a}}(z_2)$ and $\phi_{\bc{a|f}}(z_3)$
at the corners denoted by $z_2$ and $z_3$.}
{
By the same convention {as above}, we refer to this b.c. as $\bc{afaa}$.}

\item
In the 3rd case, we take an initial state composed of $L$ pairs of maximally entangled qubits (i.e. Bell pairs), where different pairs are unentangled with each other (as required by monogamy of entanglement).
Taking one qubit from each pair, we form an $L$-qubit chain (which we call the ``system''), and the rest form another $L$-qubit chain (which we call the ``environment'').
We let the ``system chain'' 
{undergo}
the circuit dynamics of depth $T$, while the ``environment chain'' is left unevolved.
By the end of the evolution, we naturally have the ``system'' at the upper edge of the rectangle,
and we assume that the ``environment'' ``lives'' on the lower edge (in a sense to be specified in Sec.~\ref{sec:conjecture}).
We further posit that the upper and lower edges are described
{again}
by the same b.c. $\bc{a}${, discussed in the previous two setups (a) and (b)}, as shown in Fig.~\ref{fig:upc_rect}(c).

In this setup, there are bcc operators 
{at}
all 4 corners to start with: $\phi_{\bc{f|a}}(z_1)$, $\phi_{\bc{a|f}}(z_2)$, $\phi_{\bc{f|a}}(z_3)$, $\phi_{\bc{a|f}}(z_4)$.
We refer to this b.c. as $\bc{fafa}$.

\item
In the 4th case, we combine the b.c. in (b) and (c) so that we have physical qubits on all four edges.
Specifically, we take the initial state as described in (c), and while evolving the ``system'',
{we}
inject physical qubits at each time step as in (b).
The physical qubits on all four edges are assumed to correspond to the same b.c., $\bc{a}$, as shown in Fig.~\ref{fig:upc_rect}(d).

In this setup, we do not have any bcc operators at any of the corners (since the b.c. do not change).
We refer to this b.c. as $\bc{aaaa}$.
}

As clarified above, at this point issues like the ``labelling'' of the boundary conditions (with $\bc{f}$ or $\bc{a}$) and ``where the physical qubits sit on the rectangle'' are meaningless until certain observables are assigned to them.
{As we will see next in Sec.~\ref{sec:conjecture}, the boundary conditions are important in defining 
{boundary}
free energies within the putative conformal field theory.}

\subsection{Statement of the conjecture and example calculations of entanglement entropy \label{sec:conjecture}}

Previous works 
{on the measurement-induced entanglement transition}
are quite suggestive of the 
{presence}
of full conformal invariance in spacetime, though the models considered differ from one another in details.
Among these are Ref.~\cite{nahum2018hybrid}, where the critical percolation description
{of the $0$th (Hartley) R\'enyi entropy}
{in circuits with random Haar gates}
was already manifestly conformally invariant;
Refs.~\cite{li1808hybrid, gullans1905purification}, where a dynamic exponent of $z=1$ was found;
and Ref.~\cite{li1901hybrid}, where the presence of conformal invariance in the steady state was numerically confirmed{, all for Clifford circuits}.
More recently in Refs.~\cite{choi2019spin, andreas2019hybrid}, concrete critical spin models which admit conformal field theory (CFT) descriptions at their critical points were proposed to describe
{the $n$th R\'enyi entropies with $n\geq 1$ in}
{hybrid quantum circuits with Haar random unitaries in the limit of infinite local Hilbert space dimension.}

Motivated by these considerations, we propose the following conjecture(s) at entanglement transitions in generic hybrid quantum circuits:
\env{enumerate}{
\item
There is an emergent CFT living on the two-dimensional finite spacetime manifold of the circuit (with certain spatial and temporal b.c.), where the real-time direction of the circuit becomes the ``imaginary time'' of the CFT.
\item
Physical qubits live on boundaries of the finite circuit, and the von Neumann entanglement entropy\footnote{Throughout the paper we consider Clifford circuits, for which all Renyi entropies are equal to the von Neumann entropy.}
of a contiguous segment $A$ of qubits is given by the change in 
{(boundary)}
free energy of the CFT in the finite geometry due to change of the b.c. inside $A$ (recall that free energies of a CFT depend crucially on the 
{specific b.c.\footnote{See, e.g., Ref.~\cite{cardy0411bcft} for a review.}.)}
}
Specifically, for
{a contiguous segment $A$ of the boundary of the rectangle with endpoints located at $z_1$ and $z_2$, which we denote by
$A=[z_1, z_2]$,}
we posit that
\env{eqnarray}{
\label{eq:EE_Z}
	S([z_1, z_2]) \equiv -\ln \frac{ Z_{\rm circuit}[{\phi(z_1) \phi(z_2)}] } {Z_{\rm circuit}},
}
where $Z_{\rm circuit}$ 
is a suitably defined 
{background}
``circuit partition function'' of the rectangle specified by boundary conditions of the circuit,
and
$Z_{\rm circuit}[{\phi(z_1) \phi(z_2)}]$ is the partition function with the
same boundary conditions as $Z_{\rm circuit}$, except that in the boundary 
segment $A=[z_1, z_2]$ the boundary condition has changed as compared
to $Z_{\rm circuit}$, which in a CFT can be accounted for by the insertion of 
{
boundary condition changing (bcc)
}
operators $\phi$ at the endpoints $z_1$ and $z_2$ of $A$.
{An expression similar to Eq.~\eqref{eq:EE_Z} first appeared in the extreme volume-law phase of Random Tensor Networks aimed at 
{describing}
gravitational Ryu-Takayanagi behavior~\cite{Hayden2016}, 
then in Random Tensor Network Models for entanglement transitions~\cite{vasseur2018rtn} {which 
are~\cite{andreas2019hybrid}
very close cousins of the entanglement transitions
in hybrid circuits discussed here,}
and shortly after in the present context of measurement-driven
entanglement transitions~\cite{li1901hybrid, choi2019spin, andreas2019hybrid}.}

We remark on 
{an apparent}
conceptual leap
{on which we briefly elaborate at the end of this paragraph:}
{While}
previously in Fig.~\ref{fig:upc_rect} the bcc operators $\phi_\bc{f|a}$ are merely placeholders to signify the change of boundary condition, in a CFT they become scaling fields that {define} the partition function;
we further assume that these fields are 
{what is called} \textit{primary}~\cite{BPZ1984}.
{These boundary scaling fields}
are the central objects of this paper, and govern the entanglement structure of the circuit through Eq.~\eqref{eq:EE_Z}.

{The expression 
Eq.~\eqref{eq:EE_Z} can be
obtained directly 
by repeating the steps 
presented in Ref.~\cite{andreas2019hybrid},
but now for the reduced density
matrix for the {random} \textit{Clifford} circuit with measurements, upon making the
only assumption that
an effective 
{statistical mechanical}
model emerges after averaging, which exhibits a conformally invariant transition in the bulk of the circuit.\footnote{Hybrid circuits with periodic, non-random (Floquet) unitaries and/or (quasi-)periodically located measurements in space and time also appear to exhibit an entanglement transition in numerics~\cite{li1901hybrid}.
Provided these are also conformal transitions, with which the numerical evidences appear to be consistent, general assumptions of this paper also apply, although a statistical mechanical model cannot be readily obtained along the lines outlined in Ref.~\cite{andreas2019hybrid}.}
We provide in this 
paper extensive evidence for the validity of this assumption
for Clifford circuits. All the remaining assumptions
made in this paper about the appearance of boundary condition changing operators follow from general properties of CFT.
In particular, any microscopic boundary condition (satisfying certain locality conditions) on a CFT 
will at long distance scales in general always turn into
a ``conformal boundary condition'' described by a (boundary) 
fixed point of the Renormalization Group.
Moreover, at a point on the boundary where two different such ``conformal boundary conditions''
meet, a boundary condition changing conformal boundary operator will appear.}
For the convenience of 
{subsequent}
discussions
{in this paper}, we summarize in Table~\ref{table:bcc} all relevant boundary conditions, bcc operators, and their operator product expansion (OPE), that will appear in later sections.

We illustrate the prescription in Eq.~\eqref{eq:EE_Z} with the $\bc{fffa}$ circuit
in Fig.~\ref{fig:upc_rect}(a),
which we choose, in the present case, to represent the ``background'' configuration of boundaries.
Because of the two bcc operators at the corners $\phi_{\bc{f|a}}(z_1)$ and $\phi_{\bc{a|f}}(z_4)$ (as defined in Table~\ref{table:bcc}), the circuit partition function
is given as
(see Fig.~\ref{fig:bc_fffa}(a)) 
\env{eqnarray}{
    \label{eq:Z_background}
    Z_{\rm circuit} = \avg{\phi_{\bc{f|a}}(z_1) \phi_{\bc{a|f}} (z_4)} Z_0,
}
where $\avg{\cdots}$ denotes the ``expectation value''
taken in an underlying $(2+0)$-dimensional CFT in the bulk of the rectangle, which
can be thought of as some suitable classical statistical mechanics system
representing the CFT;
and $Z_0$ is the partition function of this CFT living in a rectangle with free boundary condition $\bc{f}$ on all four sides.

\env{figure}{[t]
\includegraphics[width=.235\textwidth]{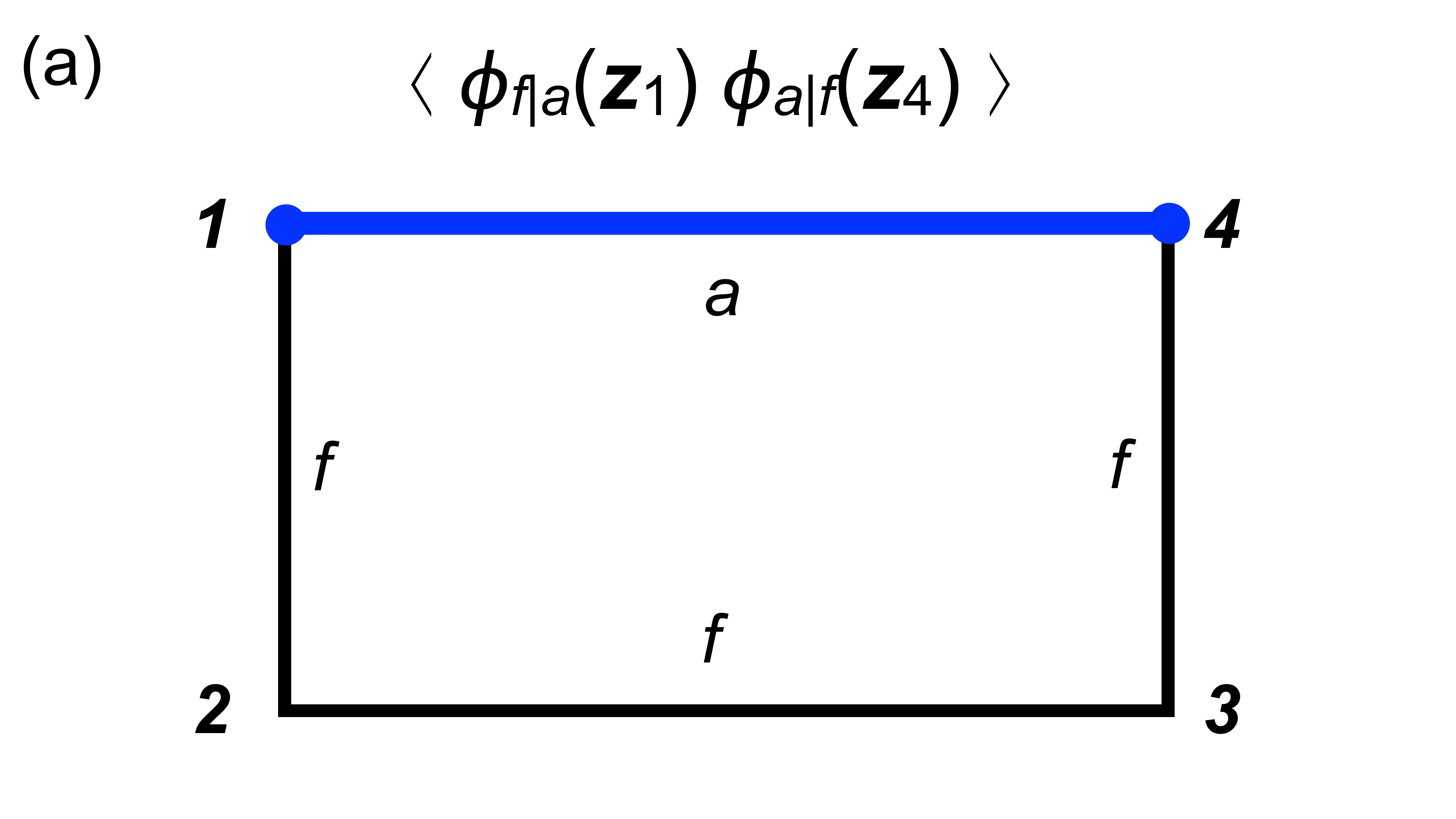}
\includegraphics[width=.235\textwidth]{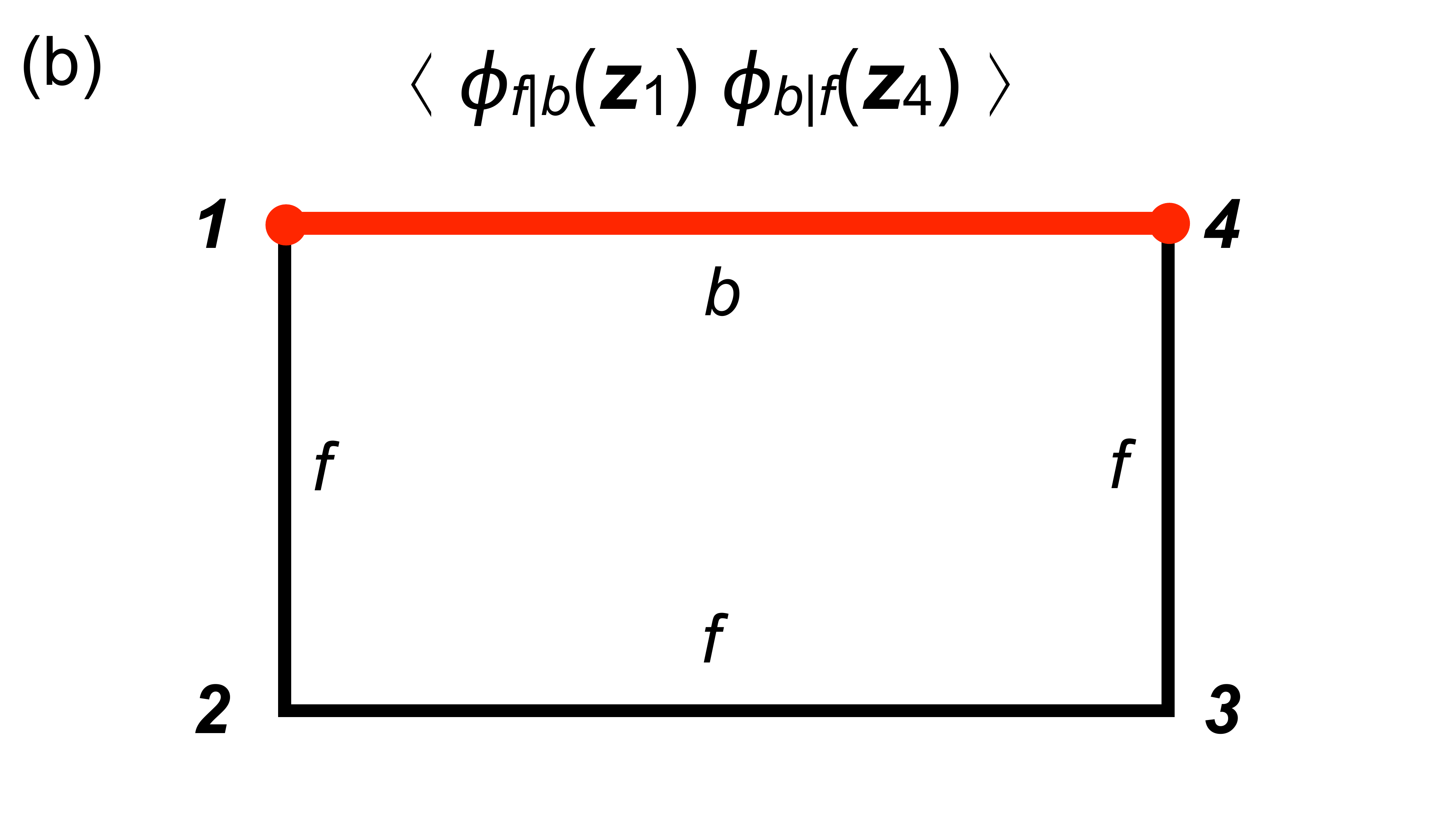}
\includegraphics[width=.235\textwidth]{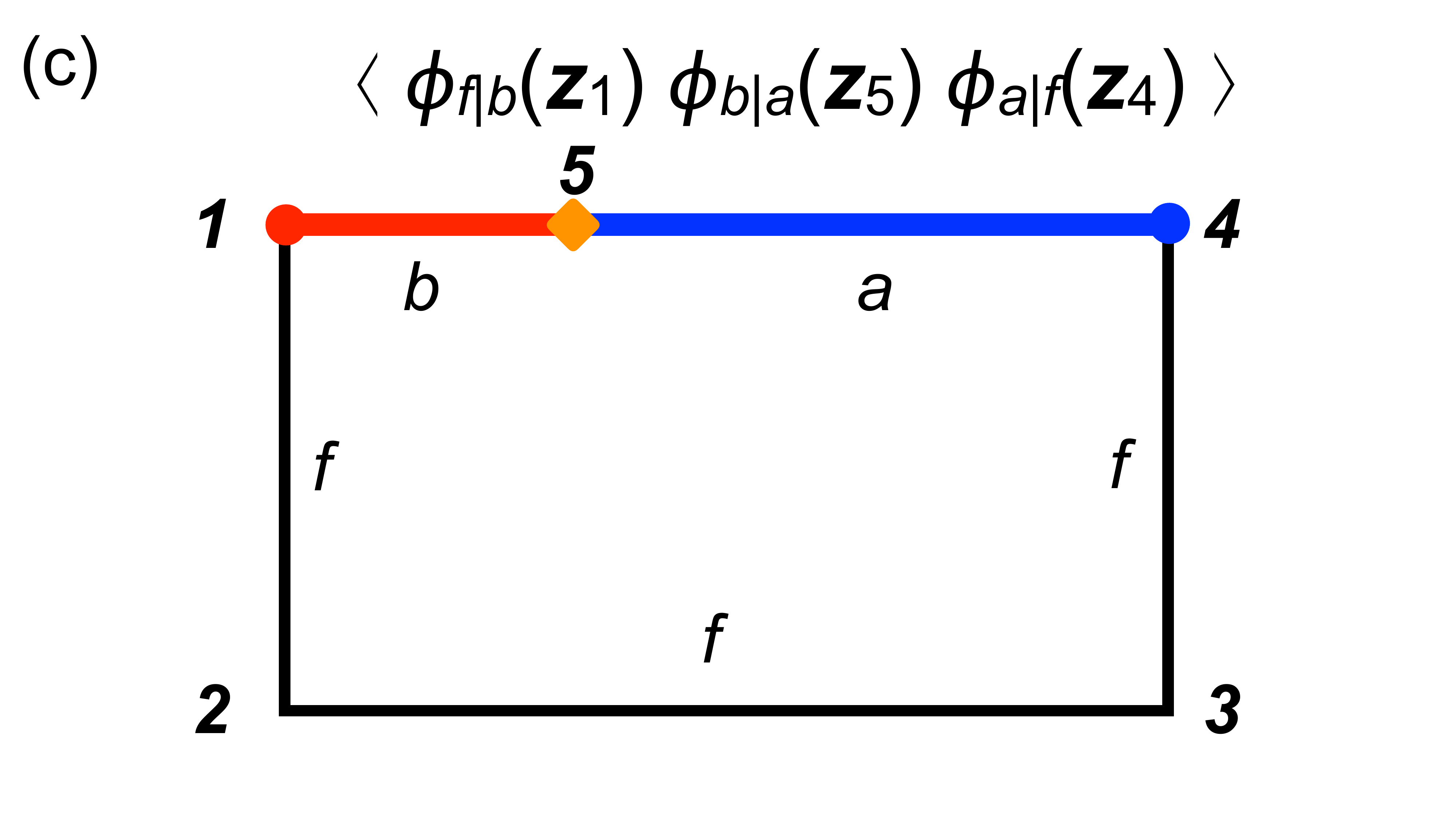}
\includegraphics[width=.235\textwidth]{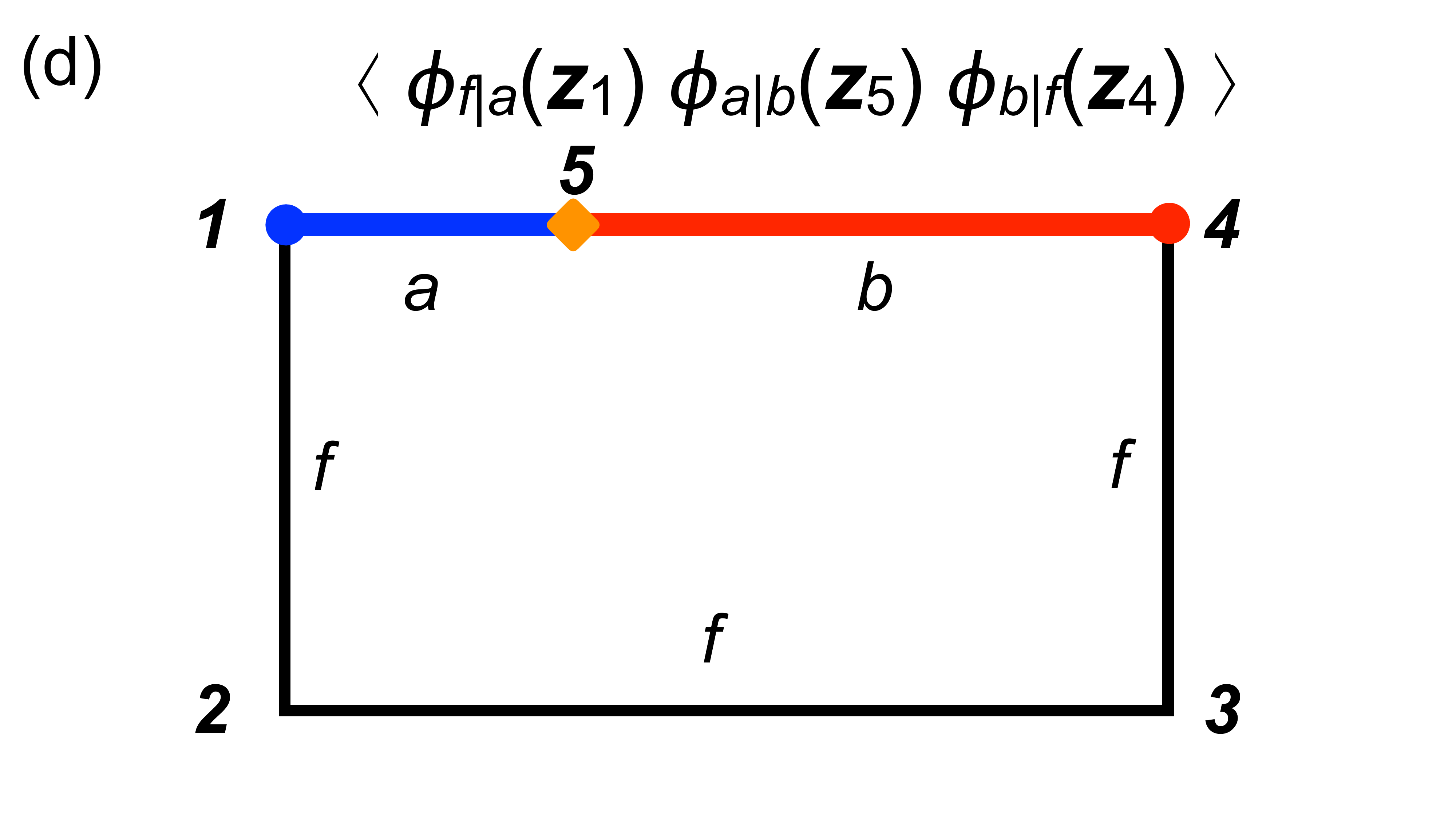}
\caption{Pictorial representations of the parition functions with bcc operators inserted at the corner and on the edge, for computations of bipartite entanglement entropies in the $\bc{fffa}$ circuit shown in Fig.~\ref{fig:upc_rect}(a).
(a) The ``background'' partition function, given by correlation function of bcc operators at $z_1$ and $z_4$ separating $\bc{f}$ and $\bc{a}$.
(b) The partition function corresponding to computation of entanglement entropy of the whole qubit chain. Since the entire system is in a pure state, the entanglement entropy should be $0$, as realized by the exchange symmetry between $\bc{a}$ and $\bc{b}$ (see Eq.~\eqref{eq:corr_fb}).
(c, d) The partition functions corresponding to the calculation of $S(A = [z_1, z_5]) = S(\ovl{A} = [z_5, z_4])$ (see Eq.~\eqref{eq:corr_fab}).
}
\label{fig:bc_fffa}
}

Next, let us consider the entanglement entropy of a contiguous segment $A$ of physical qubits within $[z_1, z_4]$. 
According to the conjecture, $S(A)$ is the change in free energy due to 
{change of}
b.c. in $A$
{from $\bc{a}$}
to yet another one, 
{denoted by $\bc{b}$, which is assumed to be of the same type
as $\bc{a}$, but different (we will be more specific below).}
Such effects are accounted for by inserting bcc operators at the endpoints of $A$,
separating boundary conditions $\bc{a}$ (outside $A$) and $\bc{b}$ (inside $A$).
We denote such an operator $\phi_{\bc{a|b}}$ (see Table~\ref{table:bcc} for its definition). 

In the simple case when $A = [z_1, z_5]$
{as depicted in Fig.~\ref{fig:bc_fffa}(c)}, 
i.e. having one of its endpoint at the corner $z_1$ (therefore
{specifying}
a single bipartition
{of the top boundary of the rectangle}
at $z_5$),
the boundaries of the rectangle is labelled by three distinct boundary conditions: $\bc{a}$ in $\ovl{A} = [z_5, z_4]$, $\bc{b}$ in $A = [z_1, z_5]$, and $\bc{f}$ elsewhere, in counter-clockwise order. 
The corresponding partition function
should therefore be given by 
the correlation function of three bcc operators 
{located}
at $z_1, z_5$ and $z_4$.
Explicitly, following Eq.~\eqref{eq:EE_Z}, the bipartite entanglement entropy $S(A)$ 
{can be written as}
\footnote{
In fact, literally following Eq.~\eqref{eq:EE_Z}, this entropy is related to the correlation function involving the inserted two bcc operators at $z_1$ and $z_5$, in addition to the existing ones at the corners $z_1$ and $z_4$.
Thus we have a four-point correlation function:
\env{eqnarray}{
	&&  S(A = [z_1, z_5]) \nn
	&=& -\ln \frac{Z_{\rm circuit}[\phi_\bc{a|b}(z_1) \phi_\bc{b|a}(z_5)]}{Z_{\rm circuit}} \nn
	&=& -\ln \frac{\avg{\phi_{\bc{f|a}}(z_1) \phi_{\bc{a|b}}(z_1) \phi_{\bc{b|a}}(z_5) \phi_{\bc{a|f}} (z_4)}}{\avg{\phi_{\bc{f|a}}(z_1) \phi_{\bc{a|f}} (z_4)}}.\nonumber 
}
In going from this to Eq.~\eqref{eq:EE_Z_z1_z5}, 
we have implicitly invoked the following OPE (to leading order; see Table~\ref{table:bcc}),
\env{eqnarray}{
    \phi_\bc{f|a}(z_1) \phi_\bc{a|b}(z_1 + \epsilon) \sim \epsilon^{-h_\bc{a|b}} \phi_\bc{f|b}(z_1) + \ldots \nonumber
}
to account for the coincidence of the left endpoint of $A$ with the corner of the rectangle, therefore
effectively reducing the four-point function to a three-point function, as expected.
Despite its apparent complexity, the physical picture is intuitive: after $\phi_\bc{f|a}(z_1)$ and $\phi_\bc{a|b}(z_1 + \epsilon)$ have fused into $\phi_\bc{f|b}(z_1)$, there are only three ``colored segments'' on the boundary, and therefore the partition function is given by a simple three point function.
}
\env{eqnarray}{
    \label{eq:EE_Z_z1_z5}
	&&  S(A = [z_1, z_5]) \nn
	&=& -\ln \frac{Z_{\rm circuit}[\phi_\bc{a|b}(z_1) \phi_\bc{b|a}(z_5)]}{Z_{\rm circuit}} \nn
	&=& -\ln \frac{\avg{\phi_{\bc{f|b}}(z_1) \phi_{\bc{b|a}}(z_5) \phi_{\bc{a|f}} (z_4)}}{\avg{\phi_{\bc{f|a}}(z_1) \phi_{\bc{a|f}} (z_4)}}.
}

{For a pure wave function, the entanglement entropies satisfy $S(A)=S(\overline{A})$, where $\overline{A}$ is the complement of
{the segment $A$ on the upper boundary of the rectangle.}
This requires that the partition function is invariant under exchanging $\bc{a}$ and $\bc{b}$;
indeed, using Eq.~\eqref{eq:EE_Z_z1_z5}, we have (see Fig.~\ref{fig:bc_fffa}(c, d))
\env{eqnarray}{
	\label{eq:corr_fab}
	&&  S(A = [z_1, z_5]) \nn
	&=& -\ln \frac{\avg{\phi_{\bc{f|b}}(z_1) \phi_{\bc{b|a}}(z_5) \phi_{\bc{a|f}} (z_4)}}{\avg{\phi_{\bc{f|a}}(z_1) \phi_{\bc{a|f}} (z_4)}} \nn
	&=& -\ln \frac{\avg{\phi_{\bc{f|a}}(z_1) \phi_{\bc{a|b}}(z_5) \phi_{\bc{b|f}} (z_4)}}{\avg{\phi_{\bc{f|a}}(z_1) \phi_{\bc{a|f}} (z_4)}} \nn
	&=& S(\ovl{A} = [z_5, z_4]).
}


In the limit when $A$ includes all the physical qubits (
{-- i.e. when $z_5=z_4$,}
see Fig.~\ref{fig:bc_fffa}(b)),
\env{eqnarray}{
	\label{eq:corr_fb}
	S([z_1, z_4]) 
	= -\ln \frac{\avg{\phi_{\bc{f|b}}(z_1) \phi_{\bc{b|f}} (z_4)}}{\avg{\phi_{\bc{f|a}}(z_1) \phi_{\bc{a|f}} (z_4)}}
	= 0,
}
as expected for a pure state.}
{Again, we have used the exchange symmetry between $\bc{a}$ and $\bc{b}$.}
{These considerations illustrate more specifically the sense in which ``the boundary condition
$\bc{b}$ is of the same type
as $\bc{a}$, but different'', as mentioned above.}

Although for simplicity, we have only considered the $\bc{fffa}$ circuit and have only taken segment $A$ to start from either $z_1$ or $z_4$ in this calculation, straightforward generalizations can be made to other cases, as we will see in Sec.~\ref{sec:res_rect}.

Our approach here is 
{``experimental'',}
though reasonably motivated by {general principles}.
{For example, we notice}
that the exchange symmetry between $\bc{a}$ and $\bc{b}$ comes about naturally from the general requirement of purity of the wavefunction.
In Sec.~\ref{sec:res_rect}, we will provide numerical evidences for CFT calculations like Eq.~\eqref{eq:corr_fab}, 
{supporting our}
conjectures,
together with the prescription for computing entanglement entropies and the assignments of boundary conditions.

\env{figure*}{[t]
\includegraphics[width=0.8\textwidth]{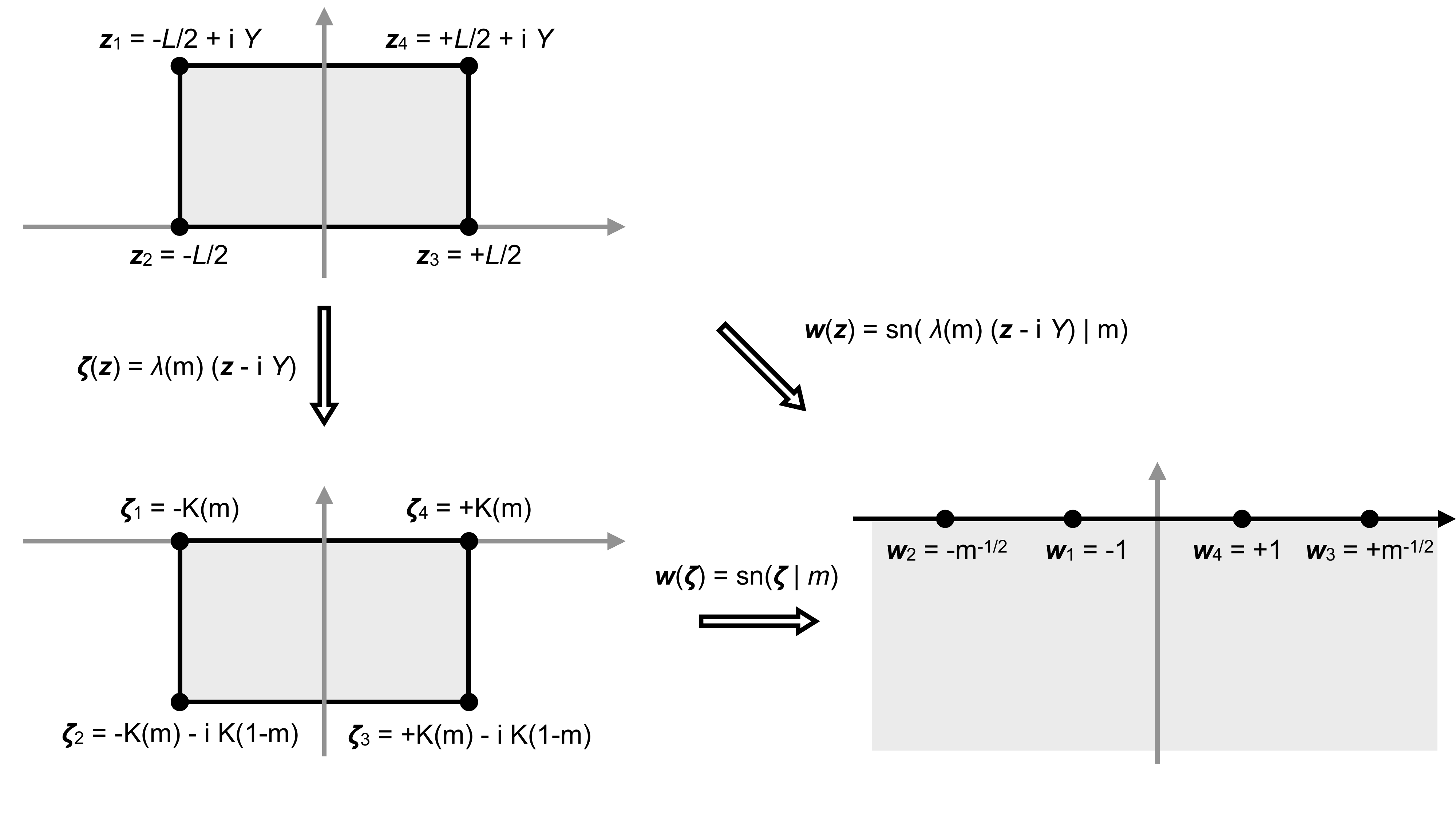}
\caption{The conformal mapping, 
from the finite rectangle to the LHP.
The parameter $m$ is chosen such that the aspect ratio match.
The boundary of the rectangle, highlighted, is mapped to the real axis of the LHP, where the 4 vertices of the rectangle map to $w_1 = -1, w_2 = -m^{-1/2}$, $w_3 = +m^{-1/2}$, and $w_4 = +1$, respectively.
}
\label{fig:cfm_sn}
}

\subsection{Finite rectangular geometry and the Schwarz-Christoffel mapping \label{sec:cfm}}

In most of this paper we will focus on
{systems as}
in Fig.~\ref{fig:upc_rect}, where the circuit manifold has the geometry of a finite rectangle
{(open boundary conditions)}; the case of cylindrical geometry 
{(periodic boundary conditions)}
is treated in Sec.~\ref{sec:pbc}.
{The former case}
is convenient because the rectangle is {simply connected}, and can thus be mapped to the lower half plane (LHP) via a conformal mapping (due to Schwarz-Christoffel)\footnote{See, e.g. Ref.~\cite{scmappingbook} for an introduction.}, allowing simple calculations of correlation functions in the rectangle (such as those in Fig.~\ref{fig:bc_fffa}), 
{due to their conformal covariance in the putative CFT (see Appendix~\ref{sec:cft_calc}).}
{Since all rectangles are conformally equivalent to the LHP, one can relate dynamics at different time scales via the conformal mapping, using the LHP as an intermediary.}
Similar ideas have been applied to crossing probabilities in two-dimensional critical percolation~\cite{cardy9111finite}.

We first address an important subtlety in mapping the circuit to a CFT in a finite rectangle.
In the circuit model, the physical qubits undergo real-time evolution, and there is no obvious space-time rotational symmetry; therefore, space and time are on separate footing,
and in particular, a circuit with $L=T$ does not necessarily correspond to a square system when viewed as a CFT.
We must therefore introduce a
{suitable}
``lattice spacing'' for both the space and time directions, $\lambda_x$ and $\lambda_t$, with $\lambda_x$ measured in
the number of qubits, and $\lambda_t$ in the number of layers.
The ``correct'' aspect ratio of the rectangular circuit when viewed as a CFT is therefore given by
\env{eqnarray}{
    \label{eq:DefinitionOfTau}
	\tau \coloneqq \frac{T/\lambda_t}{L/\lambda_x} 
	\equiv \frac{Y}{L},
}
where $Y = \(\frac{\lambda_x}{\lambda_t}\) T$ is the ``rescaled (imaginary) time'' or {\cyan ``depth''}. 
For the random Clifford circuit, 
we fix
{the ratio} $Y/T \approx 0.61$; the determination of this ratio is detailed in Sec.~\ref{sec:pbc}.
We emphasize that $Y/T$ is a bulk property and is independent of the boundary conditions.
The value $Y/T \approx 0.61$ is
{thus}
fixed for all boundary conditions of the random Clifford circuit considered in this paper.
However, $Y/T$ is non-universal and can vary from circuit to circuit; in particular, for the percolation problem that describes the zeroth R\'{e}nyi entropy in Haar random circuits, there is explicit rotational symmetry therefore $Y/T = 1$ (see Appendix~\ref{sec:perc}).


In the rest of this subsection, we detail the particular conformal mapping we use to relate the finite rectangle and the LHP, as summarized in Fig.~\ref{fig:cfm_sn}.
Points in the original rectangle are labeled by a complex coordinate,
\env{eqnarray}{
    z = x+iy,
}
where we take the convention $x \in [-L/2, L/2]$ for the position of the qubit, and $y = \frac{Y}{T} t \in [0, Y]$ the rescaled time coordinate.
As a first step, we perform a translation by $-iY$, followed by an overall scaling, to transform the $L \times Y$ rectangle (living in the complex 
{$z$-plane}) to the $2 K(m) \times K(1-m)$ ``canonical'' rectangle (living in the complex
{$\zeta$-plane}), where the overall scaling factor is 
\env{eqnarray}{
    \lambda(m) \  
   \coloneqq \ 
    2K(m)/L \overset{!}{=} K(1-m)/Y.
}
Here $K(m)$ is the complete elliptic integral of the 1st kind with parameter $m \in [0, 1]$, and $m$ is chosen such that aspect ratios match, 
\env{eqnarray}{
    \tau(m) \  
    \coloneqq \ 
    K(1-m) / 2 K(m) \overset{!}{=} Y/L.
}
It is only through this parameter $m$ that the aspect ratio (hence time) comes into the correlation functions.
We will take the convention that the four corners of the rectangle sit at~\cite{scmappingbook}
\env{eqnarray}{
	\zeta_1 &=& - K(m), \\
	\zeta_2 &=& - K(m) - i K(1-m), \\
	\zeta_3 &=& + K(m) - i K(1-m), \\
	\zeta_4 &=& + K(m).
}
In the second step, we map the canonical rectangle to the LHP via a Jacobi sn function~\cite{scmappingbook},
\env{eqnarray}{
    w(\zeta) = {\rm sn}(\zeta | m),
}
and we have
\env{eqnarray}{
	w_1 = w(\zeta_1) &=& - 1, \\
	w_2 = w(\zeta_2) &=& - m^{-1/2}, \\
	w_3 = w(\zeta_3) &=& + m^{-1/2}, \\
	w_4 = w(\zeta_4) &=& + 1.
}
Thus, the composition of these two maps, $z \to \zeta \to w$,
reads
\env{eqnarray}{
	w(z | \tau(m) = Y/L) = {\rm sn}(\lambda(m) (z-iY) | m).
}
It is useful to 
recall~\cite{scmappingbook}
the asymptotic forms of $\tau(m)$,
\env{eqnarray}{
	\label{eq:tau_asymt_form}
	\tau(m) \sim
	\env{cases}{
	\frac{\pi}{2} \(\ln \frac{16}{1-m}\)^{-1}, & \text{ as } m \to 1\ (\tau \to 0),\\
	\frac{1}{2\pi} \ln \frac{16}{m}, & \text{ as } m \to 0\ (\tau \to \infty),
	}
}
and also the asymptotic forms of the the cross ratio,
\env{eqnarray}{
	\eta = \frac{w_{12} w_{34}}{w_{13} w_{24}} \sim 
	\env{cases}{
		16 \exp(-\pi/\tau),   & \tau \to 0\nn
		1-16 \exp(-\pi \tau), & \tau \to \infty
	}
}
where $w_{ij} := w_i - w_j$ ($i,j = 1, \ldots, 4$).





\section{Results on rectangular circuits \label{sec:res_rect}}

In this section we present results of numerical simulation of the Clifford circuits defined in Sec.~\ref{sec:bcft}.
Unless otherwise noted, we will take the circuit with length $L = 512$ (measured in {the} number of qubits), and {\cyan varying the} 
depth {\cyan up to} $T = 1024$, (measured in
{the}
number of unitary layers).
The simulation uses the stabilizer formalism~\cite{gottesman9807heisenberg}, and follows the standard algorithm in Ref.~\cite{aaronson0406chp}.
The computation of entanglement entropies~\cite{Fattal2004stabilizer, hamma0406ground, hamma0409bipartite, Klappenecker2002stabilizer, Linden2013stabilizer} is done in the ``clipped gauge'', which is a particular choice of stabilizers where entanglement entropies can be efficiently computed~\cite{nahum2017KPZ, li1901hybrid}.
It is always implicit that the entanglement entropies and mutual information are computed for various subregions at each time step,
individually for each pure-state quantum trajectory, and then averaged over ensembles of trajectories.
Only a subset of data points for selected time windows are presented to avoid crowding; we have verified that other data points also collapse well onto the same curves.
The included time windows 
{range}
from early times $\tau \ll 1$ to late
times $\tau \gtrsim 1$.
Due to limited numerical precision of floating point numbers on a standard computer, we exclude from the plots data at extremely early time $\tau(m) \lesssim 0.03$, where $|1-m(\tau)| \le 10^{-16}$.
We do not think this is an important issue, but merely a technical nuisance we have yet to fully resolve.

We will always take $p_c = 0.1600$ and $Y/T = 0.61$ for all b.c., where $Y$ is the rescaled time (see Sec.~\ref{sec:cfm}).
The determination of these values are discussed in Sec.~\ref{sec:pbc}.

Some of the analytic calculations make use of standard results of simple correlation functions and 
Operator Product Expansions (OPEs)
in CFT, which are listed in Appendix~\ref{sec:cft_calc}.

Throughout the paper we compute the entanglement entropy by taking the natural logarithm on the reduced density matrix, following a convention adopted in Refs.~\cite{andreas2019hybrid, choi2019spin, Calabrese2004},
\env{eqnarray}{
    \label{eq:s_rho_ln_rho}
    S(A) \coloneqq -{\rm Tr} \rho_A \ln \rho_A.
}
We notice that this convention differs from that in Refs.~\cite{nahum2018hybrid, li1808hybrid, li1901hybrid, choi2019qec, gullans1905purification, gullans1910scalable},
where the base-2 logarithm is used.

\subsection{Circuit with boundary conditions $\bc{fffa}$
- Fig.~\ref{fig:upc_rect}(a) \label{sec:fffa}}

\subsubsection{Bipartite entanglement entropies as 3-point functions}
\label{sec:fffa-Subsection-1}
\env{figure}{[t]
\centering
\includegraphics[width=.5\textwidth]{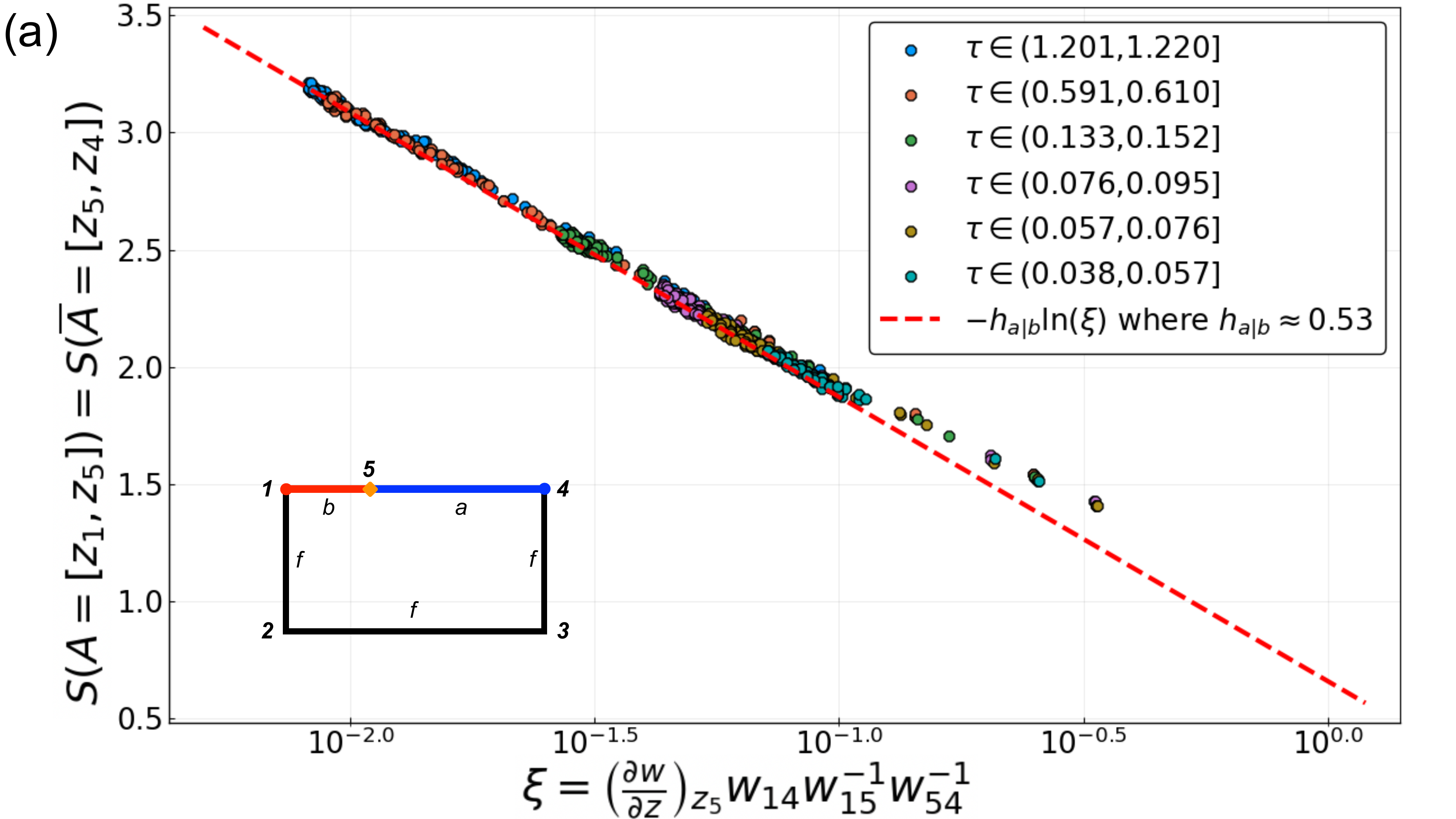}
\includegraphics[width=.5\textwidth]{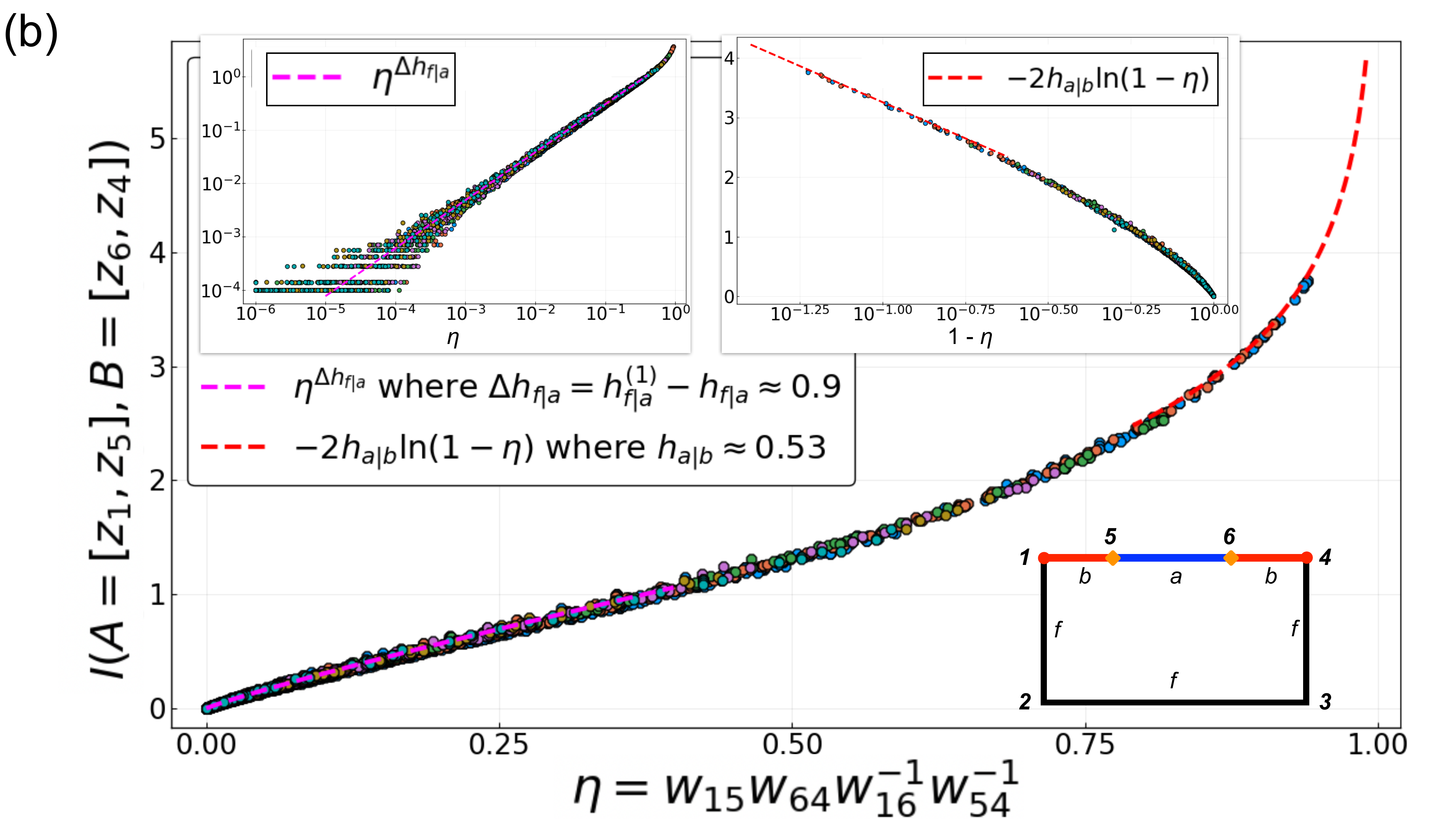}
\caption{
(a) Entanglement entropies for the $\bc{fffa}$ circuit, where the data collapse follows Eq.~\eqref{eq:EE_3pt_fun}.
{The apparent deviation of the data from the predicted form at larger values of $\xi$ is due to non-universal corrections when $z_1$ and $z_5$ are close on the lattice.}
(b) Mutual information for two subregions sitting next to the corners, where the data collapse follows Eq.~\eqref{eq:MI_4pt_fun}.
The limiting behaviors for $\eta \to 0$ and $\eta \to 1$ follow Eqs.~\eqref{eq:I_eta_0} and \eqref{eq:I_eta_1}, respectively, {\cyan and are shown in the insets}.
}
\label{fig:res_fffa}
}

Bipartite entanglement entropies within the $\bc{fffa}$ circuit (with a product initial state and open spatial b.c.; see Fig.~\ref{fig:upc_rect}(a)) 
{were}
already discussed as an example in Sec.~\ref{sec:bcft}.
This setup, as shown in Fig.~\ref{fig:bc_fffa}(c), has three bcc operators.
The simplicity of 3-point functions in CFT allows us to carry out the computation in Eq.~\eqref{eq:corr_fab} 
explicitly\footnote{\color{gr} A boundary operator $\phi_{\bc{f|b}}(z)$, if
initially  located at a position $z$ on a straight edge (say, top or side edge)
{\it away from} the corner,
is known to acquire, as it approaches the corner $z_1$, a powerlaw singularity in the distance $(z_1-z)$, because the scaling dimension of the operator is twice as large when placed at the 90-degree corner, as compared to at a straight edge. This singularity is a consequence of the conformal mapping. The same powerlaw singularity occurs in the denominator of the ratio appearing in the equation below and cancels out. The same type of cancelation occurs in all other ratios of correlation functions involving boundary operators located
directly at a corner that we consider in this paper.}
\env{eqnarray}{
&&  \exp \lz -S([z_1, z_5]) \rz 
\qquad\qquad \qquad \qquad \qquad
{\text{(Fig.~\ref{fig:res_fffa}(a))}} \nn
&=& \frac{\avg{\phi_{\bc{f|b}}(z_1) \phi_{\bc{b|a}}(z_5) \phi_{\bc{a|f}} (z_4)}}{\avg{\phi_{\bc{f|a}}(z_1) \phi_{\bc{a|f}} (z_4)}} \nn
&=& \(\frac{\pd w}{\pd z}\)_{z_5}^{h_{\bc{a|b}}}
\frac{\avg{\phi_{\bc{f|b}}(w_1) \phi_{\bc{b|a}}(w_5) \phi_{\bc{a|f}} (w_4)}}{\avg{\phi_{\bc{f|a}}(w_1) \phi_{\bc{a|f}} (w_4)}} \nn
&\propto& 
\( \frac{\(\frac{\pd w}{\pd z}\)_{z_5} w_{14}}{w_{15} w_{54} } \)^{h_{\bc{a|b}}}.
}
Thus,
\env{eqnarray}{
	\label{eq:EE_3pt_fun}
	S([z_1, z_5]) = - h_{\bc{a|b}} \ln \(\frac{\(\frac{\pd w}{\pd z}\)_{z_5} w_{14}}{w_{15} w_{54}}\) + {\rm const.}
}
The data collapse for $S([z_1, z_5])$ where $z_5 = x_5 + i Y$ with varying $x_5$ and $Y$ (that is, bipartite entanglement entropies for varying positions of the bipartition at different circuit depths) against $\xi = \frac{\(\frac{\pd w}{\pd z}\)_{z_5} w_{14}}{w_{15} w_{54}}$ is shown in Fig.~\ref{fig:res_fffa}(a).
Consistency with Eq.~\eqref{eq:EE_3pt_fun} is found, and we fit for $h_{\bc{a|b}} \approx 0.53$.\footnote{We note that $h_\bc{a|b} \approx 0.76 \ln 2$, where the value $0.76$ is consistent with Refs.~\cite{li1901hybrid, gullans1905purification}.
}

\subsubsection{Entanglement dynamics \label{sec:ee_dyn_fffa}}

The quality of the data collapse in Fig.~\ref{fig:res_fffa}(a) (together with Fig.~\ref{fig:res_fffa}(b); see below) lends strong support to our conjecture regarding the conformal invariance of the circuit,
together with our assumptions
about
the boundary conditions and the algorithm for computing the entanglement entropy.
Assuming these are indeed correct assumptions,
the 3-point functions, in turn, provide a complete description of the entanglement entropy dynamics {and mutual information dynamics, as we show in this subsection and the next.}
For example,
{as we will now show,}
Eq.~\eqref{eq:EE_3pt_fun} leads to the logarithmic 
{temporal}
growth of entanglement entropies at early times~\cite{nahum2018hybrid, li1808hybrid}, as well as the logarithmic scaling {with spatial size}
in the steady state~\cite{nahum2018hybrid, li1808hybrid, li1901hybrid}.
To see this explicitly, we focus on the two simplifying regimes when $\tau = Y/L \ll 1$ and when $\tau = Y/L \gg 1$, where we recall that $Y \propto T$ is the rescaled imaginary time
(proportional to the circuit depth).
\env{enumerate}{
\item
$\tau \ll 1$.
In this limit, the conformal mapping for $z=x + i Y$
reduces to 
\env{eqnarray}{
    \lim_{\tau \to 0} w(z)
    = \lim_{m \to 1} {\rm sn}(\lambda(m) x|m)
    = \tanh \lz \frac{\pi}{2Y} x\rz,
}
so that Eq.~\eqref{eq:EE_3pt_fun} takes the following simple form
\env{eqnarray}{
    &&  S([z_1,z_5]) \nn
    &=& - h_{\bc{a|b}} \ln \frac{\pi}{Y} + {\rm const.}\nn
    &=& h_{\bc{a|b}} \ln Y + {\rm const.}
}
This is independent of $z_5$ since when $L \gg Y$ the corners of the rectangle are infinitely far away.
\item
$\tau \gg 1$.
In this limit, the conformal mapping for 
$z=x+iY$
reduces to
\env{eqnarray}{
    \lim_{\tau \to \infty} w(z)
    = \lim_{m \to 0} {\rm sn}(\lambda(m) x|m)
    = \sin \lz \frac{\pi}{L} x\rz,
}
and Eq.~\eqref{eq:EE_3pt_fun} becomes
\env{eqnarray}{
    &&  S([z_1,z_5]) \nn
    &=& h_{\bc{a|b}} \ln \( \frac{L}{\pi} \sin\lz \frac{\pi x_{15}}{L} \rz\) + {\rm const.},
}
where $x_{15}=x_1-x_5$,
reminiscent of the Cardy-Calabrese formula~\cite{Calabrese2004}, and when $x_{15} \ll L$, reduces to $S([z_1,z_5]) = h_{\bc{a|b}} \ln x_{15}$.
}

\subsubsection{Mutual information as 4-point functions}

We take a segment away from the corners, $A = [z_5, z_6]$, where 
$z_j = x_j + i Y$
and $-L/2 = x_1 < x_5 < x_6 < x_4 = L/2$,
and compute $S([z_5, z_6])$.
(According to our prescription, this is the entanglement entropy
of the segment $A = [z_5, z_6]$ at time $y=Y$.)
Since the segment $A$ is away from the corners, this geometry involves four boundary changing operators 
{at positions $z_1, z_5, z_6, z_4$ along the upper boundary of the
rectangle};
see the inset of Fig.~\ref{fig:res_fffa}(b).
Following Eq.~\eqref{eq:EE_Z}, this is given by
\env{eqnarray}{
	\label{eq:EE_4pt_fun}
	&&  \exp \lz - S([z_5, z_6]) \rz 
	\qquad \qquad \qquad \qquad \qquad \quad
	\text{(Fig.~\ref{fig:res_fffa}(b))} \nn
	&=&
	{
    	\frac{
    		\avg{\phi_{\bc{f|a}}(z_1) \phi_{\bc{a|b}}(z_5) \phi_{\bc{b|a}}(z_6) \phi_{\bc{a|f}} (z_4)}
    	}
    	{
    		\avg{\phi_{\bc{f|a}}(z_1) \phi_{\bc{a|f}} (z_4)}
    	}
    }\nn
	&=&
	{
    	\frac{
    		\avg{\phi_{\bc{f|b}}(z_1) \phi_{\bc{b|a}}(z_5) \phi_{\bc{a|b}}(z_6) \phi_{\bc{b|f}} (z_4)}
    	}
    	{
    		\avg{\phi_{\bc{f|b}}(z_1) \phi_{\bc{b|f}} (z_4)}
    	}
    }
    \nn
	&=&
	\(\frac{\pd w}{\pd z}\)_{z_5}^{h_{\bc{a|b}}} \(\frac{\pd w}{\pd z}\)_{z_6}^{h_{\bc{a|b}}}\nn
	&&\times
    {
    	\frac{
    		\avg{\phi_{\bc{f|b}}(w_1) \phi_{\bc{b|a}}(w_5) \phi_{\bc{a|b}}(w_6) \phi_{\bc{b|f}} (w_4)}
    	}
    	{
    		\avg{\phi_{\bc{f|b}}(w_1) \phi_{\bc{b|f}} (w_4)}
    	}
    }\nn
	&\propto& \lz
	\frac{
	    \(\frac{\pd w}{\pd z}\)_{z_5} \(\frac{\pd w}{\pd z}\)_{z_6}
	}
	{
	    \(w_{56}\)^2
	}
	\rz^{h_{\bc{a|b}}} F_{\bc{fbab}}(\eta),
}
where
\env{eqnarray}{
\eta = \frac{w_{15} w_{64}}{w_{16} w_{54}}
}
is the cross ratio, and we have defined $F_{\bc{fbab}}(\eta)$ with the following convention,
\env{eqnarray}{
\label{eq:F_eta_def}
&& F_{\bc{fbab}}(\eta) \nn
&=&
{
\frac{\avg{\phi_{\bc{f|b}}(w_1) \phi_{\bc{b|a}}(w_5) \phi_{\bc{a|b}}(w_6) \phi_{\bc{b|f}} (w_4)}}
{\avg{\phi_{\bc{f|b}}(w_1)
\phi_{\bc{b|f}} (w_4)}\avg{\phi_{\bc{b|a}}(w_5) \phi_{\bc{a|b}}(w_6)}}.
}
}

Given $S([z_5, z_6])$, we are now ready to compute another quantity of physical interest, namely the mutual information between two subregions sitting next to the corners, $A = [z_1, z_5]$ and $B = [z_6, z_4]$ (illustrated in the inset of Fig.~\ref{fig:res_fffa}(b)).
We have
\env{eqnarray}{
\label{eq:DEFMutualInformationEntropies}
	&&  I([z_1, z_5], [z_6, z_4]) \nn
	&=& S([z_1, z_5]) + S([z_6, z_4]) - S([z_1, z_5] \cup [z_6, z_4])\nn
	&=& S([z_1, z_5]) + S([z_6, z_4]) - S([z_5, z_6])
}
for a pure state, so that
\env{eqnarray}{
	\label{eq:MI_4pt_fun}
	&&  \exp \lz - I([z_1, z_5], [z_6, z_4]) \rz \nn
	&\propto& \frac{1}{F_{\bc{fbab}}(\eta)} \(\frac{\eta}{1-\eta} \frac{1}{1-\eta} \)^{-h_{\bc{a|b}}},
}
where 
we have used Eq.~\eqref{eq:EE_3pt_fun}, \eqref{eq:EE_4pt_fun}, and the exchange symmetry between $\bc{a}$ and $\bc{b}$.
Thus, the mutual information is a function only of
{the cross ratio}
$\eta$, and this is supported by the data collapse shown in Fig.~\ref{fig:res_fffa}(b), where the numerical data is again obtained at different times with various
{values of}
$z_5$ and $z_6$.

We note in passing that the scaling form of the entropy $S[z_5, z_6]$ 
(illustrated in the inset of Fig.~\ref{fig:res_fffa}(b))
is fully determined by that of the mutual information
in Eq.~\eqref{eq:MI_4pt_fun}, as well
as those of $S([z_1, z_5])$ and $S([z_6, z_4])$, as
already discussed
in subsection \ref{sec:fffa-Subsection-1}.



\subsubsection{Limits of the 4-point function from Operator Product Expansion (OPE) \label{sec:fffa_OPE}}


Let us examine the limit
in which
$z_5 \to z_1$,
or $z_6 \to z_4$,
so that the crossratio
$\eta \to 0$.
In this limit, $\phi_{\bc{f|b}}(z_1)$ and $\phi_{\bc{b|a}}(z_5)$,
as well as
$\phi_{\bc{a|b}}(z_6)$
and
$\phi_{\bc{b|f}}(z_4)$,
are close to one another,
and 
it is the following OPE 
that is needed in Eq.~\eqref{eq:F_eta_def} (see Table~\ref{table:bcc}),

\env{eqnarray}{
    \label{eq:OPE_phi_psi_sub}
	&& 
	\phi_{\bc{f|b}}(w_1)
	\phi_{\bc{b|a}}(w_5)  
	\qquad \qquad \qquad \qquad \qquad \text{ (Fig.~\ref{fig:res_fffa}(b))}
	\nn
	&\sim& w_{15}^{-h_{\bc{a|b}}} \( \phi_{\bc{f|a}}(w_1) + C_\bc{f|b|a}^{(1)} w_{15}^{ h_{\bc{f|a}}^{(1)}-h_{\bc{f|a}} } \phi^{(1)}_{\bc{f|a}}(w_1) + \ldots\), \nn
}
where {we have denoted by}
{$\phi^{(1)}_{\bc{f|a}}(w_1)$}
{the}
subleading bcc operator in the 
{$\bc{f|a}$-channel with a larger scaling dimension $h_{\bc{f|a}}^{(1)} > h_{\bc{f|a}} = h_{\bc{f|b}}$.}
With this, $F_{\bc{fbab}}(\eta)$ in Eq.~\eqref{eq:F_eta_def} reads
\env{eqnarray}{
	\label{eq:F_eta_0}
	&&        F_{\bc{fbab}}(\eta) \nn
	&\propto& \(\frac{\eta}{1-\eta}\)^{-h_{\bc{a|b}}} \(1 + \# \eta^{h_{\bc{f|a}}^{(1)} -  h_{\bc{f|a}}} \), \eta \to 0.
}
{Inserting}
this equation into Eq.~\eqref{eq:MI_4pt_fun}, we obtain the mutual information as a powerlaw function of $\eta$,\footnote{Here (and in all following equations), we use the symbol $\#$ to denote an order one, nonuniversal number.}
\env{eqnarray}{
	\label{eq:I_eta_0}
	&& I([z_1, z_5], [z_6, z_4]) = I(\eta)\nn
	&\approx& \#\eta^{h_{\bc{f|a}}^{(1)} - h_{\bc{f|a}}} + h_{\bc{a|b}} \times \eta, \quad \eta \to 0.
}
{When}
$h_{\bc{f|a}}^{(1)} - h_{\bc{f|a}} < 1$, the first term 
{is
more dominant than the analytic term of order $O(\eta)$.}
From the fit in Fig.~\ref{fig:res_fffa}(b) {\cyan (see inset),} we find the powerlaw exponent
$h_{\bc{f|a}}^{(1)} - h_{\bc{f|a}} \approx 0.9$.

{Referring again to Fig.~\ref{fig:res_fffa}(b), another}
limit of interest is $z_5 \to z_6$, where $\eta \to 1$.
The following OPE appearing in Eq.~\eqref{eq:F_eta_def} is now relevant (see Table~\ref{table:bcc}),
\env{eqnarray}{
	\label{eq:OPE_psi_psi}
	&& \phi_{\bc{b|a}}(w_5) \phi_{\bc{a|b}}(w_6) \qquad \qquad \qquad \qquad \qquad \quad
	\text{(Fig.~\ref{fig:res_fffa}(b))} \nn
	&\sim& w_{56}^{-2h_{\bc{a|b}}} \( {\bf 1}_{\bc{b|b}} + C_\bc{b|a|b}^{(1)} w_{56}^{h_\bc{b|b}^{(1)}} \phi_{\bc{b|b}}^{(1)}(w_6) + \ldots \)
}
After the two operators
on the left hand side
fuse, the b.c. is $\bc{b}$ on both sides of the new operator, therefore the leading behavior is captured by the identity operator, in addition to which we also include the subleading 
operator $\phi_{\bc{b|b}}^{(1)}$,
which denotes the most relevant operator with
positive scaling dimension in the spectum\footnote{
This spectrum of operators is of course not analytically known to
us in the present theory.} of all possible boundary
operators at boundary condition $\bc{b}$, with the scaling dimension being $h^{(1)}_\bc{b|b} = h^{(1)}_\bc{a|a}$.
At the same time, the following
OPE-channel of the remaining two operators in the 4-point function
appearing
in Eq.~\eqref{eq:F_eta_def} is relevant in the limit $\eta \to 1$
(compare Fig.~\ref{fig:res_fffa}(b)),
\env{eqnarray}{
	\label{eq:OPE_psiaf_psifa}
	&& \phi_{\bc{b|f}}(w_4) \phi_{\bc{f|b}}(w_1) 
	\qquad \qquad \qquad \qquad \qquad \quad \text{(Fig.~\ref{fig:res_fffa}(b))} \nn
	&\sim& w_{41}^{-2h_{\bc{f|b}}} \( {\bf 1}_{\bc{b|b}} + C_\bc{b|f|b}^{(1)} w_{41}^{h_\bc{b|b}^{(1)}} \phi_{\bc{b|b}}^{(1)}(w_1) + \ldots \)
}
From these two OPEs, and that $h^{(1)}_\bc{b|b} = h^{(1)}_\bc{a|a}$, we obtain the following behavior of the 4-point function
(defined in Eq.~\eqref{eq:F_eta_def})
\env{eqnarray}{
\label{eq:F_eta_1}
F_{\bc{fbab}}(\eta) \propto 1 + \#\(1-\eta\)^{h_{\bc{a|a}}^{(1)}}, \eta \to 1.
}
Using this result and Eq.~\eqref{eq:MI_4pt_fun} to compute the mutual information, we find
\env{eqnarray}{
    \label{eq:I_eta_1-Prelim}
	&&  \exp \lz - I([z_1, z_5], [z_6, z_4]) \rz \nn
	&\propto& \frac{1}{ 1 + \#\(1-\eta\)^{h_{\bc{a|a}}^{(1)}} } \(\frac{\eta}{1-\eta} \frac{1}{1-\eta} \)^{-h_{\bc{a|b}}} \nn
	&\approx& \frac{(1-\eta)^{2h_{\bc{a|b}}}}{ 1 + \#\(1-\eta\)^{h_{\bc{a|a}}^{(1)}} },
}
so that 
\env{eqnarray}{
	\label{eq:I_eta_1}
	&& I([z_1, z_5], [z_6, z_4]) \nn
	&=& -2h_{\bc{a|b}} \ln (1-\eta) + \#\(1-\eta\)^{h_{\bc{a|a}}^{(1)}}, \quad \eta \to 1.
}

The leading term fits well to 
{the data in}
Fig.~\ref{fig:res_fffa}(b);
however, we cannot reliably extract $h_{\bc{a|a}}^{(1)}$ from 
{these data}
since here the leading term diverges in this limit while the subleading term goes to zero.
{Here we {\cyan mention} 
that
a different way to determine the same exponent
for different b.c.'s of the background circuit
will yield in Eqs.~\eqref{eq:FababWithaaaaCircuit},\eqref{eq:I_eta_0_aaaa}
of 
Sec.~\ref{Sec:aaaaMutualInformationFourPointFunctions} the estimate\footnote{\label{FtnLogCFT} For the reader 
interested in details, we remark here on a  subtlety:
Our CFT could be what is called a ``logarithmic CFT''[log-CFT] in which, roughly
speaking, certain powerlaws 
are not the pure powerlaws which we display in the equations of this paper, but some  of the same  powerlaws would be in
fact multiplied by a logarithm of the argument of the powerlaw. However, the presence or absence of such multiplicative logarithms
is  unlikely to be concinvingly identifiable in numerics.
For this reason we will not elaborate in this paper on the presence of
possible logarithms, such as e.g. those desribed in
Ref.~\cite{GurarieLudwigCirc2004}. In particular, the appearance of the 
scaling dimension $h_\bc{a|a}^{(1)}=2.0$ may be related to the sitation 
discussed in this Reference. We plan on coming back these
questions in future work.}
$h_\bc{a|a}^{(1)}=2.0$.}

{Note}
that
{in the limit where $z_5 \to z_6$, and thus}
$\eta \to 1$, the 
{regions (intervals)
$A=[z_1,z_5]$ and $B=[z_6,z_4]$} 
sit close to each other, so that $S({A \cup B}) \to 0$, and the mutual information becomes twice the entanglement entropy of $A$ (or $B$, which has equal entanglement entropy).
Therefore, Eq.~\eqref{eq:I_eta_1} must recover the result in Eq.~\eqref{eq:EE_3pt_fun}.
Indeed,
\env{eqnarray}{
	&&\lim_{z_6 \to z_5} I([z_1, z_5], [z_6, z_4]) \nn
	&\approx& -2h_{\bc{a|b}} \lim_{z_6 \to z_5} \ln (1-\eta) \nn
	&=& -2h_{\bc{a|b}} \lim_{z_6 \to z_5} \ln \frac{w_{56} w_{14}}{w_{16} w_{54}} \nn
	&\approx& -2h_{\bc{a|b}} \ln \frac{ \(\frac{\pd w}{\pd z}\)_{z_5} w_{14}}{w_{15} w_{54}} \nn
	&\approx& -2h_{\bc{a|b}} \ln \xi \nn
	&=& 2 S(A).
}


\subsection{Circuit with boundary conditions $\bc{afaa}$
- Fig.~\ref{fig:upc_rect}(b)\label{sec:afaa}}

\env{figure}{[t]
\centering
\includegraphics[width=.5\textwidth]{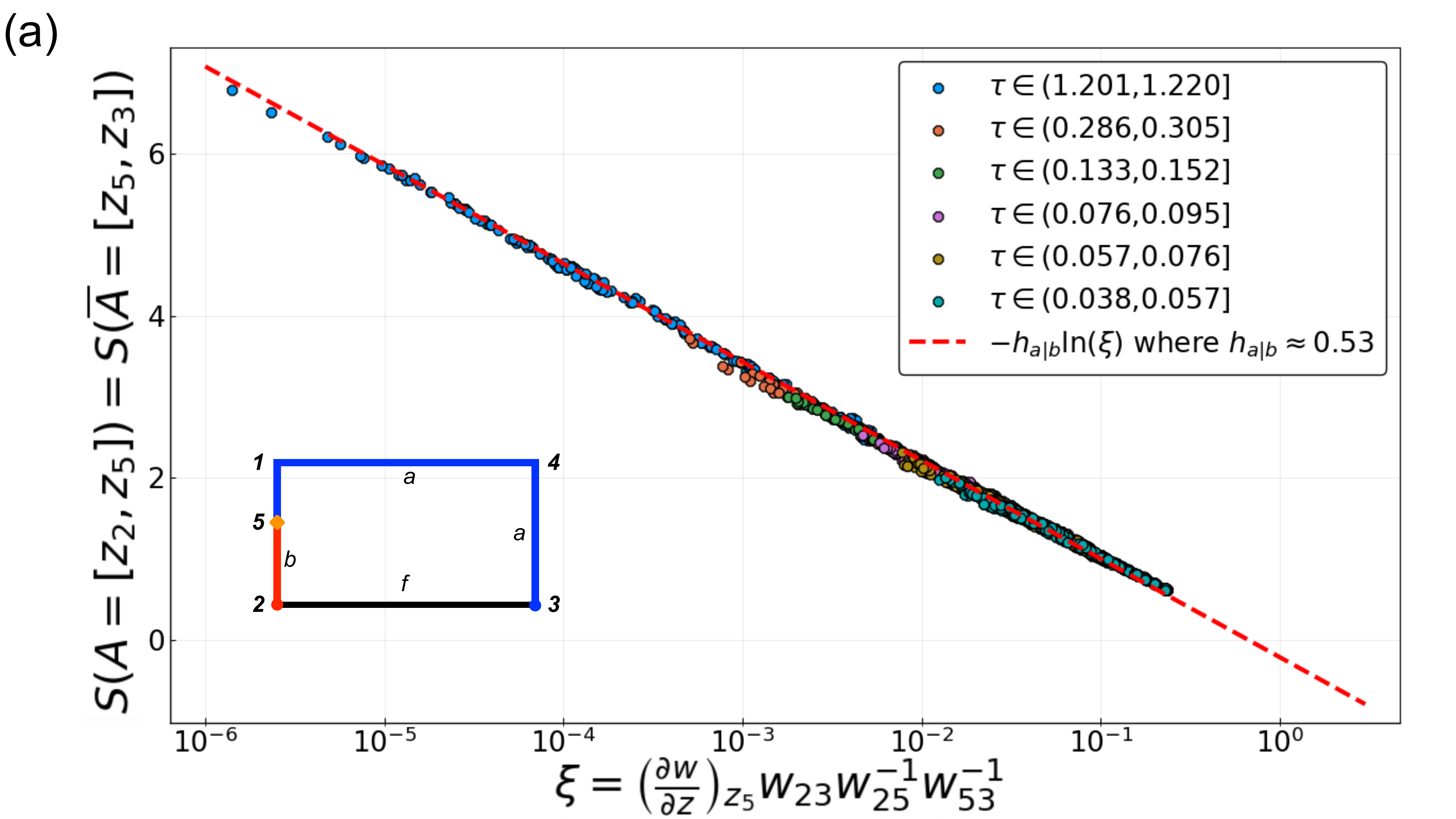}
\includegraphics[width=.5\textwidth]{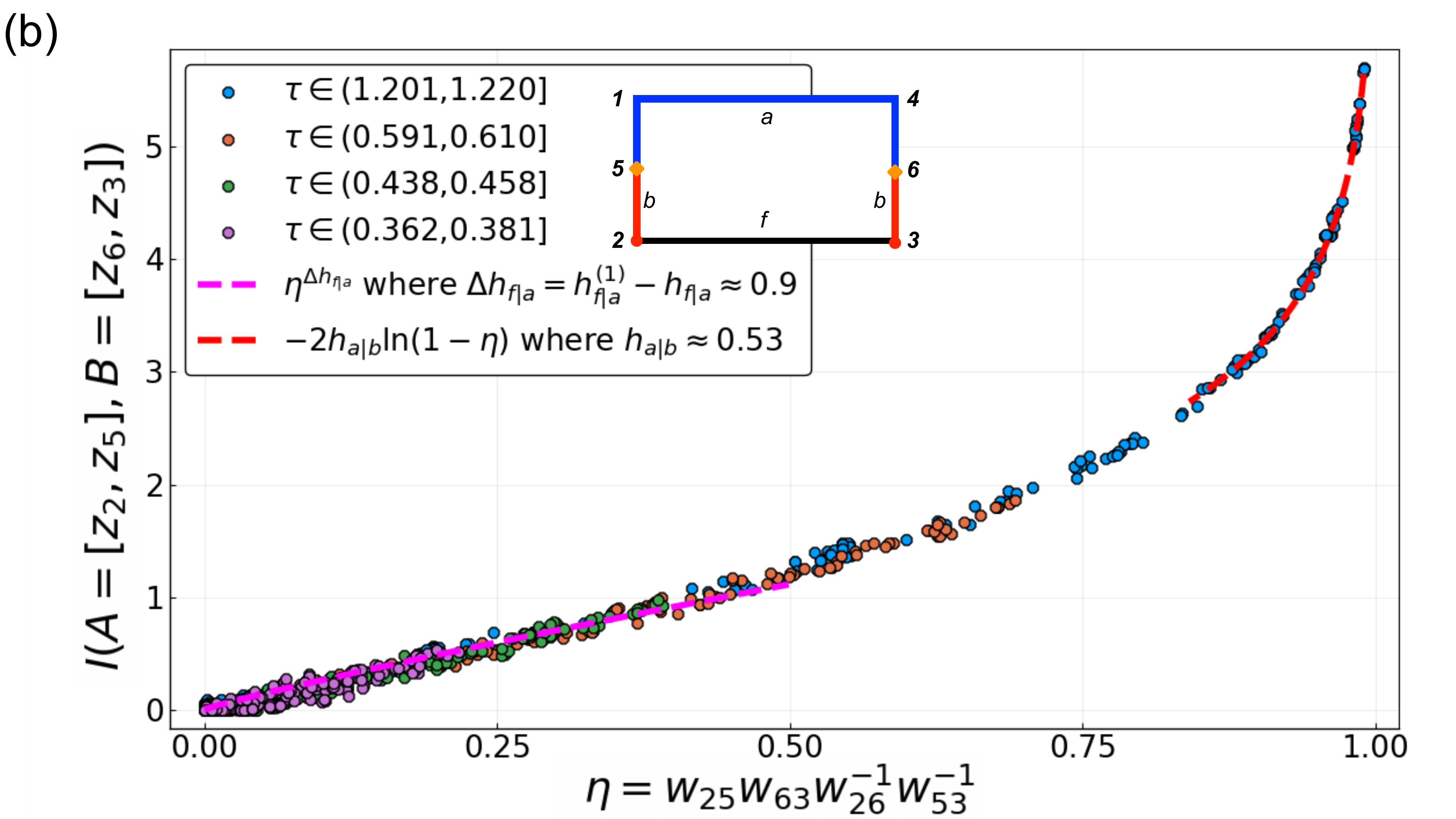}
\caption{
Numerical results for the $\bc{afaa}$ circuit.
(a) Entanglement entropies,
where $z_5$ takes different locations on either the left ($[z_2, z_1]$) or the right ($[z_4, z_3]$) side of the rectangle.
The data collapse follows Eq.~\eqref{eq:EE_3pt_fun}.
(b) Mutual information for two subregions sitting next to the corners, with $z_5 \in [z_2, z_1]$ and $z_6 \in [z_4, z_3]$.
The data collapse confirms Eq.~\eqref{eq:MI_4pt_fun}.
The limiting behaviors for $\eta \to 0$ and $\eta \to 1$ follow Eqs.~\eqref{eq:I_eta_0} and \eqref{eq:I_eta_1}, respectively.
}
\label{fig:res_afaa}
}

We briefly discuss the $\bc{afaa}$ circuit defined in Fig.~\ref{fig:upc_rect}(b).
In this setup, we still evolve from the product state, but with physical qubits injected at the left and right sides of the circuit.
The situation here is entirely similar to
{the circuit with boundary conditions
$\bc{fffa}$, discussed in the previous subsection
\ref{sec:fffa},}
except that we have moved the corner bcc operators from $z_1$ and $z_4$ ``down'' to $z_2$ and $z_3$ 
{(compare the insets of Fig.~\ref{fig:res_afaa}
with those of the previous Fig.~\ref{fig:res_fffa}).}
Accordingly, we compute the entanglement entropies and mutual information for regions 
that begin at the lower corners of the rectangle
at $z_2$ and/or $z_3$. 
This amounts to modifying Eq.~\eqref{eq:EE_3pt_fun} to
\env{eqnarray}{
    S([z_2, z_5]) = -h_\bc{a|b} \ln \( \frac{\(\frac{\pd w}{\pd z}\)_{z_5} w_{23}}{w_{25} w_{53}}\) + {\rm const.},
}
and
{to a different choice for}
the cross ratio,
\env{eqnarray}{
    \eta = \frac{w_{25} w_{63}}{w_{26}w_{53}},
}
where the forms of the mutual information in Eq.~\eqref{eq:MI_4pt_fun}, as well as its limits in Eq.~\eqref{eq:I_eta_0} and \eqref{eq:I_eta_1}, remain unchanged, since they are given by the same 4-point correlation functions.

The numerical results are given in Fig.~\ref{fig:res_afaa}, which has similar interpretations as Fig.~\ref{fig:res_fffa}; in particular, it gives consistent estimations of the scaling dimensions.
The data for $\bc{afaa}$ 
{provide further evidence for the presence of
conformal invariance,}
and justifies our assumption 
{about}
the b.c. 
{corresponding to}
physical qubits at the left and right sides of the rectangle.

We also studied yet another similar circuit with b.c. $\bc{ffaa}$ with physical qubits only on the left side, which is again consistent with $\bc{fffa}$ and $\bc{afaa}$ (data not displayed).


\subsection{Circuit with boundary conditions $\bc{fafa}$ - Fig.~\ref{fig:upc_rect}(c) \label{sec:fafa}}

\env{figure}{[t]
\includegraphics[width=.235\textwidth]{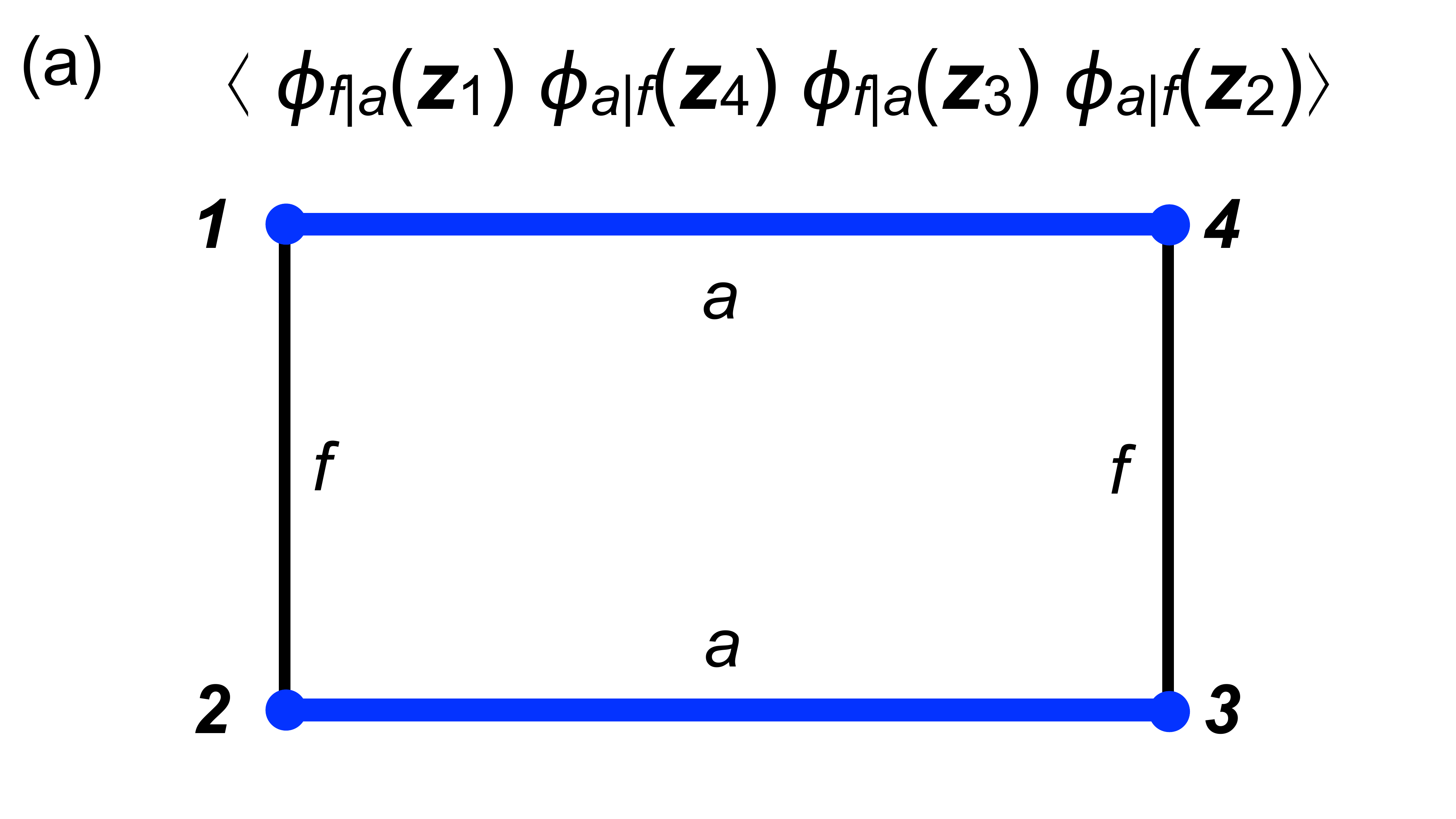}
\includegraphics[width=.235\textwidth]{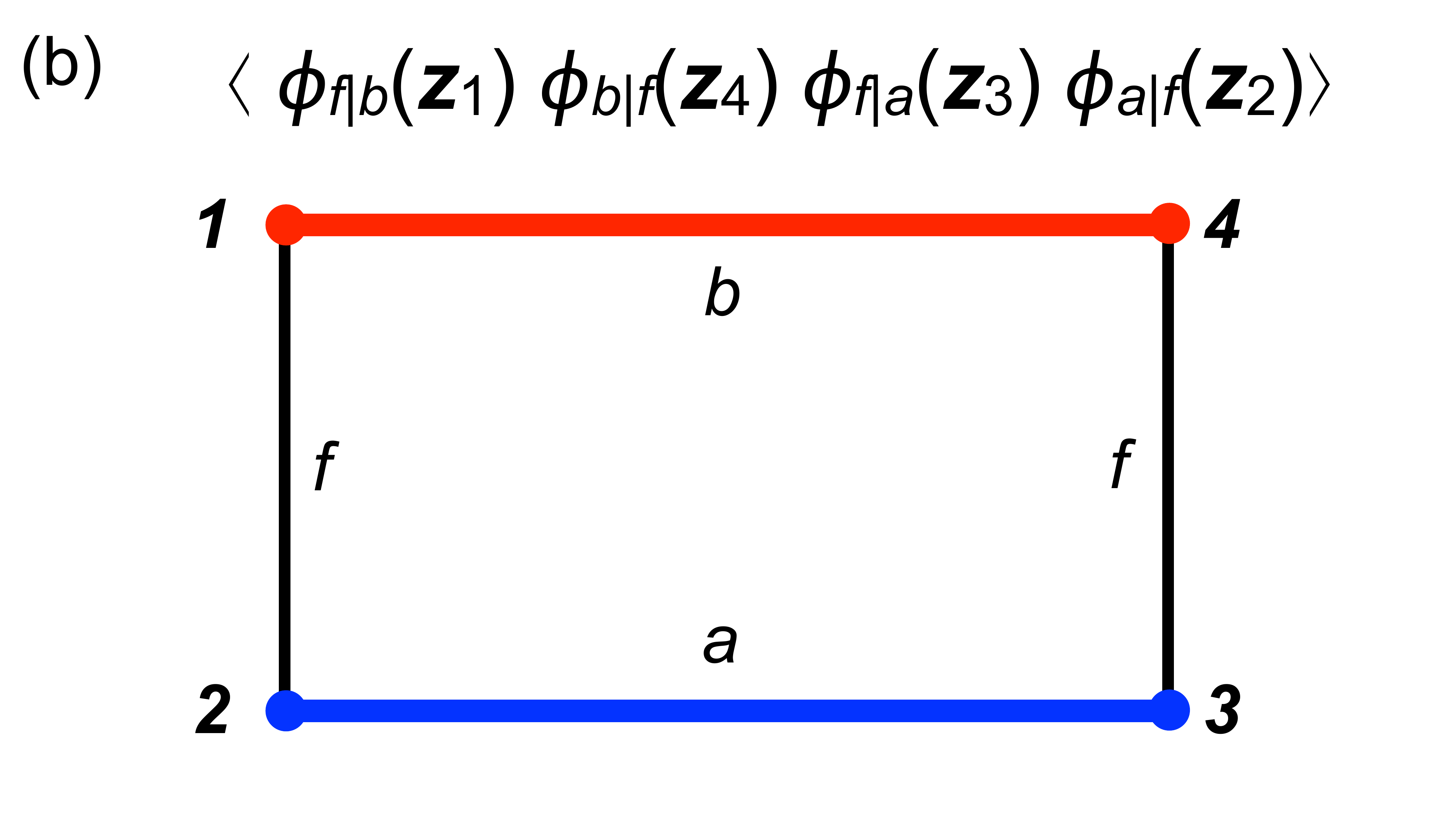}
\includegraphics[width=.235\textwidth]{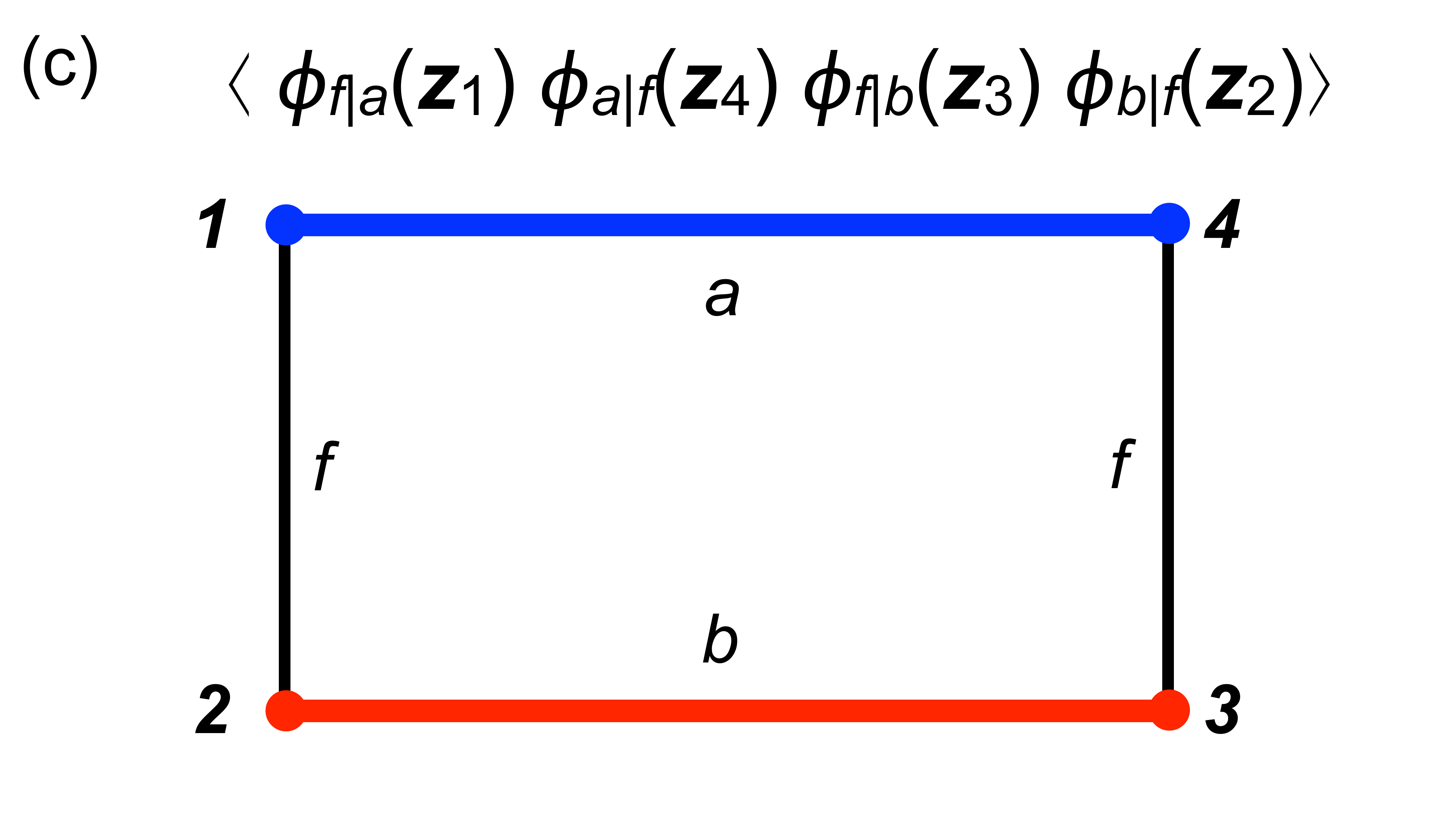}
\includegraphics[width=.235\textwidth]{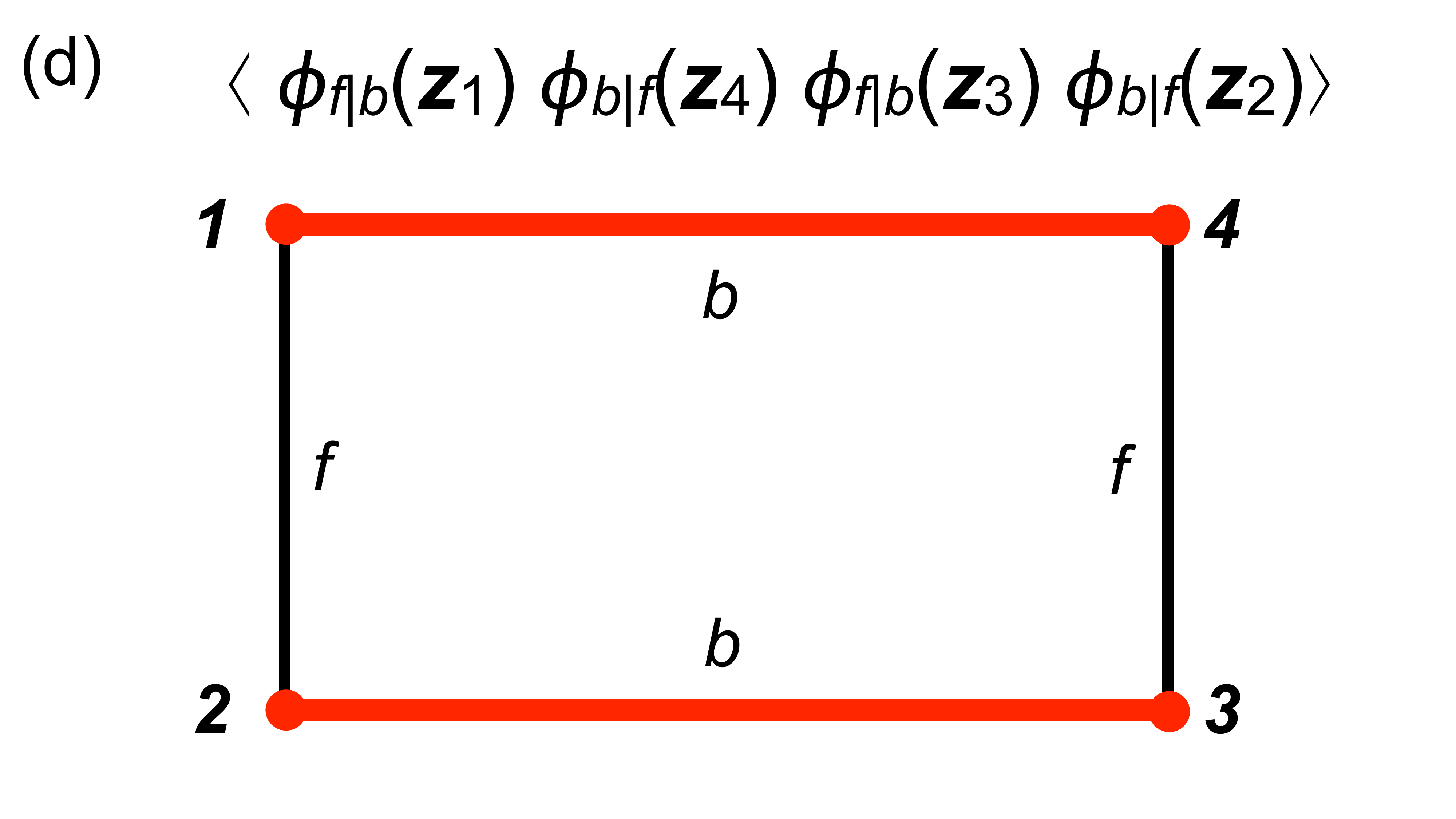}
\caption{Pictorial representations of the parition functions for the $\bc{fafa}$ circuit, with the $L$-Bell pair initial state.
(a) represents the background circuit, while (b) and (c) represents the partition functions relevant to computations of the entanglement entropy of $[z_1, z_4]$ and $[z_2, z_3]$, respectively.
(d) corresponds to the partition function for the computation of the entanglement entropy of $[z_1, z_4] \cup [z_2, z_3]$, i.e. all the physical qubits.
We notice the similarity between this figure and an illustration in Ref.~\cite{choi2019spin}.
}
\label{fig:bc_fafa}
}

We consider the $\bc{fafa}$ circuit (see Sec.~\ref{sec:bcft} and Fig.~\ref{fig:upc_rect}(c)), where the initial state consists of $L$ Bell pairs, so that 
{we have, as discussed above, two maximally entangled chains of qubits of length $L$ each,}
and only one chain is evolved under the circuit dynamics with open boundary condition (the ``system''); the other chain is left unevolved (the ``environment'').
We are interested in the entanglement entropy between the ``system'' (living on the upper boundary of the rectangle) and the ``environment'' 
(living on the lower boundary of the rectangle).
We have
$S([z_1, z_4]) = S([z_2, z_3])$
{which arises physically from the maximal entanglement of the original Bell pairs}
(compare
Eqs.~\eqref{eq:SU_vs_SL},\eqref{eq:PureState} below).
We illustrate the boundary conditions for these computations in Fig.~\ref{fig:bc_fafa}, following our general prescription in Sec.~\ref{sec:bcft}. 

The partition function for the $\bc{fafa}$ circuit reads (see Fig.~\ref{fig:bc_fafa}(a))
\env{eqnarray}{
	Z_{\rm circuit} = \avg{\phi_{\bc{f|a}}(z_1) \phi_{\bc{a|f}}(z_4) \phi_{\bc{f|a}}(z_3) \phi_{\bc{a|f}}(z_2)},
}
having the form of a 4-point correlation function of bcc operators at all four corners.
For Fig.~\ref{fig:bc_fafa}(b,c), we have
\env{eqnarray}{
	\label{eq:SU_vs_SL}
	&&  \exp \lz -S([z_1, z_4]) \rz \nn
	&=& \frac{
		\avg{\phi_{\bc{f|b}}(z_1) \phi_{\bc{b|f}}(z_4) \phi_{\bc{f|a}}(z_3) \phi_{\bc{a|f}}(z_2) }
	}
	{
		\avg{\phi_{\bc{f|a}}(z_1) \phi_{\bc{a|f}}(z_4) \phi_{\bc{f|a}}(z_3) \phi_{\bc{a|f}}(z_2) }
	} \nn
	&=& \frac{
		\avg{\phi_{\bc{f|a}}(z_1) \phi_{\bc{a|f}}(z_4) \phi_{\bc{f|b}}(z_3) \phi_{\bc{b|f}}(z_2) }
	}
	{
		\avg{\phi_{\bc{f|a}}(z_1) \phi_{\bc{a|f}}(z_4) \phi_{\bc{f|a}}(z_3) \phi_{\bc{a|f}}(z_2) }
	} \nn
	&=& \exp \lz -S([z_2, z_3]) \rz,
}
where we used the exchange symmetry between $\bc{a}$ and $\bc{b}$, as expected for a pure state, while for Fig.~\ref{fig:bc_fafa}(d),
\env{eqnarray}{
    \label{eq:PureState}
	&&  \exp \lz -S([z_1,z_4] \cup [z_2, z_3]) \rz \nn
	&=& \frac{
		\avg{\phi_{\bc{f|b}}(z_1) \phi_{\bc{b|f}}(z_4) \phi_{\bc{f|b}}(z_3) \phi_{\bc{b|f}}(z_2) }
	}
	{
		\avg{\phi_{\bc{f|a}}(z_1) \phi_{\bc{a|f}}(z_4) \phi_{\bc{f|a}}(z_3) \phi_{\bc{a|f}}(z_2) }
	} \nn
	&=& \frac{
		\avg{\phi_{\bc{f|a}}(z_1) \phi_{\bc{a|f}}(z_4) \phi_{\bc{f|a}}(z_3) \phi_{\bc{a|f}}(z_2) }
	}
	{
		\avg{\phi_{\bc{f|a}}(z_1) \phi_{\bc{a|f}}(z_4) \phi_{\bc{f|a}}(z_3) \phi_{\bc{a|f}}(z_2) }
	} \nn
	&=& 1,
}
again consistent with a pure state.

The computation in Eq.~\eqref{eq:SU_vs_SL} involves a 4-point function whose
{explicit}
form we do not know.
We can nevertheless examine the two limits
{of small and large (relative) circuit depth}, $\tau \to 0$ and $\tau \to \infty$,
as we discuss in the next two sections.
{($\tau$ is the aspect ratio of the rectangle defined in Eq.~\eqref{eq:DefinitionOfTau}}.)

Before diving into the calculations, we notice an important point, namely the symmetry between the 
{``system'', the upper edge $[z_1, z_4]$,
and the ``environment'', the lower edge $[z_2, z_3]$}
of the rectangle.
Viewed geometrically, the symmetry is merely a reflection.
Viewed as collections of qubits, the two edges are drastically different: the ``system'' qubits  actually experience the circuit dynamics, while the ``environment'' qubits are merely sitting there.
The symmetry between the two edges implies that they have identical average entanglement structures.
This means that if we take {an arbitrary subset of qubits $A$}
{of the upper edge $[z_1, z_4]$}
and its counterpart $B$,
{i.e. the subset of the lower edge $[z_2, z_3]$
which contains precisely the qubits that are initially Bell-entangled with those in $A$,}
their entanglement entropies will have the same expectation value at all times, despite that they might 
{be described by}
multi-point functions in the CFT which we do not know how to compute
{explicitly}.
In particular,
{this}
implies that at long times, when the
{upper and the lower}
edges have disentangled with each other, they will both appear ``critical''.
This is possible since the qubits in 
{lower edge}
$[z_2, z_3]$, initially unentangled with one another, can nevertheless have nontrivial entanglement structure due to the a ``entanglement swapping'' mechanism induced by local measurements performed in
{upper edge}
$[z_1, z_4]$.\footnote{For example, a possible such event ``swaps'' two inter-chain pairs for two intra-chain pairs (see Fig.~\ref{fig:EE_swap}).}
This symmetry
{has been}
checked numerically (data not displayed) and 
{can be justified in the case of the Hartley entropy in random Haar
circuits with measurements, using heuristic arguments based on its description by a ``minimal cut'' optimization problem in
percolation}~\cite{nahum2018hybrid} (see also Appendix~\ref{sec:perc} for detailed discussions).

\subsubsection{Bell-pair entanglement entropy at early times}
\label{Sect:Bell-pair-entanglement-early-times}
\env{figure}{[t]
\centering
\includegraphics[width=.5\textwidth]{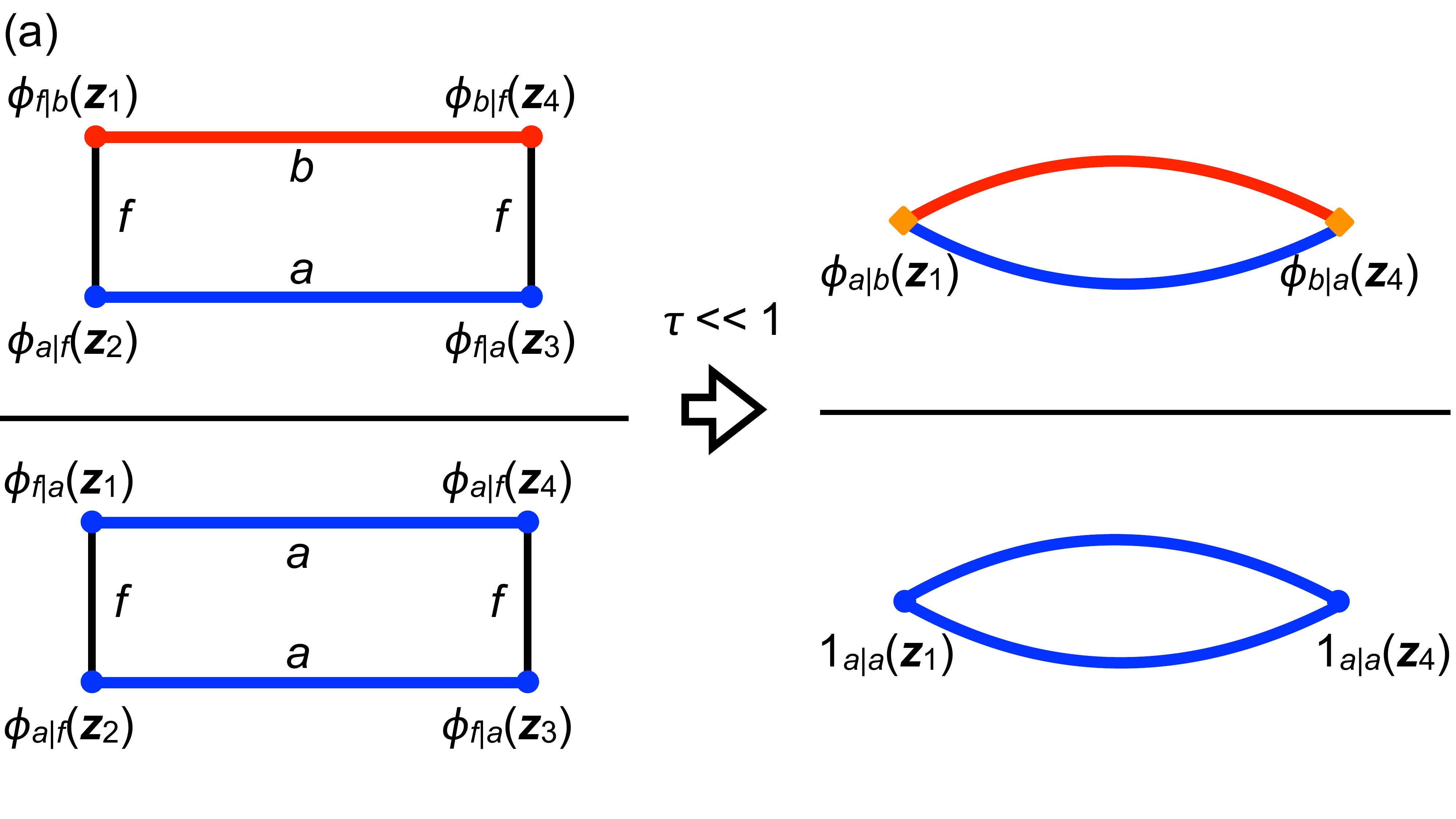}
\includegraphics[width=.5\textwidth]{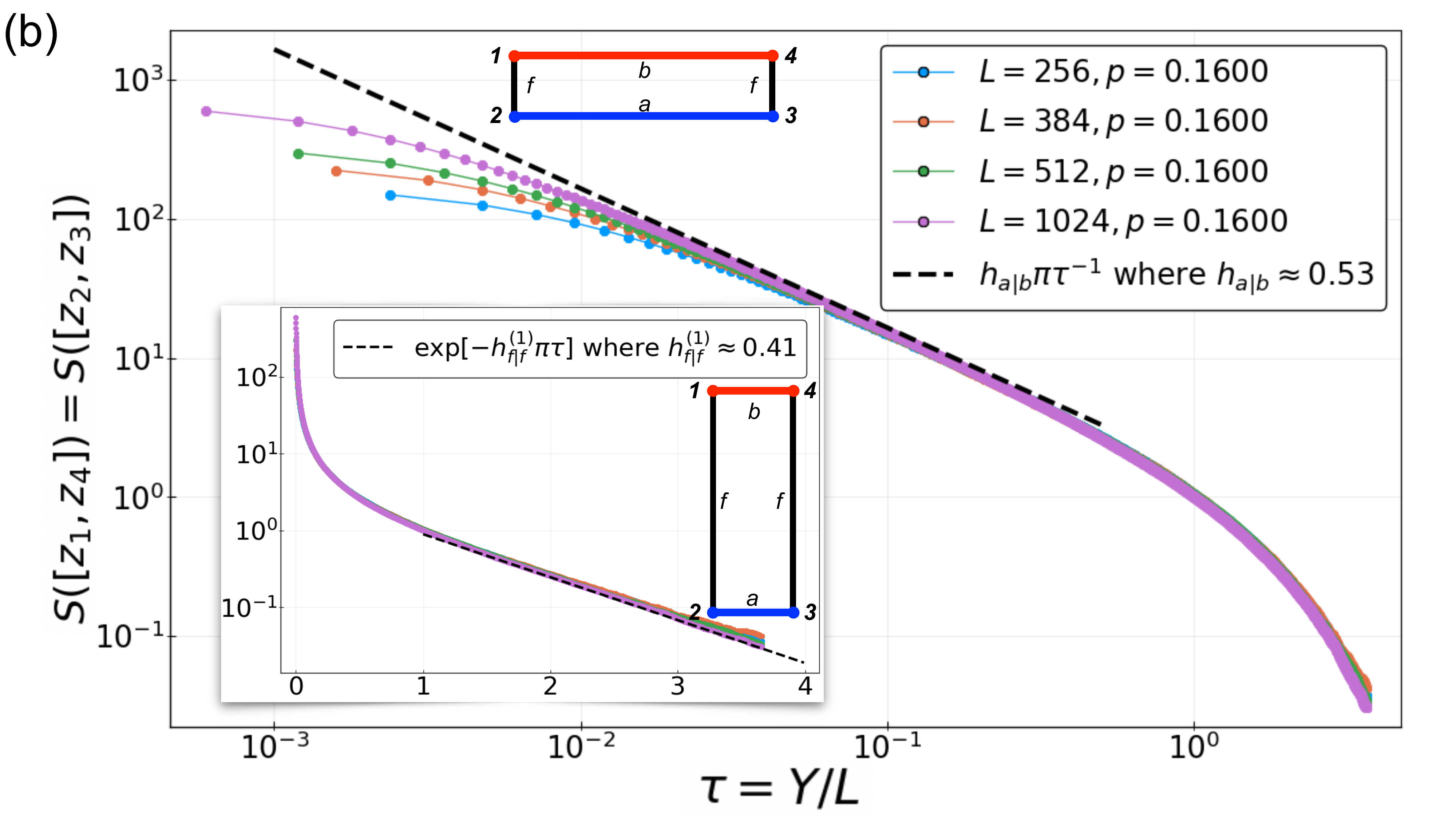}
\caption{
(a) Pictorial representations of the parition functions with bcc operators inserted at the corners, with the Bell-pair initial state, in the limit $\tau \to 0$.
The relevant OPEs are Eq.~\eqref{eq:OPE_phi_phi_aa}, \eqref{eq:OPE_phi_phi_ab}.
(b) Numerical data for $S([z_1, z_4]) = S([z_2, z_3])$, in the limits $\tau \to 0$ (main) and $\tau \to \infty$ (inset).
The data agrees well with calculations in Eq.~\eqref{eq:S_U_tau_0}, \eqref{eq:S_U_tau_1}.
We see from the data
that $S([z_1, z_4])$ is smaller than the predicted value when $\tau \lesssim 10^{-2}$.
We attribute this deviation to finite size 
{effects}.
The entanglement entropy of the system $[z_1, z_4]$ is always 
{bounded from above}
by $L \ln 2$. 
Thus, the formula must break down when $h_{\bc{a|b}} \pi \tau^{-1} > L$, or 
{$\tau < \tau_0(L) \coloneqq h_{\bc{a|b}} \pi L^{-1}$.}
This
{temporal cutoff $\tau_0(L)$}
vanishes in the thermodynamic limit; this trend is confirmed in Fig.~\ref{fig:res_fafb}(b).
}
\label{fig:res_fafb}
}

The regime of 
{a shallow depth circuit, $\tau \to 0$,} is
{illustrated}
in Fig.~\ref{fig:res_fafb}(a).
We observe that the $\tau \to 0$ limit corresponds to the $m \to 1$ limit, where $m$ is the parameter for the conformal mapping (see
{Eq.~\eqref{eq:tau_asymt_form} of}
Sec.~\ref{sec:cfm}).
{The bcc operator}
at corner $z_1$ is now very close to
{that at corner
$z_2$ (and the same is the case for the bcc operators at corners $z_3$
and $z_4$), so that they can be desribed by the OPE of these operators,
describing their ``fusion'', as }
illustrated in Fig.~\ref{fig:res_fafb}(a).
{After mapping to the lower half complex plane (LHP), the
distance between these point is precisely $w_{12} = w_{34} = m^{-1/2} - 1$, and vanishes in the limit $m \to 1$.}

We assume the following forms of the OPE to leading order (see Table~\ref{table:bcc}),
\env{eqnarray}{
	\label{eq:OPE_phi_phi_ab}
	&&\phi_{\bc{a|f}}(w_2) \phi_{\bc{f|b}}(w_1)
	\nn
    &\sim& w_{12}^{-2h_{\bc{f|a}} + h_{\bc{a|b}}} \phi_{\bc{a|b}}(w_1) + \ldots \\
    \label{eq:OPE_phi_phi_aa}
	&&\phi_{\bc{a|f}}(w_1)\phi_{\bc{f|a}}(w_2) 
	\nn 
	&\sim& w_{12}^{-2h_{\bc{f|a}}} \(\textbf{1}_\bc{a|a} + w_{12}^{h_\bc{a|a}^{(1)}}C_\bc{a|f|a}^{(1)} \phi_\bc{a|a}^{(1)}(w_2) + \ldots \),
}
Using these, we obtain Eq.~\eqref{eq:SU_vs_SL} in the limit
$z_1 \to z_2$, $z_3 \to z_4$ (compare Fig.~\ref{fig:res_fafb}(a))
\env{eqnarray}{
	\label{eq:SU_OPE_tau_0}
	&&  \exp \lz -S([z_1, z_4])\rz 
	\qquad \qquad \qquad \qquad \qquad
	\text{(Fig.~\ref{fig:res_fafb}(a))}
	\nn
	&=& \frac{
		\avg{\phi_{\bc{a|f}}(z_2) \phi_{\bc{f|b}}(z_1) \phi_{\bc{b|f}}(z_4) \phi_{\bc{f|a}}(z_3) }
	}
	{
		\avg{\phi_{\bc{a|f}}(z_2) \phi_{\bc{f|a}}(z_1) \phi_{\bc{a|f}}(z_4) \phi_{\bc{f|a}}(z_3) }
	} \nn
	&\propto&  \frac{
		w_{12}^{-2h_{\bc{f|a}} + h_{\bc{a|b}}}
		w_{34}^{-2h_{\bc{f|a}} + h_{\bc{a|b}}}
		\avg{\phi_{\bc{a|b}}(w_1) \phi_{\bc{b|a}}(w_4)}
	}
	{
		w_{12}^{-2h_{\bc{f|a}}}
		w_{34}^{-2h_{\bc{f|a}}}
	} \nn
	&\propto&
		w_{12}^{h_{\bc{a|b}}} w_{34}^{h_{\bc{a|b}}} \avg{\phi_{\bc{a|b}}(w_1) \phi_{\bc{b|a}}(w_4)}
	\nn
	&\propto& (w_{12} w_{34})^{h_{\bc{a|b}}} \nn
	&\propto& (m^{-1/2} - 1)^{2h_{\bc{a|b}}},
}
where we 
{used the fact that $w_{14} \to 2$ (a constant) in that limit.}
Using
the asymptotic form of $\tau$ in Eq.~\eqref{eq:tau_asymt_form}, 
we 
{obtain the asymptotic behavior
$m \sim 1 - 16 \exp(-\frac{\pi}{2\tau})$}
where the second term is small as $\tau \to 0$.
{Using this in the previous equation yields the following asymptotic behavior of the entropy in the limit $\tau\to 0$ of
a shallow-depth circuit}
\env{eqnarray}{
	\label{eq:S_U_tau_0}
	&&  \exp \lz -S([z_1, z_4])\rz \nn
	&\propto& (m^{-1/2} - 1)^{2h_{\bc{a|b}}} \nn
	&\propto& \exp\lz -\frac{\pi}{2\tau} \rz^{2h_{\bc{a|b}}} \nn
	&\propto& \exp \lz -\frac{h_{\bc{a|b}} \pi}{\tau} \rz,
}
implying
\env{eqnarray}{
\label{eq:EntropyEarlyTimesFirstDerivation}
S([z_1, z_4])= S([z_2, z_3])= h_{\bc{a|b}} \pi \tau^{-1},
\ \ \ (\tau\to 0),
}
a form first obtained numerically in Ref.~\cite{gullans1905purification}.
The fit in Fig.~\ref{fig:res_fafb}(b) gives $h_{\bc{a|b}} \approx 0.53$, consistent with estimation of $h_{\bc{a|b}}$ in the previous section.

Alternatively,
the
{asymptotic $\tau^{-1}$ behavior of the entropy as $\tau \to 0$}
can be understood in terms of 
{the}
transfer matrix formalism.
Here we take the 
\textit{spatial}
direction to be the ``direction of propagation''
{of the transfer matrix}, and 
{denote the generator of 
translations
in this direction by 
$H_{\bc{ab}}$. Specifically, 
$H_{\bc{ab}}$
denotes the Hamiltonian of the CFT in question, defined on an interval of length  $Y$ {(compare Eq.~\eqref{eq:DefinitionOfTau})}, with
boundary conditions $\bc{a}$ and $\bc{b}$ at the two ends of the interval.}
The 
{(finite size) spectrum of energies $E_\bc{ab}$ of the Hamiltonian
}
$H_{\bc{ab}}$ 
{is known~\cite{CARDY1986} in any CFT
to take on the form}
\env{eqnarray}{
    \label{eq:H_ab}
    E_{\bc{ab}} = E_0 +\frac{\pi (h^{(j)}_{\bc{a|b}}+n)}{Y}.
}
Here
$n \geq 0$ is an integer, and
$h^{(j)}_{\bc{a|b}}$, where $j=0, 1, 2, \ldots$, denotes the 
spectrum\footnote{Here we choose for simplicity a notation suitable for a discrete
spectrum.}
of scaling dimensions (in increasing order) of all possible primary bcc operators that occur when the boundary condition changes
from $\bc{a}$ to $\bc{b}$.
The smallest such scaling dimension corresponding the $j=0$,
we denoted previously  by $h_{\bc{a|b}}$,
i.e. $h_{\bc{a|b}}=h^{(j=0)}_{\bc{a|b}}$.
(The quantity $E_0$ cancels out
in the observables of interest to us, and is not needed in the sequel.)


A special case of the above situation is the case where the two boundary conditions are the same, $\bc{a}=\bc{b}$. In this case the (finite size) spectrum
takes the form
\env{eqnarray}{
    \label{eq:H_aa}
    E_{\bc{a|a}} = E_0 +\frac{\pi (h^{(j)}_{\bc{a|a}}+n)}{Y}.
}
As before,  $n \ge 0$ is  an integer, and
$h^{(j)}_{\bc{a|a}}$, with $j=0, 1, 2, \ldots$ 
denotes 
the spectrum of
scaling dimensions of all possible primary bcc operators that occur
at a {\it given} boundary condition $\bc{a}$.
The smallest such scaling dimension corresponding the $j=0$, is the identity operator, i.e.
$\phi^{(j=0)}_{\bc{a|a}} = {1}$ corresponding to
$h^{(j=0)}_{\bc{a|a}}=0$.


The partition function 
of the rectangle is written in the usual manner in terms of
the transfer matrix $\exp(-H_{\bc{ab}} \times L)$ and a state
$\ket{\bc{f}}$ representing the vertical  boundary  of the rectangle
(compare Fig.~\ref{fig:res_fafb}(a)) with free boundary condition $\bc{f}$,
as
the amplitude
\env{eqnarray}{
    Z_{\bc{ab}} = \bra{\bc{f}} \exp(-H_{\bc{ab}} \times L) \ket{\bc{f}}.
}
Upon inserting a complete set of eigenstates, one sees that in the limit
$ L \gg Y$,  both $Z_{\bc{ab}}$ and $Z_{\bc{aa}}$ are dominated by their
respective lowest energy eigenvalues
$h^{(j=0)}_{\bc{a|b}}=h_{\bc{a|b}}$ and $n=0$, as well as
$h^{(j=0)}_{\bc{a|a}}=0$ and $n=0$, yielding the following asymptotic form of
the ratio 
\env{eqnarray}{
    \frac{Z_{\bc{ab}}}{Z_{\bc{aa}}} \sim \exp \(-h_{\bc{a|b}} \pi L/Y \).
}
The
{resulting}
entanglement entropy
{thus behaves asymptotically as} (recall from Eq.~\eqref{eq:DefinitionOfTau} that $\tau=Y/L$)
\env{eqnarray}{
    S([z_1, z_4]) = -\ln \frac{Z_{\bc{ab}}}{Z_{\bc{aa}}} \sim h_{\bc{a|b}} \pi \tau^{-1},
    \quad (\tau \to 0).
}

\subsubsection{Bell-pair entanglement entropy at late times}

{For a very deep circuit where}
$\tau \to \infty$,
{corresponding to} $w_{14} \to 0$ and $w_{23} \to 0$, we now have the cross ratio 
\env{eqnarray}{
\eta=\frac{w_{12}w_{34}}{w_{24}w_{13}}\to 1.
}
In this limit, to compute 4-point correlation functions defined in Eq.~\eqref{eq:SU_vs_SL}, 
{we need the}
the vacuum channel OPE in Eq.~\eqref{eq:OPE_phi_phi_aa}, 
where we now include a subleading term, 
\env{eqnarray}{
    \label{eq:ope_phi_phi_sub}
	&& \phi_{\bc{f|a}}(w_1) \phi_{\bc{a|f}}(w_4)
	\qquad \qquad \qquad \qquad \qquad \quad
	\text{(Fig.~\ref{fig:res_fafb}(b))}
	\nn
	&\sim& w_{14}^{-2h_{\bc{f|a}}} \( \textbf{1}_{\bc{f|f}} + 
	C_\bc{f|a|f}^{(1)}
w_{14}^{ h_{\bc{f|f}}^{(1)}} \phi_{\bc{f|f}}^{(1)} (w_1) + \ldots \) \\
	&& \phi_{\bc{f|b}}(w_1) \phi_{\bc{b|f}}(w_4)\nn
	&\sim& w_{14}^{-2h_{\bc{f|a}}} \( \textbf{1}_{\bc{f|f}} + 
	C_\bc{f|b|f}^{(1)}
    w_{14}^{h_{\bc{f|f}}^{(1)}} \phi_{\bc{f|f}}^{(1)} (w_1) + \ldots\)
}
where $\phi_{\bc{f|f}}^{(1)} (w_1)$ 
denotes the most relevant subleading operator that does not change this boundary condition (i.e. ``which appears in the  $\bc{f|f}$-channel''). 
{Here, $ C_\bc{f|a|f}^{(1)}$ and
$ C_\bc{f|b|f}^{(1)}$ denote OPE coefficients of the corresponding
BCC operators,
where\footnote{{To phrase this in a more general language, consider the case where labels $\bc{A}, \bc{B}, ...$ take values in a set specifying $M$ different boundary conditions of type 
$\bc{a}$, $\bc{b}$, ...,  i.e.
$\bc{A}, \bc{B} \in \{\bc{a}, \bc{b}, ... \}$. Permutation symmetry of these $M$ boundary conditions implies, under the condition listed below, the following generalized form of the
OPE considered
in Eq.~\eqref{eq:ope_phi_phi_sub}:
$\psi_{\bc{f|A}}(w_1) \psi_{\bc{A|f}}(w_4)\sim $ 
$w_{14}^{-2h_{\bc{f|A}}} \textbf{1}_{\bc{f|f}} +$ 
$ C_\bc{f|f}^{(1)} \ 
w_{14}^{-2h_{\bc{f|A}} + h^{(1)}_{\bc{f|f}}} \psi^{(1);A}_{\bc{f|f}} (w_4)$.
Under permutations of the $M$ boundary conditions, the left hand side forms a representation
of the permutation group $S_M$ of $M$
objects which is known to
decompose into a sum of the totally symmetric one-dimensional and the $(M-1)$-dimensional
irreducible representation. This decomposition is reflected on the right hand side:
The set of operators on the right hand side satisfy
$\sum_{\bc{A}=1}^M \psi^{(1);A}_{\bc{f|f}}=0$ and transform in the $(M-1)$-dimensional representation.
In writing this OPE we have assumed that the first subleading operator beyond the identity operator is the operator
$\psi^{(1);\bc{A}}_{\bc{f|f}}$ transforming in the $(M-1)$ irreducible representation as opposed to another (singet)
operator,}
besides the identity operator, which transforms
in the one-dimensional (totally symmetric) representation. 
{
We note that the linear dependency condition immediately implies the following condition for the two point function
$\langle\psi^{(1);\bc{a}}_{\bc{f|f}}\psi^{(1);\bc{a}}_{\bc{f|f}}\rangle
+(M-1) 
\langle\psi^{(1);\bc{a}}_{\bc{f|f}}\psi^{(1);\bc{b}}_{\bc{f|f}}\rangle =0
$ where permutation symmetry was used. This implies that the
generalizations of
Eq.~\eqref{eq:Ffafb}
and Eq.~\eqref{eq:Ffafa} below to $M$ permutation symmetric boundary conditions
are not equal, which is a necessary condition
for obtaining a non-trivial result in the subsequent equation
Eq.~\eqref{eq:S_U_tau_1}, which is confirmed by our numerics.
At the same time, had the first subleading operator in the above OPE been the totally symmetric one-dimensional representation, the first subleading terms in
Eq.~\eqref{eq:Ffafb}
and Eq.~\eqref{eq:Ffafa} would be equal, in contrast to our numerical results.
[Our assumption is thus confirmed by the numerics.]
- We can now immediately recover
the formulation  presented in
Eq.~\eqref{eq:ope_phi_phi_sub} upon specializing to the case of $M=2$ boundary conditions of this type,
 i.e. $\bc{A}, \bc{B} \in \{\bc{a}, \bc{b}\}$: In this case the linear dependency condition reads
$\psi^{(1);\bc{a}}_{\bc{f|f}}+ \psi^{(1);\bc{b}}_{\bc{f|f}}=0$. Upon
making the identifications
$C_\bc{f|f}^{(1)}\psi^{(1);\bc{a}}_{\bc{f|f}}\equiv$
$C_\bc{f|a|f}^{(1)}\phi^{(1)}_{\bc{f|f}}$
as well as
$C_\bc{f|f}^{(1)}\psi^{(1);\bc{b}}_{\bc{f|f}}=$
$(-1) C_\bc{f|f}^{(1)}\psi^{(1);\bc{a}}_{\bc{f|f}}\equiv$
$C_\bc{f|b|f}^{(1)}\phi^{(1)}_{\bc{f|f}}$,
we recover Eq.~\eqref{eq:ope_phi_phi_sub} with
$C_\bc{f|b|f}^{(1)}=$ $(-1)C_\bc{f|a|f}^{(1)}$.
}
}
in general $ C_\bc{f|a|f}^{(1)} \not = C_\bc{f|b|f}^{(1)}$.}
The same OPE in Eq.~\eqref{eq:ope_phi_phi_sub} holds for $\phi_\bc{f|a}(z_3) \phi_\bc{a|f}(z_2)$, that also appears in Eq.~\eqref{eq:SU_vs_SL}.

By using these
{OPEs, one can express the leading behavior of}
the 4-point function in Eq.~\eqref{eq:SU_vs_SL}
{in the limit $\eta\to 1$, in terms of the limiting behavior of the following two functions}
\env{eqnarray}{
    \label{eq:Ffafb}
	&& F_{\bc{fafb}}(\eta)=\frac{\avg{
	\phi_{\bc{f|a}}(w_3)
	\phi_{\bc{a|f}}(w_2) \phi_{\bc{f|b}}(w_1) \phi_{\bc{b|f}}(w_4)  }}{\avg{\phi_{\bc{a|f}}(w_2)\phi_{\bc{f|a}}(w_3)\rangle\langle \phi_{\bc{f|b}}(w_1) \phi_{\bc{b|f}}(w_4) }} \nn
	&=&1+ 
C_{\bc{fafb}} 
	(1-\eta)^{h_{\bc{f|f}}^{(1)}} ,\quad \eta\to 1.
}
where $C_{\bc{fafb}}=$
$C_\bc{f|a|f}^{(1)}C_\bc{f|b|f}^{(1)}$, and
\env{eqnarray}{
    \label{eq:Ffafa}
	&& F_{\bc{fafa}}(\eta)=\frac{\avg{
	\phi_{\bc{f|a}}(w_3)
	\phi_{\bc{a|f}}(w_2) \phi_{\bc{f|a}}(w_1) \phi_{\bc{a|f}}(w_4)  }}{\avg{\phi_{\bc{a|f}}(w_2)\phi_{\bc{f|a}}(w_3)\rangle\langle \phi_{\bc{f|a}}(w_1) \phi_{\bc{a|f}}(w_4) }} \nn
	&=&1+
C_{\bc{fafa}}
	(1-\eta)^{h_{\bc{f|f}}^{(1)}} ,\quad \eta\to 1.
}
where $C_{\bc{fafa}}=$
$C_\bc{f|a|f}^{(1)}C_\bc{f|a|f}^{(1)}$.
Inserting the above results into Eq.~\eqref{eq:SU_vs_SL}, we obtain
\env{eqnarray}{
    \label{eq:fafb_over_fafa}
	&& \exp[-S([z_1, z_4])] = \frac{F_{\bc{fafb}}(\eta)}{F_{\bc{fafa}}(\eta)} \nonumber\\
	&\approx& 1- (C_{\bc{fafa}} - C_{\bc{fafb}}) (1-\eta)^{h_{\bc{f|f}}^{(1)}},\quad 1-\eta\to 0.
}
Since $\eta= 1 - 16 \exp(-\pi \tau)$ in the limit $\tau \to \infty$ (Eq.~\eqref{eq:tau_asymt_form}),
we can show that 
\env{eqnarray}{
	\label{eq:S_U_tau_1}
	S([z_1, z_4])
	\propto (1-\eta)^{h_{\bc{f|f}}^{(1)}}
	\propto \exp(-h_{\bc{f|f}}^{(1)} \pi \tau).
}
From our fit in Fig.~\ref{fig:res_fafb}(b)(inset), we have the conformal dimension $h_{\bc{f|f}}^{(1)} \approx 0.41$.
{We note that the exponential decay in Eq.~\eqref{eq:S_U_tau_1} is understood as a consequence of crossover to a quasi-one-dimensional system as $Y \gg L$, where every correlation function falls off exponentially, with the correlation length set by $L$.
}



\subsection{Circuit with boundary conditions $\bc{aaaa}$ - Fig.~\ref{fig:upc_rect}(d)\label{sec:aaaa}}

\subsubsection{Entanglement entropies as 2-point functions}

\env{figure}{[t]
\includegraphics[width=.5\textwidth]{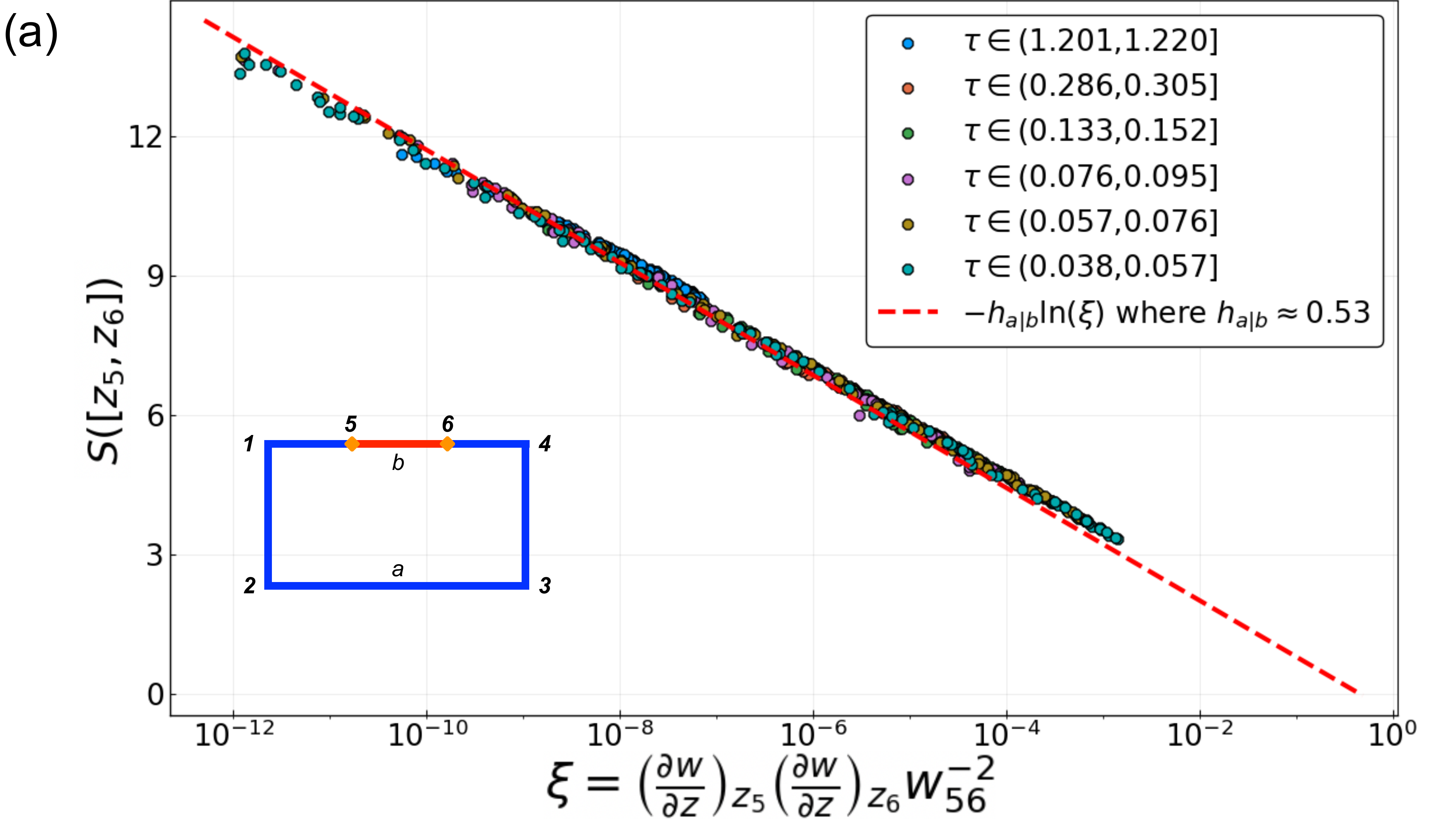}
\includegraphics[width=.5\textwidth]{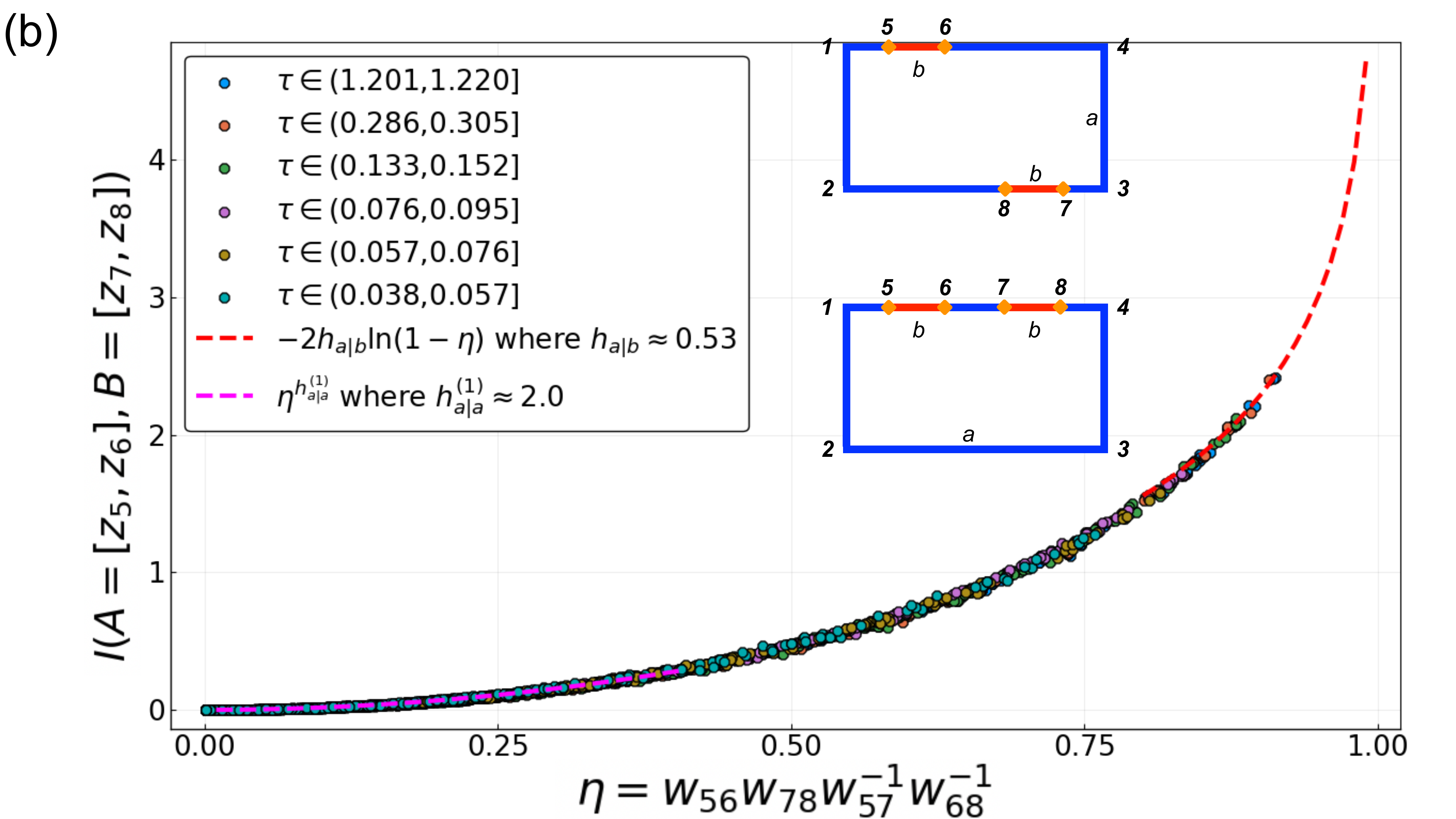}
\caption{
Numerical results for the $\bc{aaaa}$ circuit.
(a) Entanglement entropy fitted to 2-point functions according to Eq.~\eqref{eq:aaaa_sa_2pt}.
{Here we take $z_5, z_6 \in [z_1, z_4]$ for simplicity.}
(b) Mutual information fitted to 4-point functions according to Eq.~\eqref{eq:aaaa_I_4pt}.
The two intervals are either both on the same side $[z_1, z_4]$, or on opposite sides $[z_1, z_4]$ and $[z_2, z_3]$.
The limiting behaviors are given in Eqs.~\eqref{eq:I_eta_0_aaaa}, \eqref{eq:I_eta_1_aaaa}.
}
\label{fig:res_aaaa}
}

As mentioned in Sec.~\ref{sec:bcft}, the $\bc{aaaa}$ circuit has physical qubits on all four edges of the rectangle, therefore the background partition function of the circuit is defined without any boundary condition changing operators; see Fig.~\ref{fig:upc_rect}(d).
This is convenient since now the entanglement entropy of a contiguous subregion is given by a 2-point function, which has a simple form.
(Recall that in contrast, for boundary conditions
of the rectangle
of type $\bc{fffa}$ and $\bc{fafa}$, the entanglement entropies map to (more complicated) 3- or higher-point functions.)
In terms of the conformal mapping, the entanglement entropy of an interval $[z_5, z_6]$
reads for the present boundary conditions
\env{eqnarray}{
    && \exp \lz -S([z_5, z_6]) \rz
    \qquad \qquad \qquad \qquad \qquad
    \text{(Fig.~\ref{fig:res_aaaa}(a))}\nn
    &=&
    \(\frac{\pd w}{\pd z}\)_{z_5}^{h_{\bc{a|b}}}
    \(\frac{\pd w}{\pd z}\)_{z_6}^{h_{\bc{a|b}}}
    \avg{\phi_{\bc{a|b}}(w_5) \phi_{\bc{b|a}}(w_6)} \nn
    &\propto&
    \lz
    \frac{
        \(\frac{\pd w}{\pd z}\)_{z_5}
        \(\frac{\pd w}{\pd z}\)_{z_6}
    }
    {
        \(w_{56}\)^2
    }
    \rz^{h_{\bc{a|b}}},
}
hence
\env{eqnarray}{
    \label{eq:aaaa_sa_2pt}
    S([z_5, z_6]) = -h_{\bc{a|b}} \ln \lz \frac{
        \(\frac{\pd w}{\pd y}\)_{z_5}
        \(\frac{\pd w}{\pd y}\)_{z_6}
    }
    {
        \(w_{12}\)^2
    }
    \rz + {\rm const.}\nonumber\\
}
The computed entanglement entropy and fit to the 2-point function is shown in Fig.~\ref{fig:res_aaaa}, where we took $h_{\bc{a|b}} = 0.53$.

\subsubsection{Mutual information as 4-point functions}
\label{Sec:aaaaMutualInformationFourPointFunctions}
We compute the mutual information of two subregions to further confirm the conformal symmetry.
We take the two subregions to be the intervals $A=[z_5,z_6]$ and $B=[z_7,z_8]$ 
which sit at various positions, either both on the upper edge, or with one on the upper edge and the other on the lower edge of the rectangle, as shown in the insets of Fig.~\ref{fig:res_aaaa}(b).
The mutual information is expressed in terms of the 4-point correlation function of the same bcc operators as 
\env{eqnarray}{
    \label{eq:aaaa_I_4pt}
    &&  \exp\lz -I([z_5, z_6], [z_7, z_8]) \rz
    \qquad \qquad \qquad \quad \quad
    \text{(Fig.~\ref{fig:res_aaaa}(b))}
    \nn
    &=& \frac{
        \avg{\phi_{\bc{a|b}}(z_5)\phi_{\bc{b|a}}(z_6)}
        \avg{\phi_{\bc{a|b}}(z_7)\phi_{\bc{b|a}}(z_8)}
    }
    {
        \avg{\phi_{\bc{a|b}}(z_5)\phi_{\bc{b|a}}(z_6)\phi_{\bc{a|b}}(z_7)\phi_{\bc{b|a}}(z_8)}
    } \nn
    &=& \frac{
        \avg{\phi_{\bc{a|b}}(w_5)\phi_{\bc{b|a}}(w_6)}
        \avg{\phi_{\bc{a|b}}(w_7)\phi_{\bc{b|a}}(w_8)}
    }
    {
        \avg{\phi_{\bc{a|b}}(w_5)\phi_{\bc{b|a}}(w_6)\phi_{\bc{a|b}}(w_7)\phi_{\bc{b|a}}(w_8)}
    } \nn
    &\equiv& 
    \frac{1}{F_{\bc{abab}}(\eta)}
}
where we used the crossratio
\env{eqnarray}{
    \eta\equiv\frac{w_{56}w_{78}}{w_{57}w_{68}}.
}
The numerical results are shown in Fig.~\ref{fig:res_aaaa}(b), where we find $I([z_5, z_6], [z_7, z_8])$ collapses well to a function only of $\eta$.

The limiting behaviors in $\eta\to 0$ and $\eta \to 1$ can be similarly obtained by considering the appropriate OPE{,
namely Eq.~\eqref{eq:OPE_psi_psi}}, in a fashion 
parallel to Sec.~\ref{sec:fffa_OPE}.
\env{itemize}{
\item
Limit
$z_5 \to z_6, \ z_7 \to z_8$, in which $\eta \to 0$.
{Using twice the OPE in Eq.~\eqref{eq:OPE_psi_psi}
(once for $z_5 \to z_6$ and once for $z_7 \to z_8$),}
leads to the following form
\begin{align}
\label{eq:FababWithaaaaCircuit}
    F_{\bc{abab}}(\eta\to 0)=1+\# \eta^{h_{\bc{a|a}}^{(1)}},
\end{align}
{and therefore we obtain,}
upon making use of Eq.~\eqref{eq:aaaa_I_4pt},
\env{eqnarray}{
    \label{eq:I_eta_0_aaaa}
    I([z_5, z_6], [z_7, z_8]) \propto \eta^{h_{\bc{a|a}}^{(1)}}, \eta \to 0,
}
where we 
extract\footnote{Comments regarding features of logarithmic
CFTs [Log-CFT], analogous to those made in footnote~\ref{FtnLogCFT} of Sec.~\ref{sec:fffa_OPE}, could be made here. Again, because of the inability
to determine the presence of absence of correponding logarithms multiplying
powerlaws we do not elaborate here on these possible features.}
$h_{\bc{a|a}}^{(1)} \approx 2.0$ from the plot
{in Fig.~\ref{fig:res_aaaa}(b)}, consistent with Ref.~\cite{li1901hybrid}.\footnote{The same numerical
value for this exponent was found in the mutual information
for the Hartley entropy for circuits with Haar unitaries
obtained in Ref.~\cite{nahum2018hybrid}.
The same comments concerning logarithms multiplying powerlaws,
as in the previous footnote, can be made here.}

\item
Limit $z_6 \to z_7, \ z_5 \to z_8$, in which $\eta \to 1$.
{Using again the relevant OPE Eq.~\eqref{eq:OPE_psi_psi},
we obtain}
\env{eqnarray}{
&&  \exp\lz -I([z_5, z_6], [z_7, z_8]) \rz \nn
&\propto& \frac{
        \(\frac{1-\eta}{\eta}\)^{2h_{\bc{a|b}}}
    }
    {
        1 + \# \(\frac{1-\eta}{\eta}\)^{h_{\bc{a|a}}^{(1)}}
    },
}
thus
\env{eqnarray}{
    \label{eq:I_eta_1_aaaa}
    && I([z_5, z_6], [z_7, z_8]) \nn
    &\approx& -2h_{\bc{a|b}} \ln\(1-\eta\) + \# \(1-\eta\)^{h_{\bc{a|a}}^{(1)}}, \eta \to 1
}
which has the same form as that in Eq.~\eqref{eq:I_eta_1} and
{the leading behavior $\ln (1-\eta)$ dependence} is verified in Fig.~\ref{fig:res_aaaa}(b).
}

\subsubsection{Entanglement dynamics and the absence of entanglement lightcone}

\env{figure}{[t]
\centering
\includegraphics[width=.5\textwidth]{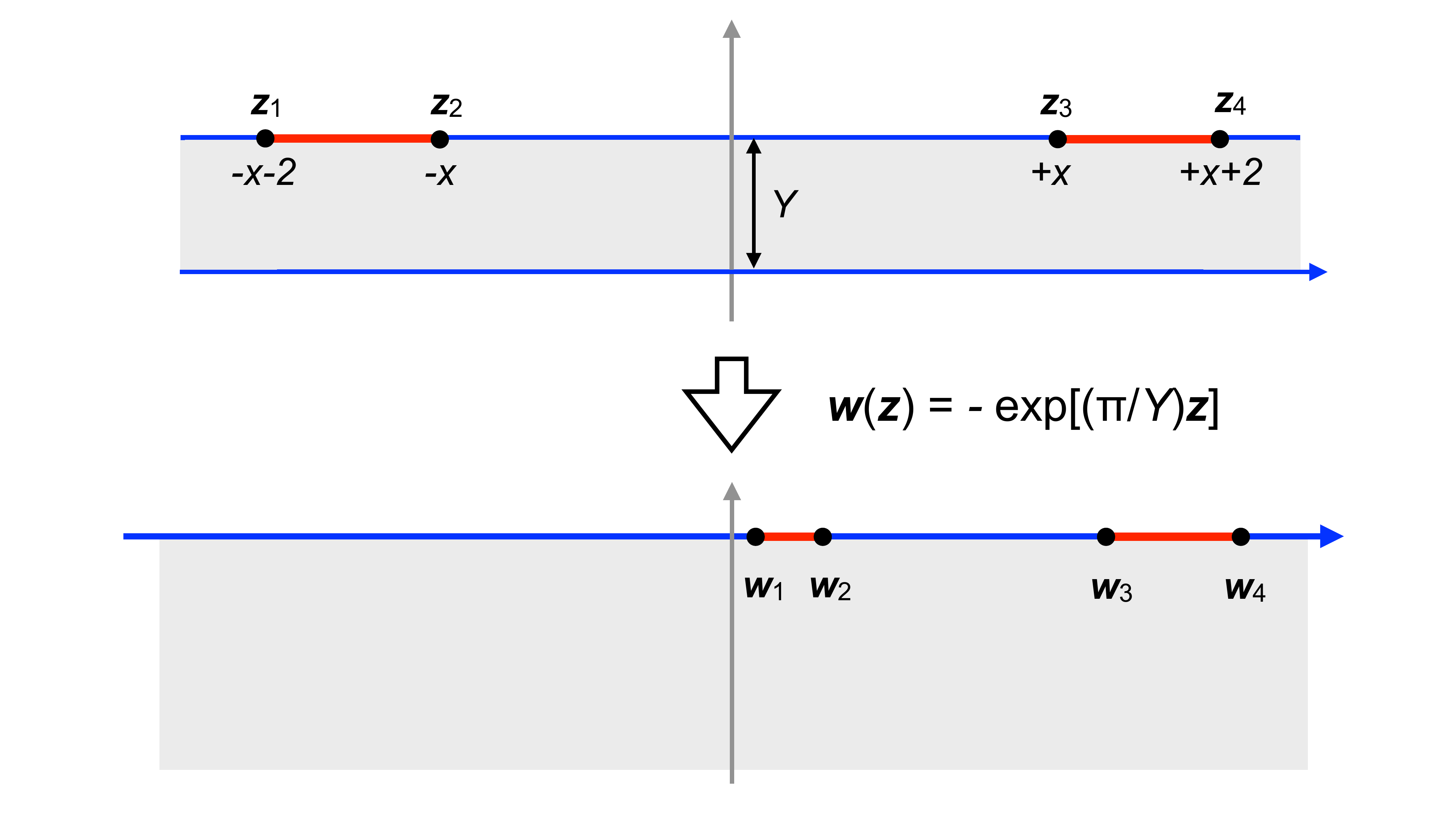}
\caption{
The conformal mapping from the infinite strip with finite width $Y$ to the LHP, allowing calculation of entanglement entropies and mutual information for finite segments.
The infinite strip is obtained by taking the thermodynamic limit ($ L \to \infty$) of the $\bc{aaaa}$ circuit.
}
\label{fig:cfm_exp}
}

As in Sec.~\ref{sec:ee_dyn_fffa}, based on the consistency between the numerics and CFT calculations, we try to obtain an analytic understanding of the entanglement dynamics for $\bc{aaaa}$ using the conformal mapping.
The simplicity of this boundary condition allows us to compute the entanglement entropy in an infinitely large system by first taking $L \to \infty$, where the corners are now 
{unimportant. (Note that, in contrast, for rectangles with boundary
conditions $\bc{fffa}$ the corners are always important because of the bcc operators present.)}  

In Fig.~\ref{fig:cfm_exp} we show a infinite-length system
{($L=\infty$)} with finite 
$Y \propto T${, i.e. an infinite strip}.
The conformal mapping from the infinite strip to the LHP takes the form 
\env{eqnarray}{
    w(z) = -\exp \lz \frac{\pi}{Y} z \rz,
}
where {the upper and lower edge of} 
the strip map to the positive and
negative real axis, respectively.
Using this map, the
entanglement entropy of a finite
interval
$A = [z_1, z_2]$ now can be easily computed,
\env{eqnarray}{
    &&  S([z_1, z_2]) \nn
    &=& -h_{\bc{a|b}} \ln \lz \frac{
        \(\frac{\pi}{Y}\)^2
    }
    {
        \cosh\(\frac{\pi}{Y} z_{12}\) - 1
    } 
    \rz + {\rm const.}\nn
    &\approx& \env{cases}{
        2h_{\bc{a|b}} \ln Y + \frac{h_{\bc{a|b}} \pi}{Y} z_{12}, & Y \propto T \ll z_{12} \\
        2h_{\bc{a|b}} \ln z_{12}, & Y \propto T \gg z_{12}
    }
}
Interestingly, at early times we see a $\ln Y$ growth in addition to the volume law of the entropy (due to the maximal entanglement in the initial state) which ``purifies'' as $\frac{z_{12}}{Y}$ in a similar fashion as Eq.~\eqref{eq:S_U_tau_0},
while at late times the entanglement entropy crosses over to the familiar logarithmic form.
Notice that $2h_{\bc{a|b}}$ has the meaning of the ``coefficient of the log'', found to be approximately $2h_{\bc{a|b}} \approx 1.6 \ln 2$ in Refs.~\cite{li1901hybrid, gullans1905purification}.
It is immediate from the computation that for two intervals $[z_1, z_2]$ and $[z_3, z_4]$ which sit next to each, i.e. where $z_2=z_3$,
their mutual information dynamics becomes
\env{eqnarray}{
    &&        I([z_1, z_2], [z_2, z_3]) \nn
    &\approx&
    \env{cases}{
        2 h_{\bc{a|b}} \ln Y, & Y \propto T \ll z_{12}, z_{23} \\
        2 h_{\bc{a|b}} \ln \(\frac{z_{12} z_{23}}{z_{13}}\), & Y \propto T \gg z_{12}, z_{23}
    }
}
where the early time dynamics is reminiscent of the $\ln Y$ growth of bipartite mutual information in Ref.~\cite{gullans1905purification}.

The dynamics of mutual information of two distant regions is more interesting in that the two regions can share non-zero mutual information with infinite speed.
Consider again the setup in Fig.~\ref{fig:cfm_exp}, where we take two finite intervals
(both of size 2 in this case), separated by a distance $r = 2x$, in an \textit{infinite} system after a circuit evolution of finitely many layers, where the qubits are at the $y=Y$ boundary.
The mutual information between these two intervals follows Eq.~\eqref{eq:I_eta_0_aaaa} in the limit $\eta \to 0$, that is
\env{eqnarray}{
    \label{eq:I_eta_aaaa}
	I(\eta) \sim \eta^{h_\bc{a|a}^{(1)}}, \text{ where } \eta = \frac{w_{12} w_{34}} {w_{13} w_{24}} = \frac{\sinh^2(\frac{\pi}{Y})}{\sinh^2\lz \frac{\pi}{Y} (1+x) \rz}.\nn
}
It is obvious that $I(\eta)$ is nonzero for arbitrarily small but finite
values of
$Y/x$, indicating an infinite entangling speed, in contrast to a finite light speed in a local unitary circuit model, i.e. one in which random projective measurements are \textit{absent}.
More generally, it can be shown that there do not exist finite constants $B, C, v$ such that~\cite{lieb1972finite}
\env{eqnarray}{
\label{eq:LR_bound}
	I(\eta) \le B \exp\lz -C(x-vT) \rz, \text{ for all $x$ and $T$}.
}
In particular, this inequality is violated by Eq.~\eqref{eq:I_eta_aaaa} in the regime $x \gg v T \gg 1$.

The infinite entanglement speed is a direct consequence of conformal invariance, where time is identified as the vertical spatial dimension.
Intuitively, the long-range correlations at the critical point are present for an arbitrarily narrow strip $Y \ll L$, or equivalently, for an arbitrarily shallow circuit.
Physically, this is possible since we have introduced local, unitarity-breaking measurements, leading to an ``entanglement swapping'' mechanism (see Fig.~\ref{fig:EE_swap}) that survives in a 
random
many-body system.\footnote{We note in passing that in Clifford circuits, the growth of the stabilizers is necessarily non-local and there must be no lightcone at the critical point, in any gauge, as required by the long-range mutual information; therefore, a hydrodynamic description with local rules of stabilizer growth cannot be accurate.
Natural extension can be made away from critical point:
with a finite correlation length $\xi$, there will be a maximal velocity as $\xi / \lambda_t$, where $\lambda_t$ is the temporal lattice spacing; this velocity diverges as we approach the critical point (compare discussion in Sec.~\ref{sec:discussion_implication_nonunitary}).}

\section{Periodic boundary condition \label{sec:pbc}}

\subsection{Numerical results}
\label{SubSection:PerodicBCNUmericalResults}
\env{figure}{[t]
\includegraphics[width=.5\textwidth]{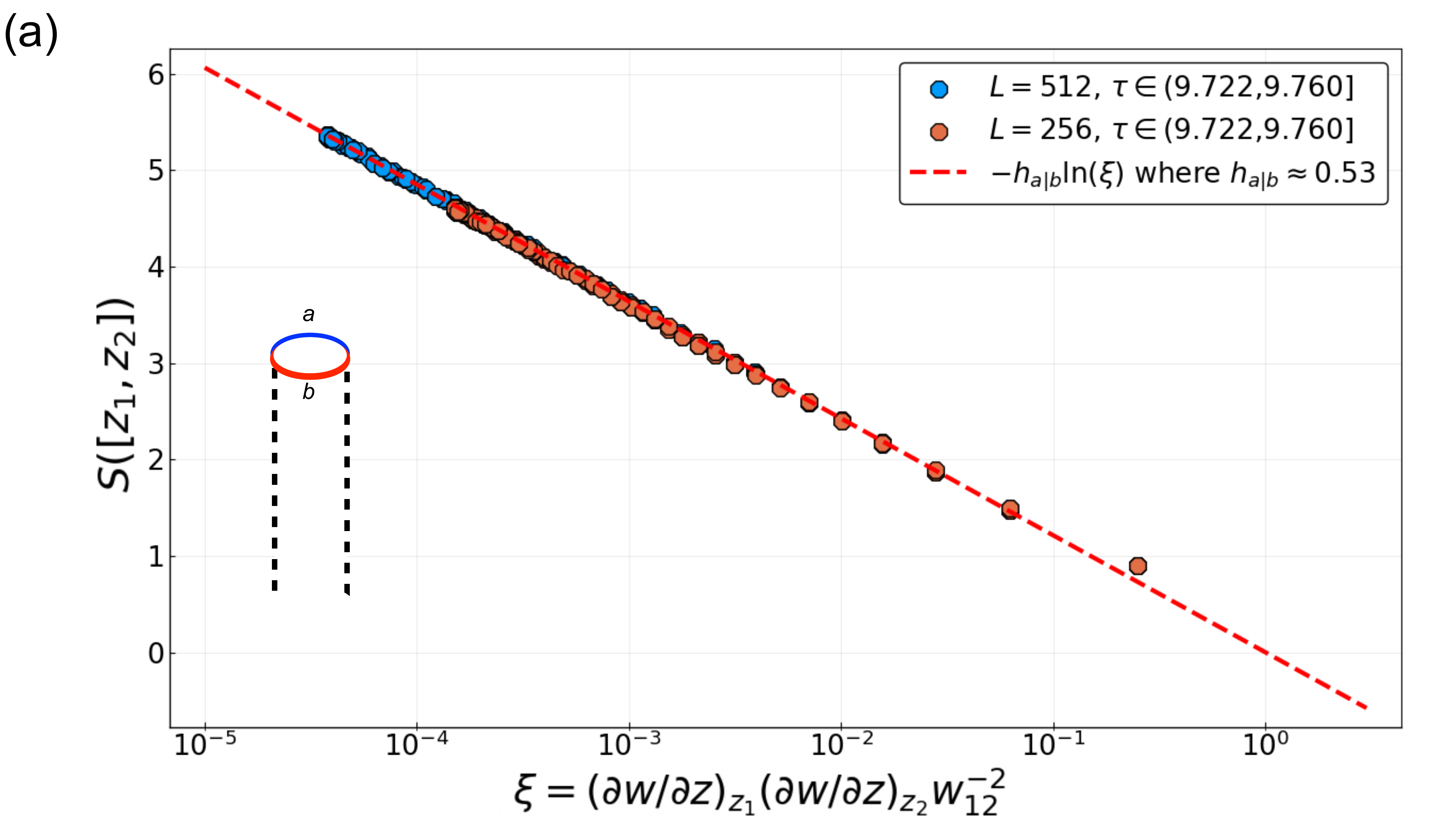}
\includegraphics[width=.5\textwidth]{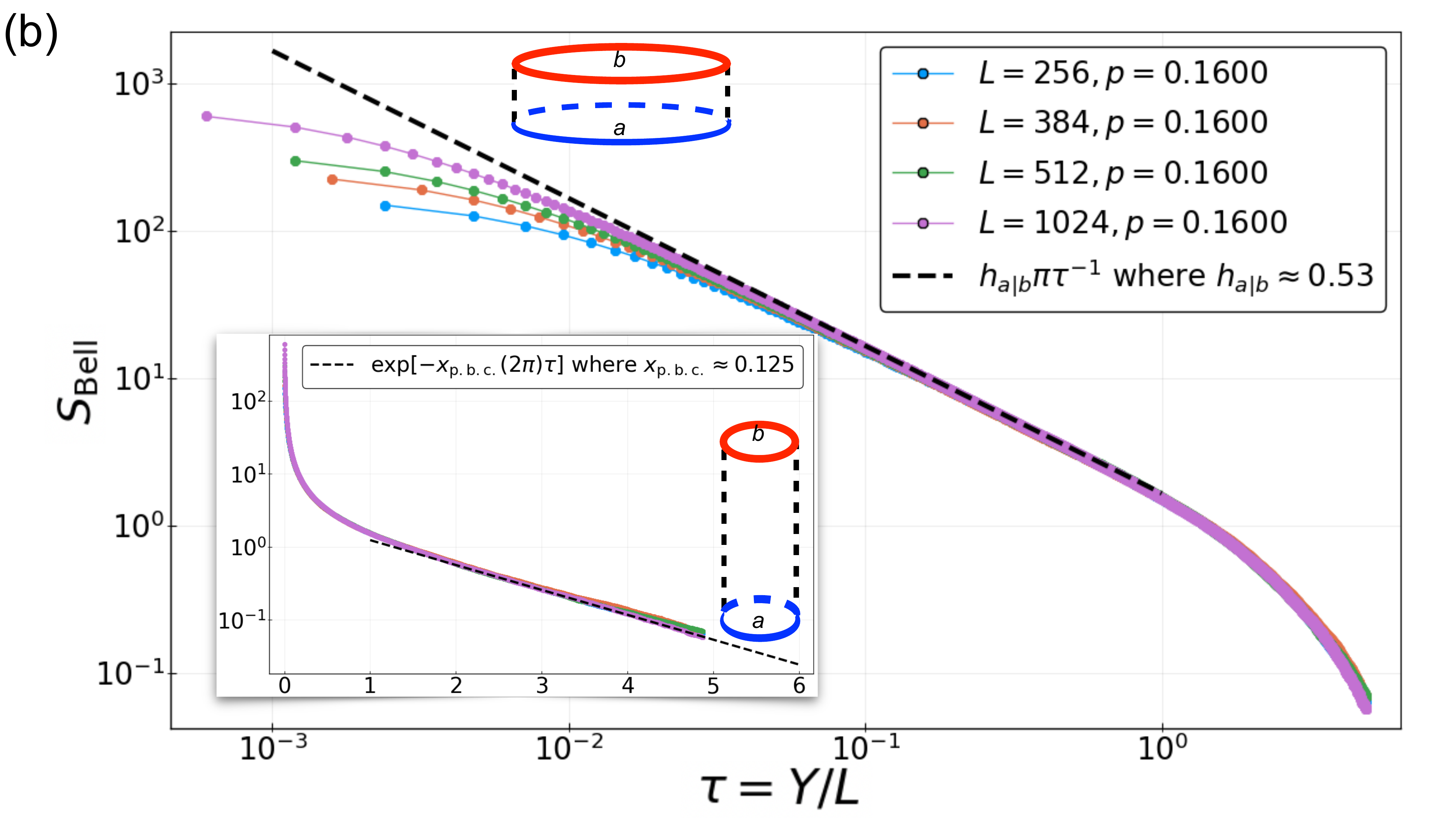}
\caption{
(a)
Starting from a product initial state with periodic spatial b.c., the entanglement entropy in the steady state $\tau \to \infty$ is mapped to a 2-point function.
(b)
Starting from a Bell-pair initial state with periodic spatial b.c., the entanglement entropy of either the {upper and lower} 
edge is predicted to also have the form $S_{\sf Bell} = h_{\bc{a|b}} \pi \tau^{-1}$ in the  $\tau \to 0$ limit, as in Eq.~\eqref{eq:S_U_tau_0_pbc}.
In the limit $\tau \to \infty$, 
{we fit to $S_{\sf Bell} = \exp\(-x_{\rm p.b.c.} \(2 \pi\) \tau \)$ with $x_{\rm p.b.c.} \approx 0.125$,} as in Eq.~\eqref{eq:def_x_pbc}.
}
\label{fig:res_pbc}
}

In this section we consider circuits with periodic spatial b.c.,
which therefore have cylindrical geometry.
This is not quite as simple as a rectangle, since a finite cylinder is topologically distinct from the LHP, and a conformal mapping
{to the latter}
is not available.
{Therefore, the dynamics of entanglement and mutual information is in general more difficult to discuss
as compared to a circuit of rectangular geometry. However,
several simplifications occur in suitable limits 
to be discussed below.}

One simplification occurs in the ``late time limit'', where the depth of the circuit is much larger than the number of qubits, $Y \propto T \gg L$, 
which is when the qubit chain is already in its steady state.
{This limit can be described by a}
semi-infinite cylindrical circuit, 
{which in turn} can then be mapped to the LHP via the following conformal map,
\env{eqnarray}{
\label{eq:conformal_mapping_semi_infinite_cylinder}
z \mapsto w(z) = \tan(\pi z / L).
}
{This leads to the following form of the entanglement entropy in the steady state,}
\env{eqnarray}{
\label{eq:EE_pbc_2PT}
    && S([z_1, z_2]) 
    \qquad \qquad \qquad \qquad \qquad \qquad \qquad
    \text{(Fig.~\ref{fig:res_pbc}(a))}\nn
    &=&-h_{\bc{a|b}} \ln \lz
    \frac{
    \(\frac{\pd w}{\pd z}\)_{z_1}
    \(\frac{\pd w}{\pd z}\)_{z_2}
    }
    {
    (w_{12})^2
    }\rz + {\rm const.},
}
and the collapse of $S([z_1, z_2])$ against $\xi = \frac{
    \(\frac{\pd w}{\pd z}\)_{z_1}
    \(\frac{\pd w}{\pd z}\)_{z_2}
    }
    {
    (w_{12})^2
    }$
is shown in Fig.~\ref{fig:res_pbc}(a), where we again find $h_{\bc{a|b}} = 0.53$.

In the steady state it is also possible to compute the mutual information of two 
non-overlapping intervals, where the OPE in Eq.~\eqref{eq:OPE_psi_psi} is now relevant, and the limiting behavior of $I(\eta)$ 
{end up being the same as that in}
Eqs.~\eqref{eq:I_eta_0_aaaa} and \eqref{eq:I_eta_1_aaaa}.
We again find 
{the same same values of the critical exponents}
$h_{\bc{a|a}}^{(1)} = 2.0$ and $h_{\bc{a|b}} = 0.53$ (data not shown) {that were found in Sec.~\ref{sec:aaaa}.
The value for both exponents
are consistent with that found in
Ref.~\cite{li1901hybrid}.}

Another simplification
{occurs in the limits $Y \ll L$ and $Y\gg L$ for
the Bell entanglement entropy for the $L$ Bell pair initial state.} 
{This is analogous to corresponding limit of the  rectangular circuit with
$\bc{fafa}$ b.c.'s,
discussed in
Sec.~\ref{sec:fafa} above,
except that the qubit chains now have periodic b.c.'s:}

\env{enumerate}{
\item
We first consider $Y \ll L$, in parallel to Sec.~\ref{sec:fafa}.
Using the transfer matrix formalism by treating the spatial 
{direction}
as the ``direction of propagation''
{of the transfer matrix}, the partition function for this setup is given by
\env{eqnarray}{
    Z_{\bc{ab}} = {\rm Tr} \exp \( -H_{\bc{ab}} \times L \),
}
where the trace accounts for the periodic b.c., and $H_{\bc{ab}}$ is the same Hamiltonian 
{as that
in Eq.~\eqref{eq:H_ab} (compare the two insets of
Fig.~\ref{fig:res_pbc}(b)).}
As $L \gg Y$, $Z_{\bc{ab}}$ is again given by the ground state energy of $H_{\bc{ab}}$.
{A similar}
reasoning applies to $Z_{\bc{aa}}$.
Combining these
{results, we obtain}
\env{eqnarray}{
    \exp[-S_{\sf Bell}]=\frac{Z_{\bc{ab}}}{Z_{\bc{aa}}} \sim \exp(-h_{\bc{a|b}} \pi L / Y),
}
thus (recall from Eq.~\eqref{eq:DefinitionOfTau} that $\tau=Y/L$)
\env{eqnarray}{
    \label{eq:S_U_tau_0_pbc}
    S_{\sf Bell} \sim h_{\bc{a|b}} \pi \tau^{-1},
}
{which is the same result as that in}
Eq.~\eqref{eq:S_U_tau_0}.

\item
As $Y \gg L$, we take the temporal direction as the ``direction of propagation'' of the transfer matrix,
{with initial and final states now denoted by}
$\ket{\bc{a}}$ and $\ket{\bc{b}}$.
We have
\env{eqnarray}{
    &&  Z_{\bc{ab}} \nn
    &=& \bra{\bc{a}} \exp(-H_{\rm p.b.c.} \times Y ) \ket{\bc{b}} \nn
    &=& e^{-\epsilon_0 Y}
    \lz
        \braket{\bc{a}}{0} \braket{0}{\bc{b}}
        + \braket{\bc{a}}{1} \braket{1}{\bc{b}} e^{-(\epsilon_1 - \epsilon_0) Y}
        + \ldots
    \rz \nn
}
and similarly
\env{eqnarray}{
    \label{eq:BulkExponents}
    &&  Z_{\bc{aa}} \nn
    &=& \bra{\bc{a}} \exp(-H_{\rm p.b.c.} \times Y ) \ket{\bc{a}} \nn
    &=& e^{-\epsilon_0 Y}
    \lz
        \braket{\bc{a}}{0} \braket{0}{\bc{a}}
        + \braket{\bc{a}}{1} \braket{1}{\bc{a}} e^{-(\epsilon_1 - \epsilon_0) Y}
        + \ldots
    \rz \nn
}
{where $H_{\rm p.b.c.}$ is the Hamiltonian of the
underlying CFT with {\it periodic} b.c.'s., whose excitations
energies are known to be related to scaling dimensions of the
{\it bulk} CFT.
Specifically, we have denote by
$\epsilon_0$ and $\epsilon_1$ the energies of the lowest
and first excited states $|0\rangle$ and $|1\rangle$ of $H_{\rm p.b.c.}$ which
have non-vanishing overlap with both, the final and initial states.
Due to conformal symmetry, the so-defined excitation energy
has the form
{$\epsilon_1-\epsilon_0\equiv 2\pi x_{\rm p.b.c.}/L$,}
where $x_{\rm p.b.c.}$ is a critical exponent of the bulk 
CFT.\footnote{
Here $x_{\rm p.b.c.}=h + {\bar h}=2h$ where $h={\bar h}$
due to translational invariance of the initial and final states.
$h$ is the scaling dimension (conformal weight)
of a primary field in the {\it bulk} CFT.}
}
Therefore we have
\env{eqnarray}{
    \label{eq:def_x_pbc}
    S_{\sf Bell} = \exp\(-x_{\rm p.b.c.} \(2\pi\) \tau\),
}
where $x_{\rm p.b.c.}$ is a scaling dimension
in the {\it bulk} Clifford CFT (i.e. a universal quantity),
and which in general will not coincide with
{\it boundary} scaling dimension $h_\bc{f|f}^{(1)}$
(discussed, e.g., in Fig. \ref{fig:res_fafb}(b)).
}

The results of the numerical computation of $S_{\sf Bell}$ are shown in Fig.~\ref{fig:res_pbc}(b), where we find $h_{\bc{a|b}} \approx 0.53$ and 
{$x_{\rm p.b.c.} \approx 0.125$.}

\subsection{Determination of $p_c$ and $Y/T$}

Due to the simplicity of the periodic b.c. at late times, namely the 
absence
of corner operators and therefore the simple form of Eq.~\eqref{eq:EE_pbc_2PT}, we use this setup for determining $p_c$.
Specifically, we choose $p_c$ such that the plot in Fig.~\ref{fig:res_pbc}(a) fits best to a straight line; this gives us $p_c = 0.1600 \pm 0.0003$, as well as $h_\bc{a|b} \approx 0.53$.
We further define $Y/T$ (where $\tau = Y/L = (Y/T)(T/L)$)  such that Fig.~\ref{fig:res_pbc}(b) fits best to $S_{\sf Bell} \approx h_\bc{a|b} \pi \tau^{-1}$ at small $\tau$.
This gives us $Y/T \approx 0.61$.

$p_c$ and $Y/T$ are the only tuning parameters in our fitting scheme.
Once they are obtained from Fig.~\ref{fig:res_pbc}, they are fixed for all random Clifford circuits in this paper, for which we have found good data collapse.
This confirms our anticipation that $Y/T$ and $p_c$ are b.c.-independent properties of the bulk.


\section{Discussion and outlook\label{sec:discussion}}

\subsection{Summary}

In this paper we presented 
{extensive numerical evidence supporting the presence of
conformal symmetry at the measurement-driven entanglement transition
in the random Clifford quantum circuit,}
via identifying entanglement entropies of
{the}
circuit with {boundary}
free energies of the 
{conformal field theory in the bulk of the circuit}
in response to changes of boundary condition.
With this identification, the critical dynamics of entanglement and mutual information can be understood from analytic computations of
{correlations of boundary condition changing (bcc) operators whose functional form is highly constrained
by conformal symmetry,}
{and we verify explicitly the specific constraint forms of these correlations in our numerics.}
{Moreover, by fitting numerical results for such
correlation function to their functional form predicted by
conformal symmetry, we are able to extract numerical values for}
scaling dimensions of several bcc operators for the circuit with
{\textit{several} sets of \textit{different} boundary conditions,}
and find a remarkable agreement.
These {results}
constitute a consistent charaterization of the Clifford CFT
{underlying the circuit at criticality}.

Crucial to our analysis is the  interpretation of the temporal direction of the circuit as the vertical spatial dimension of the CFT, which then allows a conformal mapping among circuits of different  {aspect ratios of the (space-time)  rectangle,} relating dynamics at different time scales.
{This interpretation of ``time'' was implicit, or has been anticipated in previous works on the entanglement transition~\cite{nahum2018hybrid, li1808hybrid, li1901hybrid, gullans1905purification, choi2019spin, andreas2019hybrid}.}

Conformal symmetry combined with the standard Schwarz-Christoffel conformal map gives analytical control over various finite-size scaling behaviors in the rectangular geometry of the critical circuit.
The circuit depth $T$ corresponds to the size of the ``Euclidean time'' coordinate and thus $T$ or the spatial size $L$ sets the correlation length in the quasi-one-dimensional geometry of a narrow strip when $\tau \ll 1$, or $\tau \gg 1$, respectively, {\cyan with $\tau \propto T/L$}.
This naturally explains the logarithmic growth of the entanglement entropy from an initial product state, as well as the purification dynamics of mixed state~\cite{gullans1905purification}
($h_\bc{a|b} \pi \tau^{-1}$ at small $\tau$, and $e^{-h^{(1)}_\bc{f|f} \pi \tau}$ at large $\tau$).
Other interesting scaling behaviors discussed in this paper can be understood in a similar fashion: they follow directly from conformal invariance.

An immediate consequence of the ``imaginary time''
{and criticality}
is the 
{absence of a}
lightcone in the dynamics of the entanglement structure
{of the circuit}, as highlighted by the infinite speed at which two distant finite regions develop nonzero mutual information (whereas the entire system is in the thermodynamic limit).
This is only possible in the presence of measurements that break unitarity of the time evolution, via a mechanism similar to entanglement swapping.
It is interesting to notice that while measurements reduce entanglement entropy on average, they sometimes ``trade'' short-range entanglement for long-range entanglement, which then helps stabilizing the volume-law phase.
This provides a view of the volume-law phase complementary to the quantum error correction argument in Ref.~\cite{choi2019qec}.

Although we have established our results exclusively 
{for}
the Clifford circuit, our approach builds upon general principles such as conformal invariance and reasonable assumptions about the boundary conditions, without assuming detailed knowledge of the universality class.
Therefore, most of these conclusions 
{will thus clearly immediately}
generalize to entanglement dynamics in other hybrid unitary-measurement circuits for all the R\'{e}nyi entropies, including the $n\geq 1$ R\'enyi entropies
of Haar circuits~\cite{andreas2019hybrid}, as well the $n=0$
(Hartley) R\'enyi entropy in the same circuits~\cite{nahum2018hybrid} (see also Appendix~\ref{sec:perc}).

\subsection{Restatement of the central assumptions}

{In this subsection we restate the central assumptions underlying our work and the underlying logic, which are as follows: 
\env{enumerate}{[(i)]
\item
We assume there is an emergent CFT
describing the two-dimensional space-time in the bulk of the circuit
at the entanglement transition. We provide extensive numerical verification of this assumption in this paper.
\item
Furthermore, we assume that various
boundary conditions on the circuit described 
at  microscopic scales in terms of specific configurations of qubits at the boundary, are described
at the transition by conformally invariant boundary conditions
(as long as the former {\cyan does not} possess non-local
entanglement). This is
a general feature of CFT, and emerges  ultimately from thinking of such
boundary conditions in a Renormalization Group picture. 
\item
Subsequently,
we assume that entanglement properties (such as R\'enyi entropies) of the critical circuit are described by imposition of different boundary conditions which,
by item (ii) above, can be viewed as conformal boundary conditions.
Note that the connection between  entanglement entropies and imposition of different boundary conditions
(for bipartite entanglement properties, different in boundary region $A$, as opposed to at its complement) originates from
Ref.~\cite{vasseur2018rtn} and its sequel~\cite{andreas2019hybrid},
and does not really require an assumption.
Indeed, by repeating the steps presented in Ref.~\cite{andreas2019hybrid}, but now for the reduced density matrix for the \textit{Clifford} circuit, one
directly obtains the central relation Eq.~\eqref{eq:Z_background} between
entanglement entropies and the ratio
of two partition {\cyan functions} of the circuit, one where different (numerator)
and one where the same (denominator) boundary conditions are imposed on the
circuit in region $A$ and its complement. Physically, 
as  first stressed in this context in Ref.~\cite{vasseur2018rtn},
the (negative) logarithm of such a ratio of partition functions corresponds to a difference of
boundary free energies.
In other words,  entanglement entropies
of the circuit are described by (differences of) boundary free energies.
\item
Then, at each point on the boundary where \textit{different}
conformal boundary conditions meet (for the bipartite situation the
endpoints of region $A$), a conformal boundary condition changing
(bcc) operator appears. The leading (lowest scaling dimension)
operator appearing at a boundary change is primary in standard CFT, and
we make the (probably weak) assumption that this is also the case in
the (complicated) CFT describing the circuit. (This assumption
is verified numerically in the work presented here.)
\item
Finally, we make important assumptions central to our work 
about the nature of boundary conditions which we
denote by
$\bc{f, a, b}$, and which are defined (in
Sec.~\ref{model})
at the microscopic (lattice) scale in terms of specific
properties of qubits and their physical properties.
The central objects of our work are then bcc operators changing
between different such qubit-based boundary conditions, such
as e.g. $\phi_\bc{f|a}$ or $\phi_\bc{a|b}$, and the entanglement properties
described by correlation functions of several such bcc operators,
as detailed in the main text for many different situations of physical
interest.
Assumptions about the nature of these microscopically defined boundary
conditions, whose validity we confirm through numerics, are necessary
for the \textit{Clifford} circuits since, in contrast to \textit{random Haar}
circuits, there is no explicit Statistical Mechanics model available
for the former in terms of which an explicit microscopic formulation
of these boundary conditions can be formulated.}
}

\begin{table*}
\centering
\begin{tabular}{c | c | c | c}
\hline
 Operator scaling dimension & A) Clifford CFT & 
 B) $S_0$ in Haar (percolation)
 &
 C) $S_{n \ge 1}$ in Haar as $q \to \infty$ (percolation)
 \\
\hline \hline
 $h_{\bc{a|b}}$
 \footnote{
 We adopt the same convention for all three cases, namely always taking the natural logarithm ($\ln$) in defining the entropy, as specified in Eq.~\eqref{eq:s_rho_ln_rho}.
 The $\ln(2)$ {factors} for A) and B) come from the fact that ``qubits'' (with local Hilbert space dimension $q=2$) are used for constructing the circuits, both in this paper and 
in Ref.~\cite{nahum2018hybrid}
}
 & $0.76\ln(2) \approx 0.53$
 &$\frac{\sqrt{3}}{2\pi} \ln(2) \approx 0.191$~\cite{FirstPassageMath,nahum2018hybrid} 
 &
 $\frac{1}{6} \approx 0.167$~\cite{andreas2019hybrid}
 \\ \hline
 $h^{(1)}_{\bc{f|a}} - h_\bc{f|a}$ & $0.9$ & $0.8$ & - \\ \hline
 $h_\bc{f|f}^{(1)}$ 
 \footnote{In general, this quantity is the lowest dimension boundary operator at the $\bc{f}$ boundary condition (see Eq.~\eqref{eq:H_aa} in Sec.~\ref{Sect:Bell-pair-entanglement-early-times}), also denoted by $\eta_\parallel / 2$ in Ref.~\cite{gullans1910scalable}.
 In 2D percolation the $\bc{f}$ boundary condition is the {\it free} boundary condition of the spins of the $Q$-state Potts model whose $Q\to 1$
 limit describes percolation. Here, the lowest dimension boundary operator at that {\it free} boundary condition is the boundary spin operator, known to have scaling dimension $\frac{1}{3}$.}
 & 0.41 & $\frac{1}{3} \approx 0.333$ & $\frac{1}{3} \approx 0.333$
 \\ \hline
 $h_\bc{a|a}^{(1)}$ & $2.0$\footnote{The appearance of the scaling dimension $2$ here might possibly be related to logarithmic features of the underlying  CFT~\cite{GurarieLudwigCirc2004}. \label{footnote_1}} 
 & $2~{}^{\ref{footnote_1}}$\cite{nahum2018hybrid} & 
  - 
 \\ \hline
 $x_{\rm p.b.c.}$
 \footnote{The scaling dimension $x_{\rm p.b.c.}$ of the {\it bulk}
 operator was defined in Eq.~\eqref{eq:BulkExponents} of Sec.~\ref{SubSection:PerodicBCNUmericalResults}.
 In 2D  critical percolation, it corresponds to the known scaling dimension $\frac{5}{48}$ of the {\it bulk} spin operator.}
 & 0.125 & $\frac{5}{48}\approx 0.104$~\cite{intro_perc_theory} & $\frac{5}{48}\approx 0.104$ \\ \hline
\end{tabular}
\caption{
A comparison of operator scaling dimensions between the Clifford CFT and critical percolation.
Results of A) was already summarized in Table~\ref{table:bcc}, and reproduced here for comparison.
B) refers to the first-passage percolation description of Hartley entropy ($S_0$) in random Haar circuits~\cite{nahum2018hybrid} (where the numerical results are obtained in Appendix~\ref{sec:perc}), and C) refers to the percolation description of $S_{n\ge 1}$ derived in the limit of infinite on-site Hilbert space dimension ($q$) in random Haar circuits~\cite{andreas2019hybrid}.}
\label{table:compare_clifford_percolation}
\end{table*}

\subsection{Universality class of the transition and relationship with critical 2D percolation}

{The universality class of the transition is an interesting question.
For the measurement-induced transition in Haar random circuits with
(finite)
on-site Hilbert space dimension $q$, all
$n$th R\'enyi entropies with $n\geq 1$ are described by a known
statistical mechanics model~\cite{choi2019spin,andreas2019hybrid} in the bulk of the circuit (the R\'enyi
entropies with different $n \geq 1$ being described by different
boundary observables on the same bulk which therefore become critical
at the same value of the tuning parameter, the space-time density of
measurements $p$). On the other hand, the $0$th R\'enyi (Hartley)
entropy is described by a different statistical mechanics
model which becomes critical at a different (higher) value of the density
of measurements.

This statistical mechanics model  describing all
$n$th R\'enyi entropies with $n\geq 1$ turns out~\cite{choi2019spin,andreas2019hybrid} to be
exactly solvable in the limit of infinite onsite Hilbert space dimension
$q\to \infty$, possessing a critical point in the universality class
of two-dimensional percolation. This limit provides a starting point
for a systematic access to the so-far not analytically
understood generic transition at finite $q$, which is the infrared
limit of a renormalization group flow out of percolation by a single
relevant perturbation which emerges because~\cite{vasseur2018rtn,andreas2019hybrid} a finite onsite
Hilbert space dimension $q$ turns out to
(explicitly) break a symmetry that is present when
$q=\infty$.   On the other hand, the $0$th R\'enyi (Hartley)
entropy for any onsite Hilbert space dimension $q$ 
is described~\cite{nahum2018hybrid} by ``minimal cut paths'' in
two-dimensional percolation
(argued to be described by ``first passage percolation'').}

Clifford circuits have only been accessible numerically, but can be studied for very large system sizes.
Recently in Ref.~\cite{gullans1910scalable}, several operator dimensions in Clifford CFT were found to {have numerical {values} close to their counterparts in} percolation, while recognizing some do not.
In a particular setup,
one scaling dimension 
was extracted by looking at the early-time purification dynamics of a single ``reference qubit'', {and further identified with the lowest scaling dimension 
{of the boundary spin operator at}
a \textit{free} boundary in critical percolation, $\frac{\eta_\parallel}{2} = \frac{1}{3}$}.
{We revisit this setup in Appendix~\ref{sec:refQ}, where we
{denote this scaling dimension by}
$h_\bc{f|f}^{(1)}$ (defined in Table~\ref{table:bcc} and in Sec.~\ref{sec:fafa}) whose value appears to be distinct from
{$\frac{\eta_\parallel}{2}=\frac{1}{3}$}.}
A more thorough comparison between Clifford CFT and critical percolation is summarized in Table~\ref{table:compare_clifford_percolation}, which further highlights their differences.
It might perhaps be conceivable that the
appearance of scaling dimensions 
observed in Refs.~\cite{gullans1910scalable, huse1911tripartite}
with values
close to percolation, could be due to a possible proximity of
a percolation fixed point in a generalized phase diagram.


\subsection{Outlook}

\subsubsection{Extensions within the current framework}

Besides going to even larger systems as mentioned above, it would be interesting to also extend the current fitting algorithm off the critical point, and to extract critical exponents such as $\nu$ and $\beta$ \cite{gullans1910scalable, huse1911tripartite}.

It is satisfying that the trivial product state and the $L$-Bell pair state map to 
conformal boundary conditions in the CFT formalism.
Exploring other quantum states (such as a maximally entangled {Page} state~\cite{Page1993}, {which has non-local entanglement})
and 
{attempting to}
{fit them into the current framework}
{would} be an interesting direction.


\subsubsection{Implications of non-unitary dynamics}
\label{sec:discussion_implication_nonunitary}

{The emergence of conformal symmetry}
in hybrid circuits is perhaps in itself not surprising given 
{previous numerical work on Clifford circuits~\cite{li1808hybrid, li1901hybrid}
as well as analytical results on Haar circuits~\cite{nahum2018hybrid, choi2019spin, andreas2019hybrid}.
What is surprising is the way the time dimension fits in the CFT picture, and the consequences 
{that emerge from the fact that the real time coordinate ends up acting as imaginary time.}
Therefore, this type of hybrid dynamics is in a 
class distinct from {ordinary unitary dynamics.}

Although we have established the imaginary time using conformal invariance that is only present at the 
critical point, one can generate a finite (bulk)
correlation length by detuning from the critical point (by letting
$p\not = p_c$).
Certainly, as long as one remains within the scaling limit where the correlation length is much larger than the microscopic lattice scale, the physics is expected to be the standard deformation to a theory with exponentially decaying correlations.
Therefore, one also expects that real time to still act as imaginary time.
This can be seen explicitly in the 2D {\cyan statistical mechanics} 
lattice model describing Haar unitary circuits with measurements~\cite{andreas2019hybrid}.\footnote{Detuning from criticality
only affects the local Boltzmann weights, and thus does not change
the fact that physical (real) time acts as one of two
spatial coordinates of the lattice on which the 2D {\cyan statistical mechanics} 
model is defined.}
However, since all correlations fall off exponentially away from the critical point~\cite{nahum2018hybrid, li1901hybrid}, 
it is only on length scales short compared the correlation length that the measurement-induced quantum non-locality and violations of the Lieb-Robinson bound 
{will be} manifest.

Going beyond the current model, it is {possible} that imaginary time is a general consequence of 
{non-unitarity, and might not}
be restricted to this family of unitary-measurement circuits (see e.g. Ref.~\cite{ueda1709nonlocal}).
It will be interesting if concrete examples of criticality in unitarity-breaking dynamics can be found to confirm this expectation, {possibly identifying other universality classes}.

\subsubsection{Experimental relevance}

As addressed in Refs.~\cite{li1901hybrid, choi2019spin, gullans1910scalable}, the experimental cost of directly accessing the entanglement transition grows exponentially in the product of system size and circuit depth, {since one has to post-select on all the measurement outcomes (which are intrinsically probabilistic, following Born's rule) to produce multiple copies of any wavefunction in order to measure the entanglement entropy (see also footnote~\ref{footnote:ee_swap}), or to estimate variances of correlation functions~\cite{li1901hybrid}.}
Our findings in this paper suggest that the critical behavior is already present at early stages of the circuit evolution,
and one does not have to evolve the circuit all the way to saturation to measure the entanglement entropy; an early time measurement would suffice.
In principle, it can slightly alleviate the experimental challenge.
Yet we have not been able to identify a general {\cyan experimental} protocol that allows efficient access to the transition.

In the special case of Clifford circuits, the quantum state is a ``stabilizer error-correcting code'' at all times~\cite{gottesman9807heisenberg}, for which the two possible post-measurement states resulting from a Pauli measurement are related to one another via a single Pauli string operator, that can be efficiently computed given the knowledge of the stabilizer representation of the state~\cite{aaronson0406chp}.
Thus, one can fix a choice of all the unitary gates $\mc{U}$ and measurement placements $\mc{X}$ in the circuit, as well as all the measurement outcomes $\mc{M}$, and replicate the stabilizer code state resulting from the hybrid circuit evolution $(\mc{U}, \mc{X}, \mc{M})$, by simulating the $(\mc{U}, \mc{X})$ circuit while ``correcting'' the ``errors'' -- measurement outcomes that differ from their counterparts in $\mc{M}$ -- with the application of one ``error correcting'' Pauli string operator (mentioned above) immediately after each error occurs.
This replication algorithm runs in polynomial time; therefore, entanglement entropies can be efficiently measured, at the cost of keeping track of the time evolution of all the stabilizers (a polynomial-time and polynomial-space overhead).

On the other hand, purity of ``reference qubits''~\cite{gullans1910scalable}, as well as the quantum Fisher information~\cite{choi2019qec}, might enable indirect access to the transition in polymonial time on near-term quantum computing platforms~\cite{negnevitsky1804repeated, minev1803catch, google2019supremacy}.


\section*{Acknowledgements}

We acknowledge helpful discussions with Soonwon Choi, Michael Gullans, Victor Gurarie, Tim Hsieh, David Huse, Chao-Ming Jian, Sagar Vijay, Yi-Zhuang You, and Tianci Zhou.
{AWWL also thanks
Chao-Ming Jian, Andrew Potter, Romain Vasseur and
Yi-Zhuang You for collaborations on related topics.}
This work 
{was}
supported in part by the Heising-Simons Foundation (YL and MPAF),
by a postdoctoral fellowship from the Gordon and Betty Moore Foundation under the EPiQS initiative (Grant GBMF4304) at the Kavli Institute for Theoretical Physics (XC), by the DARPA DRINQS program (XC),
and by the National Science Foundation under Grant No. DMR-1309667 (AWWL).
Use was made of computational facilities purchased with funds from the National Science Foundation (CNS-1725797) and administered by the Center for Scientific Computing (CSC). The CSC is supported by the California NanoSystems Institute and the Materials Research Science and Engineering Center (MRSEC; NSF DMR-1720256) at UC Santa Barbara.

\appendix


\section{Review of some elementary results in CFT \label{sec:cft_calc}}

In this Appendix we summarize
{very briefly a number of very basic properties pertaining to}
correlation functions and the operator product expansion
(OPE)
of \textit{primary fields} in CFT~\cite{Polyakov:1970xd}.
Notice that in a boundary CFT, 
{only the (say) holomorphic part of an operator appears, and all
correlation functions below are holomorphic.}

\env{itemize}{
\item
Two-point function:
\env{eqnarray}{
	\avg{\phi_1(w_1) \phi_2 (w_2)}
	=
	\env{cases}{
		c_{12} w_{12}^{-2h}, \text{ if } h_1 = h_2 = h. \\
		0, \text{ if } h_1 \neq h_2.
	}
}
\item
Three-point function:
\env{eqnarray}{
	&&  \avg{\phi_1(w_1) \phi_2(w_2) \phi_3(w_3)} \nn
	&=& c_{123} w_{12}^{-(h_1 + h_2 - h_3)} w_{23}^{-(h_2 + h_3 - h_1)} w_{13}^{-(h_3 + h_1 - h_2)}.
}
\item
Four-point function:
\env{eqnarray}{
	&&  \avg{\phi_1(w_1) \phi_2(w_2) \phi_3(w_3) \phi_4(w_4)} \nn
	&=& F(\eta) \prod_{i < j} w_{ij}^{h/3 - h_i - h_j},
}
where $h = \sum_i h_i$, and $\eta = \frac{w_{12} w_{34}}{w_{13} w_{24}}$ is the cross ratio.
In the case when $h_1 = h_4, h_2 = h_3$, it simplifies to 
\env{eqnarray}{
	&&  \avg{\phi_1(w_1) \phi_2(w_2) \phi_3(w_3) \phi_4(w_4)} \nn
	&=& \tilde{F}(\eta) w_{14}^{-2 h_1} w_{23}^{-2h_2}.
}
\item
Correlation functions are covariant under conformal mappings (in this case $w(z)$),
\env{eqnarray}{
	&&   \avg{\phi_1(z_1) \ldots \phi_n(z_n)} \nn
	&=& \lz \prod_{j=1}^n \( \frac{\pd w}{\pd z} \)_{z_j} \rz^{h_j} \avg{\phi_1(w_1) \ldots \phi_n(w_n)}.
}
\item
The operator product expansion accounts for the short-distance behavior of two operators.
It usually takes the following form,
\env{eqnarray}{
    && \lim_{w_2 \to w_1} \phi_i(w_1) \phi_j(w_2) \nn
    &\propto& \(w_{12}\)^{-h_i-h_j} \sum_{k}
    \(w_{12}\)^{h_k}
    C_{ijk}\,
    \phi_k(w_1),
}
where both sides have the same dimension under global scale transformations (dilations).
The numbers $C_{ijk}$ are ``boundary OPE coefficients''.
The operators $\phi_k$ are usually organized in increasing order of their scaling dimensions $h_k$.
Throughout the paper we have being using the following shorthand notation
\env{eqnarray}{
    \phi_i(w_1) \phi_j(w_2) \sim \(w_{12}\)^{-h_i-h_j} \sum_{k} \(w_{12}\)^{h_k} \phi_k(w_1).
}
}

\section{Purification dynamics of reference qubits in the Clifford Circuit\label{sec:refQ}}

\env{figure}{[b]
\includegraphics[width=.5\textwidth]{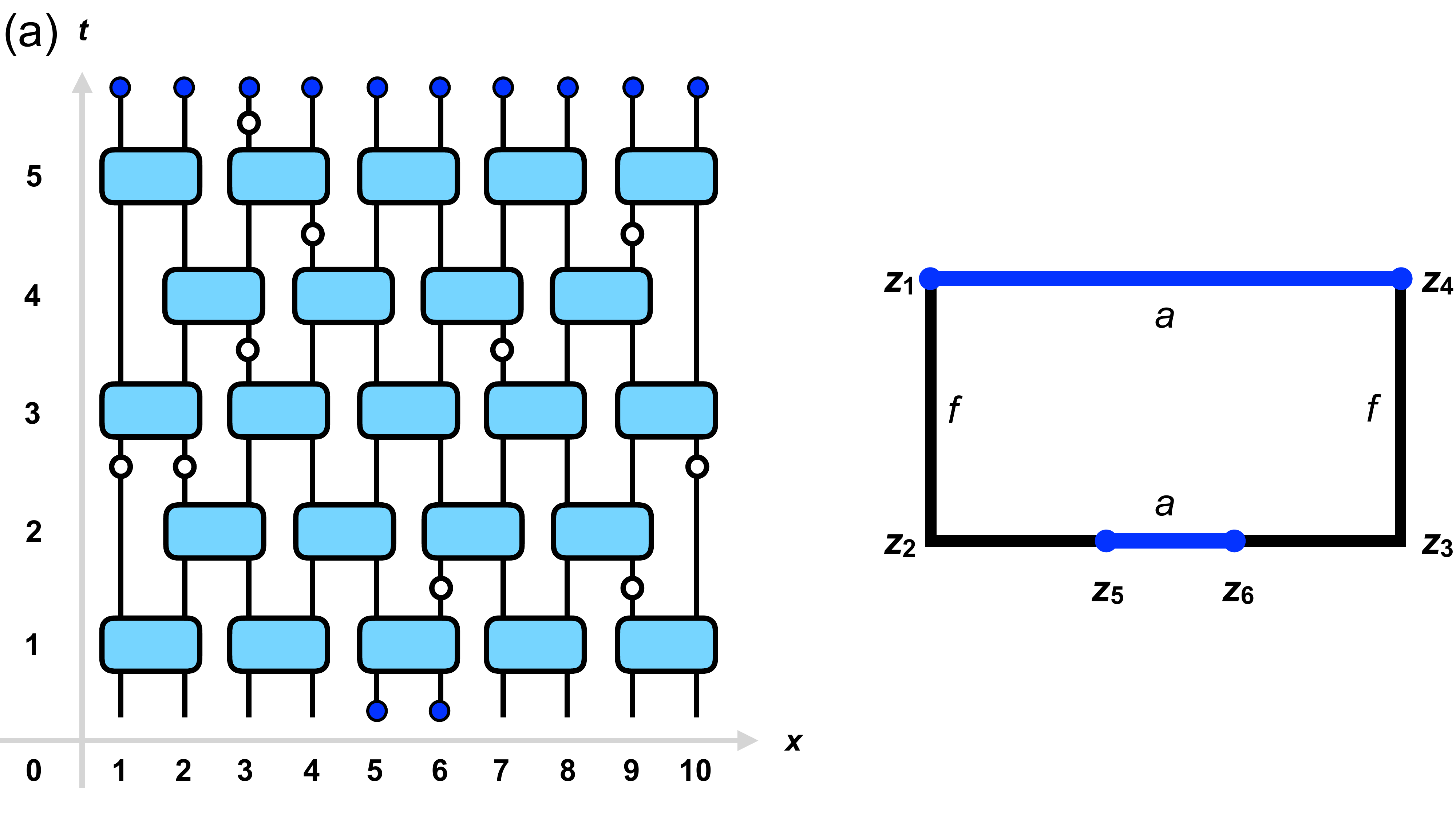}
\includegraphics[width=.5\textwidth]{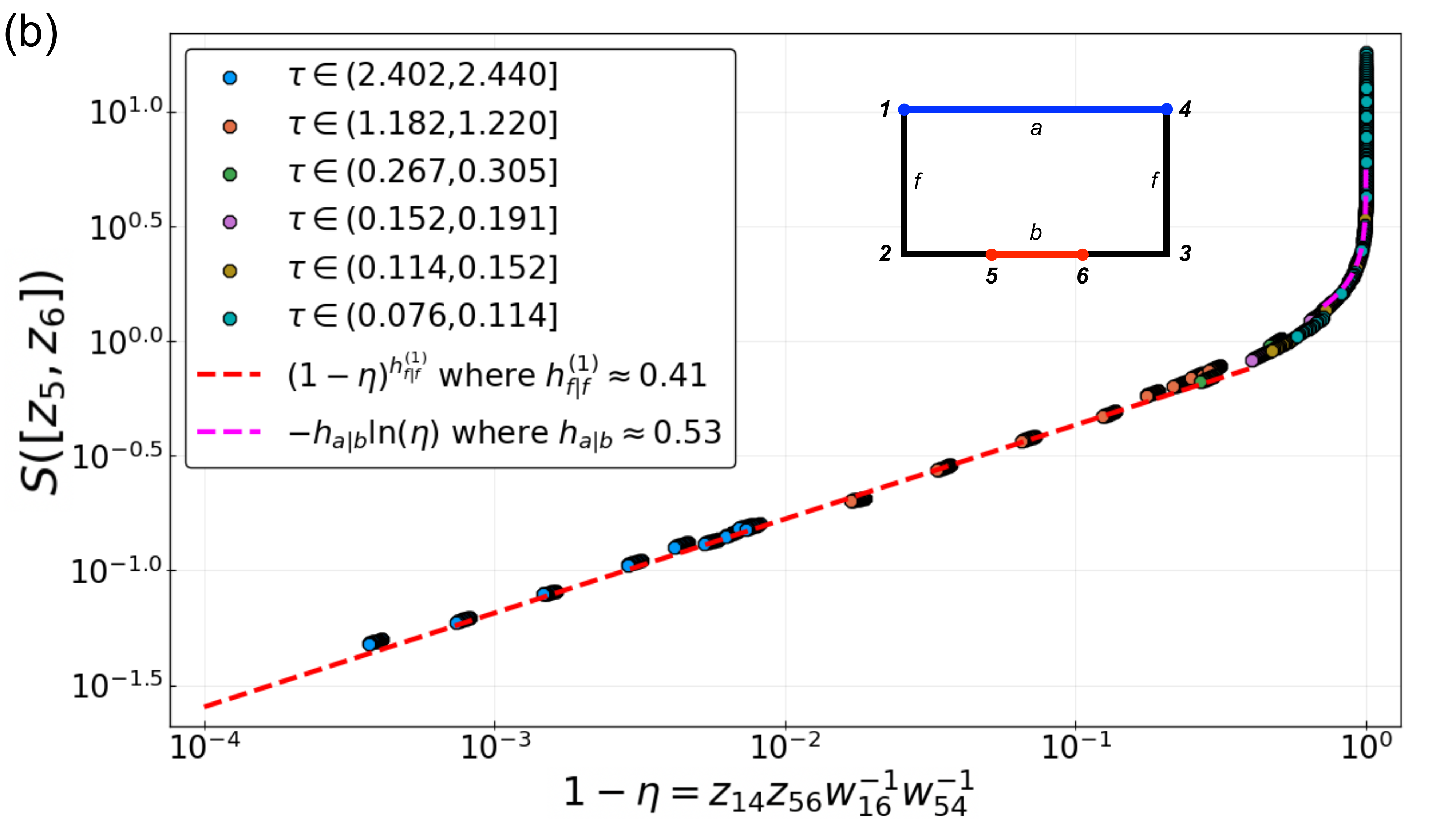}
\caption{
(a) The circuit considered in Appendix~\ref{sec:refQ}, where the description is given in the text.
This is an generalization of one setup introduced in Ref.~\cite{gullans1910scalable}.
(b) Collapsing $S([z_5, z_6])$ to the cross ratio, following Eqs.~(\ref{eq:S_eta_refQ}, \ref{eq:S_eta_refQ_limit}).
The data is obtained for various $z_{56}$ and various circuit depths.
}
\label{fig:refQ}
}

\env{figure*}{[t]
\subfigure[]{
\includegraphics[width=.48\textwidth]{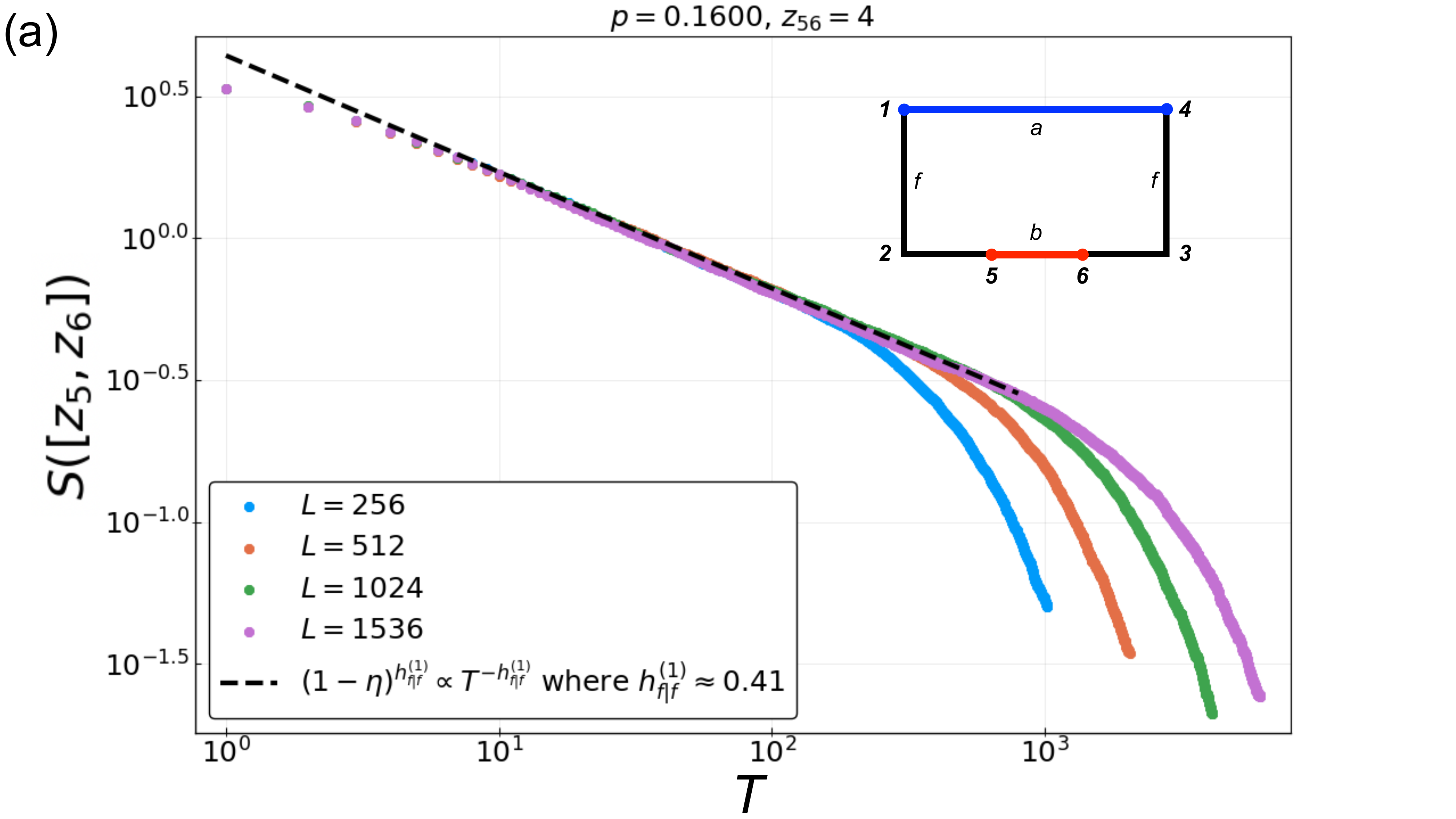}
}
\subfigure[]{
\includegraphics[width=.48\textwidth]{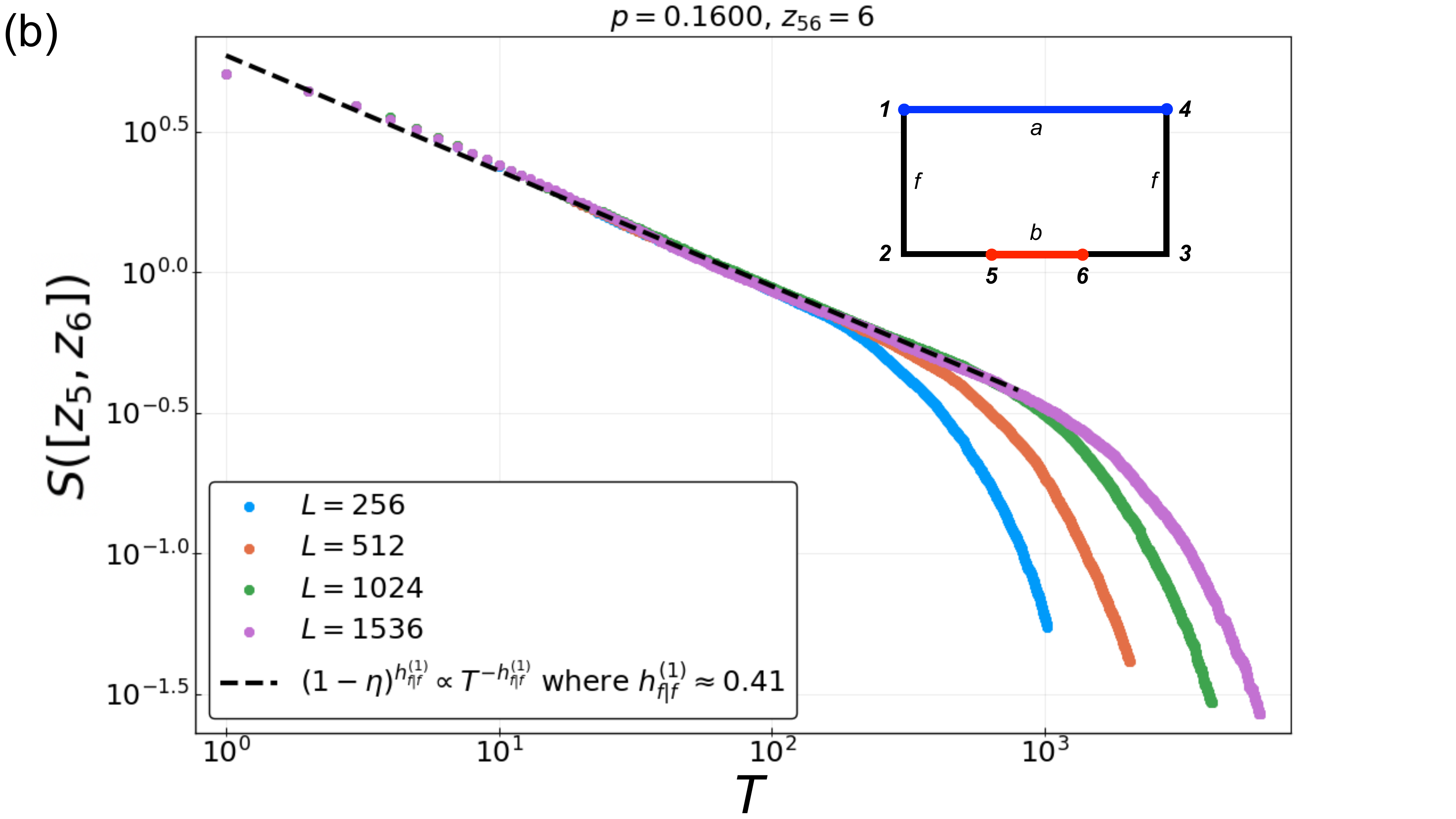}
}
\subfigure[]{
\includegraphics[width=.48\textwidth]{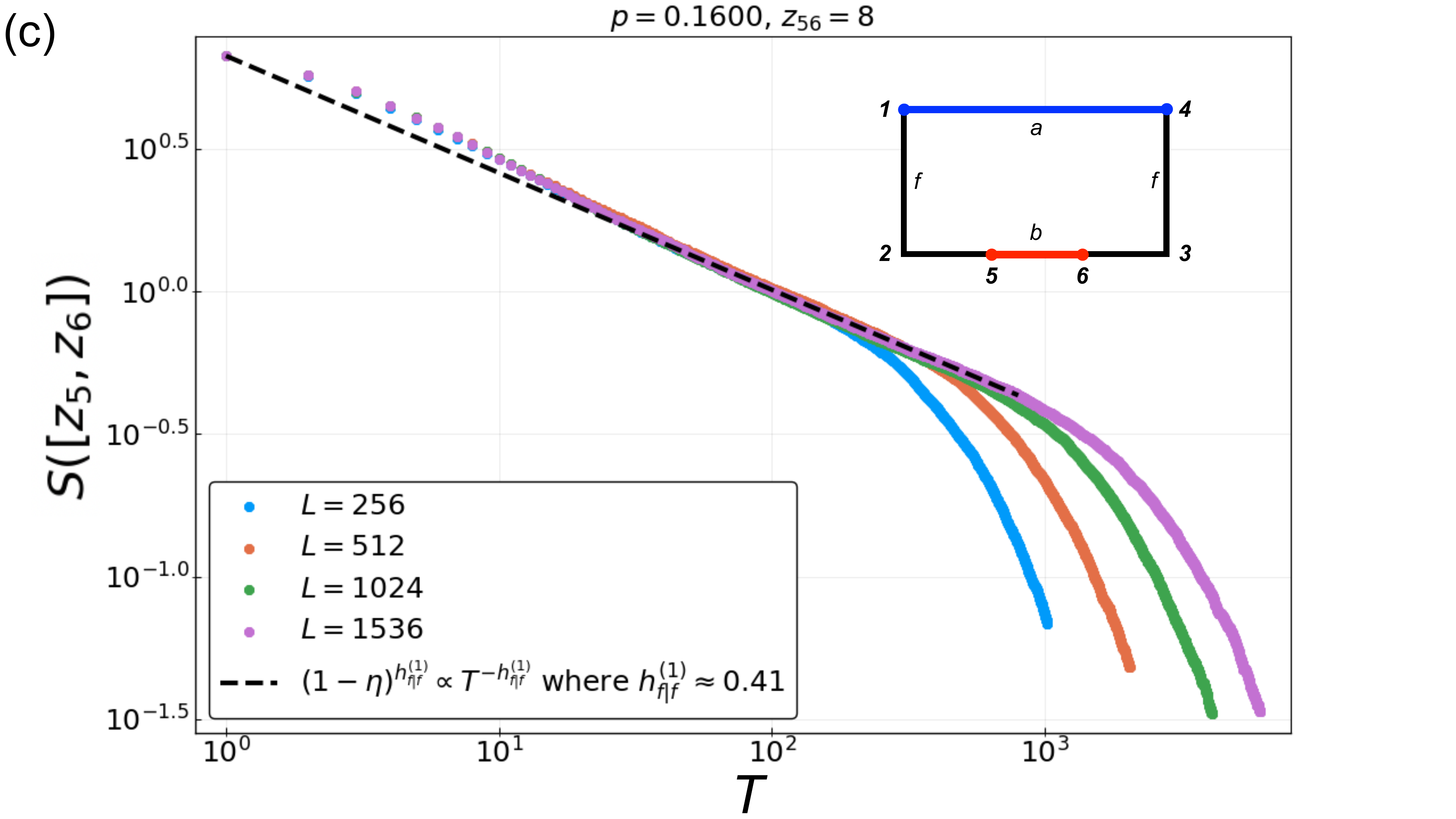}
}
\subfigure[]{
\includegraphics[width=.48\textwidth]{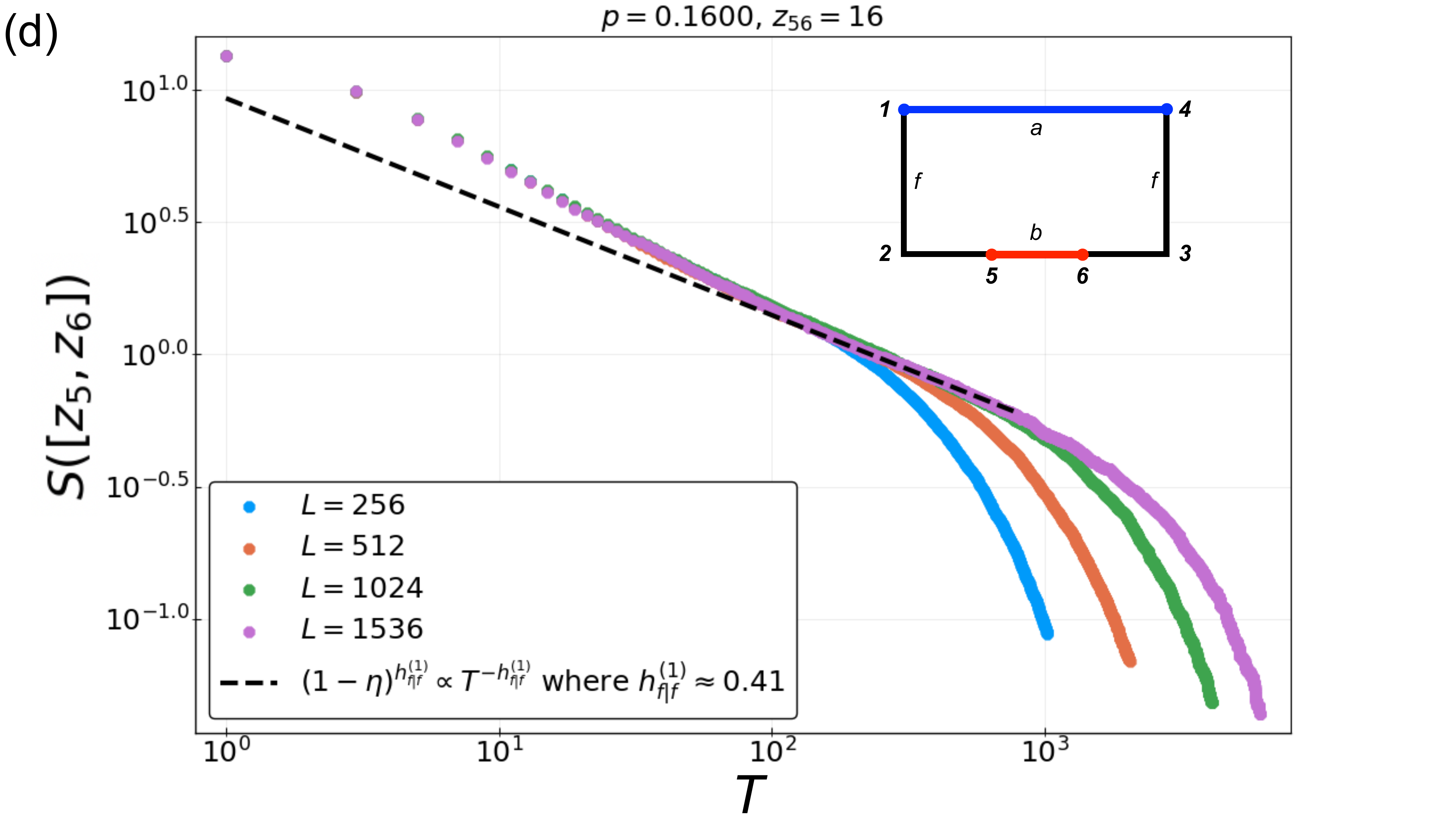}
}
\caption{
Early time data of $S(A = [z_5, z_6])$, with $|A| = z_{56}$ in $\{4, 6, 8, 16\}$.
The data matches well with Eq.~\eqref{eq:refQ_early_time} at intermediate times $z_{56} \ll T \ll L$, which we fit for $h_\bc{f|f}^{(1)} = 0.41$.
As $|A|$ increases, the allowed time window for fitting to the powerlaw shrinks.
}
\label{fig:refQ_power_law}
}

In this Appendix we consider yet another 
{boundary condition on the Clifford circuit  discussed in the main part 
this paper, which is similar to the circuit}
introduced in Ref.~\cite{gullans1910scalable}.
The setup is as follows.
\env{itemize}{
\item
One starts with a chain of $L$ qubits in a product state.
\item
One picks a contiguous segment $A$ 
{containing a number of $|A|$}
qubits
from 
{this}
chain, and 
{entangles}
each of them with an 
{extra, additional
``reference qubit''}
{with which it forms a maximally}
entangled Bell pair.
There are therefore $|A|$ Bell pairs, each containing one
{``reference qubit''}
and one ``system qubit'', in addition to the 
{remaining $L-|A|$ ``system qubit'' of the original chain of $L$ qubits.}
In Ref.~\cite{gullans1910scalable}, $|A|$ is always taken
{to be unity, $|A|=1$.}
\item
One then evolves the 
{``system'', by which we mean the}
{original chain of $L$ qubits (i.e. the $|A|$
``system qubits'' as well as the remaining $L-|A|$ qubits of
the original chain, but {\it not} the ``reference qubits'')}
with the critical hybrid 
circuit.\footnote{Note that the case where $|A|=L$, i.e. where
the ``system qubits'' are all the $L$ qubits of the original chain,
was the $\bc{fafa}$ circuit previously discussed in Fig.~\ref{fig:upc_rect}(c) and Sec.~\ref{sec:fafa}.
In that previous discussion the ``system qubits'' were referrred to as ``the system'', whereas the ``reference qubits'' were referred to as ``the environment''. The current situation is thus a generalization of this previously considered setup, and everything said in this Appendix is a  natural extension of the discussion of that previous discussion in the main text.}
\item
The quantity of interest is the entanglement entropy between 
{``the reference qubits'' and ``the system'', a quantity
denoted by $S_Q$}
in Ref.~\cite{gullans1910scalable}.
}
This circuit is illustrated Fig.~\ref{fig:refQ}(a).
Following our 
{conventions in Fig.~\ref{fig:upc_rect}(c),}
we postulate that there are now $|A|$ physical qubits living on the lower edge of the rectangle (on $A= [z_5, z_6] \subset [z_2, z_3]$), 
{indicated by solid blue dots}, and $L$ qubits living on the top (on $[z_1, z_4]$), {also indicated by solid
blue dots, implying}
the b.c. 
{shown in the same figure.}
{Note that this is again the b.c. of type
$\bc{fafa}$ (as in Fig.~\ref{fig:upc_rect}(c)),}
whereas the entanglement between
{``the system'' and ``the reference qubits''}
is again given by the difference in free energy between
{boundary conditions of types}
$\bc{fafb}$ and $\bc{fafa}$, 
{completely analogous to the discussion in}
Sec.~\ref{sec:fafa}.
{In fact, when $|A|=L$, this is exactly the
circuit discussed in in Sec.~\ref{sec:fafa} (as already mentioned in
the previous footnote).}

Explicitly, $S(A = [z_5, z_6])$ is given by
\env{eqnarray}{
    \label{eq:S_eta_refQ}
    && \exp[ -S(A = [z_5, z_6]) ] \nn
    &=&
    \frac{
		\avg{\phi_{\bc{f|a}}(z_1) \phi_{\bc{a|f}}(z_4) \phi_{\bc{f|b}}(z_6) \phi_{\bc{b|f}}(z_5) }
	}
	{
		\avg{\phi_{\bc{f|a}}(z_1) \phi_{\bc{a|f}}(z_4) \phi_{\bc{f|a}}(z_6) \phi_{\bc{a|f}}(z_5) }
	}\nn
	&=& \frac{F_\bc{fafb}(\eta)}{F_\bc{fafa}(\eta)},
}
where the $F$ functions are those defined in Eq.~\eqref{eq:fafb_over_fafa}, and \env{eqnarray}{
\eta = \frac{w_{15} w_{64}}{w_{16} w_{54}}
}
is the relevant cross ratio.
The data collapse of $S(A=[z_5, z_6])$ against $\eta$, computed 
{by varying $z_{56}$ and the}
circuit depth, is shown in Fig.~\ref{fig:refQ}(b).
The quality of the collapse supports our assumption about the b.c., and the behavior of the collapsed function in the two limits ($\eta \to 0$ and $\eta \to 1$) are consistent with Eqs.~(\ref{eq:S_U_tau_0}, \ref{eq:S_U_tau_1}), namely,
\env{eqnarray}{
    \label{eq:S_eta_refQ_limit}
    S([z_5, z_6]) = \env{cases}{
        -h_\bc{a|b} \ln \eta, & \eta \to 0,\\
        (1 - \eta)^{h_\bc{f|f}^{(1)}}, & \eta \to 1.
    }
}
{In particular, the numerical estimates
for the exponents $h_\bc{a|b}$ and $h_\bc{f|f}^{(1)}$,
extracted
from this analyis, are fully consistent with those obtained
previously for the same exponents in the main text.}

We can now use this result to obtain an analytic understanding of the behavior of $S(A=[z_5, z_6])$.
We focus on 
{the}
regime, $|A| = z_{56} \ll Y \ll L$.
In order to simplify the calculation of $\eta$, we adopt a different 
{convention for}
the conformal mapping, where
\env{eqnarray}{
    w_1 &=& w(z_1) = -m^{-1/2}, \\
    w_2 &=& w(z_2) = -1, \\
    w_3 &=& w(z_3) = +1, \\
    w_4 &=& w(z_4) = +m^{-1/2}.
}
This is related to the 
{previous convention}
defined in Fig.~\ref{fig:cfm_sn} by a global 
{(``fractional linear'')}
conformal transformation,
under which $\eta$ is invariant.
We further focus on the case when $z_5$ sits at the center of the system, where 
\env{eqnarray}{
    w_5 &=& w(z_5) = 0, \nn
    w_6 &=& w(z_6) \approx \(\frac{\pd w}{\pd z}\)_{z_5} z_{56} = \frac{K(1-m)}{Y} z_{56}.
}
Given $z_{56} \ll Y \ll L$, the cross ratio can be shown to be (using Eq.~\eqref{eq:tau_asymt_form})
\env{eqnarray}{
    1 - \eta \propto \frac{\pi z_{15}}{Y}, \quad \frac{z_{56}}{L} \ll \tau \ll 1.
}
Therefore, at early times,
\env{eqnarray}{
    \label{eq:refQ_early_time}
    S([z_5, z_6]) \propto (1-\eta)^{h_\bc{f|f}^{(1)}} \propto Y^{-h_\bc{f|f}^{(1)}} \propto T^{-h_\bc{f|f}^{(1)}}.
}
This behavior is directly observed in Fig.~\ref{fig:refQ_power_law}, where $|A| = z_{56}$ takes values in $\{4, 6, 8, 16\}$, where we find $h_\bc{f|f}^{(1)} = 0.41$.
In this particular case, it is preferable to keep $z_{56}$ small, while going to rather large system sizes, because of the constraint $z_{56} \ll Y \ll L$.
When $Y$ is comparable to $L$, the decay is exponential, as the circuit starts to crossover to a quasi-one-dimensional system (similar to Sec.~\ref{sec:fafa}).
\footnote{Notice that the powerlaw form $T^{-h_\bc{f|f}^{(1)}}$ in Eq.~\eqref{eq:refQ_early_time} does not depend on $Y/T$, and therefore should be regarded as an estimation of $h_\bc{f|f}^{(1)}$ independent of that in Fig.~\ref{fig:refQ}.
Off the critical point, there should still be a time window $z_{56} \ll T \ll \xi$ for which Eq.~\eqref{eq:refQ_early_time} applies, and therefore this estimation of $h_\bc{f|f}^{(1)}$ is also expected to be insensitive to the choice of $p_c$.
This expectation is numerically confirmed but the results are not displayed here.
}

{Extending Eq.~\eqref{eq:refQ_early_time} into the volume law phase $p < p_c$ in the late time limit when $T \gg \xi$ (the correlation length), the time $T$ should be replaced by $\xi$.
Therefore, Eq.~\eqref{eq:refQ_early_time} gives steady state value of $S([z_5, z_6])$,
\env{eqnarray}{
     \lim_{T \to \infty} S([z_5, z_6])
     \propto \xi^{-h_\bc{f|f}^{(1)}} \propto |p-p_c|^{\nu \times h_\bc{f|f}^{(1)}}.
}
This means that the reference qubit can only purify to a finite nonzero value when measurements are below the critical 
{rate, i.e. $ p < p_c$.}
In Ref.~\cite{gullans1910scalable}, $h_\bc{f|f}^{(1)}$ is identified with $\frac{\eta_\parallel}{2}$, therefore $\nu \times h_\bc{f|f}^{(1)}$ can be identifed with $\beta_\parallel$, following a standard hyperscaling relation $\beta_\parallel = \frac{1}{2}\(d-2+\eta_\parallel\)\nu$ in $d = 2$. 
Therefore, $S([z_5, z_6])$ acquires the meaning of an order parameter.
}

In Ref.~\cite{gullans1910scalable}, a different value of $h_\bc{f|f}^{(1)} \approx 0.33$ is extracted from $S([z_5, z_6])$ with $z_{56} = 1$, for a slightly different location of the transition ($p_c \approx 0.1590$) and with periodic spatial b.c..
Within our setup, we also find $h_\bc{f|f}^{(1)} \approx 0.33$
{to be}
a reasonable fit for $z_{56} = 1$, but not
{so}
for $z_{56} > 1$.
This is possibly due to the following subtleties with the one-qubit-purification data:
\env{itemize}{
\item
Statistical error.
In a Clifford circuit, all the entanglement entropies
{(when measured in units of $\ln 2$)}
are integers, and when $z_{56} = 1$, the 
{entropy}
$S([z_5, z_6])$ jumps discretely 
{between $1$ and $0$}
in a single realization {of disorder}
of the circuit.
Therefore, one must
{sample a large number }
of disorder realizations in order to 
{arrive at}
a good resolution for 
{the expectation value of the entropy.}
The smallness of this quantity 
{at small values of $z_{56}$}
also makes it more susceptible to satistical fluctuations.
\item
{Effects arising from}
finite subsystem size.
$S([z_5, z_6])$ always starts
{for small circuit depth}
with the value $\ln 2$, as given by the number of reference qubits.
Numerically, this initial value is below the predicted form in Eq.~\eqref{eq:refQ_early_time}, therefore one must wait for a while ($T^*$) before $S([z_5, z_6])$ matches on to Eq.~\eqref{eq:refQ_early_time}.
Before $T^*$, the purification will be slower than predicted, thereby giving a smaller estimation of $h_\bc{f|f}^{(1)}$.
$T^*$ presumably depends on the details of the model, as well as on $z_{56}$.
}
Due to these subtleties, we are hesitant to extract $h_\bc{f|f}^{(1)}$ from $S([z_5, z_6])$ with $z_{56} = 1$, and are instead more comfortable
{using values}
when $z_{56} \ge 4$.
These issues, however, should be resolved with a larger disorder ensemble and even larger system sizes,
{but this is beyond}
the scope of the current work.
Despite these issues, $S([z_5, z_6])$ (or $S_Q$) should still be viewed as an order
{parameter, which will represent}
a possible experimental probe of the transition.

Ref.~\cite{gullans1910scalable} also presents results of growth of mutual information between two disjoint reference qubits.
In the current framework, these would correspond to 6- or higher-point functions, for which the calculations require detailed knowledge of the CFT (although in certain limits they reduce to simpler, 4-point functions).
We have not attempted to analyze these.

\section{The scaling dimension $h_\mathsfit{f|f}^{(1)}$ from ``localizable entanglement'' \label{sec:localizable_entanglement}}

\begin{figure}[b]
    \centering
    \includegraphics[width=.5\textwidth]{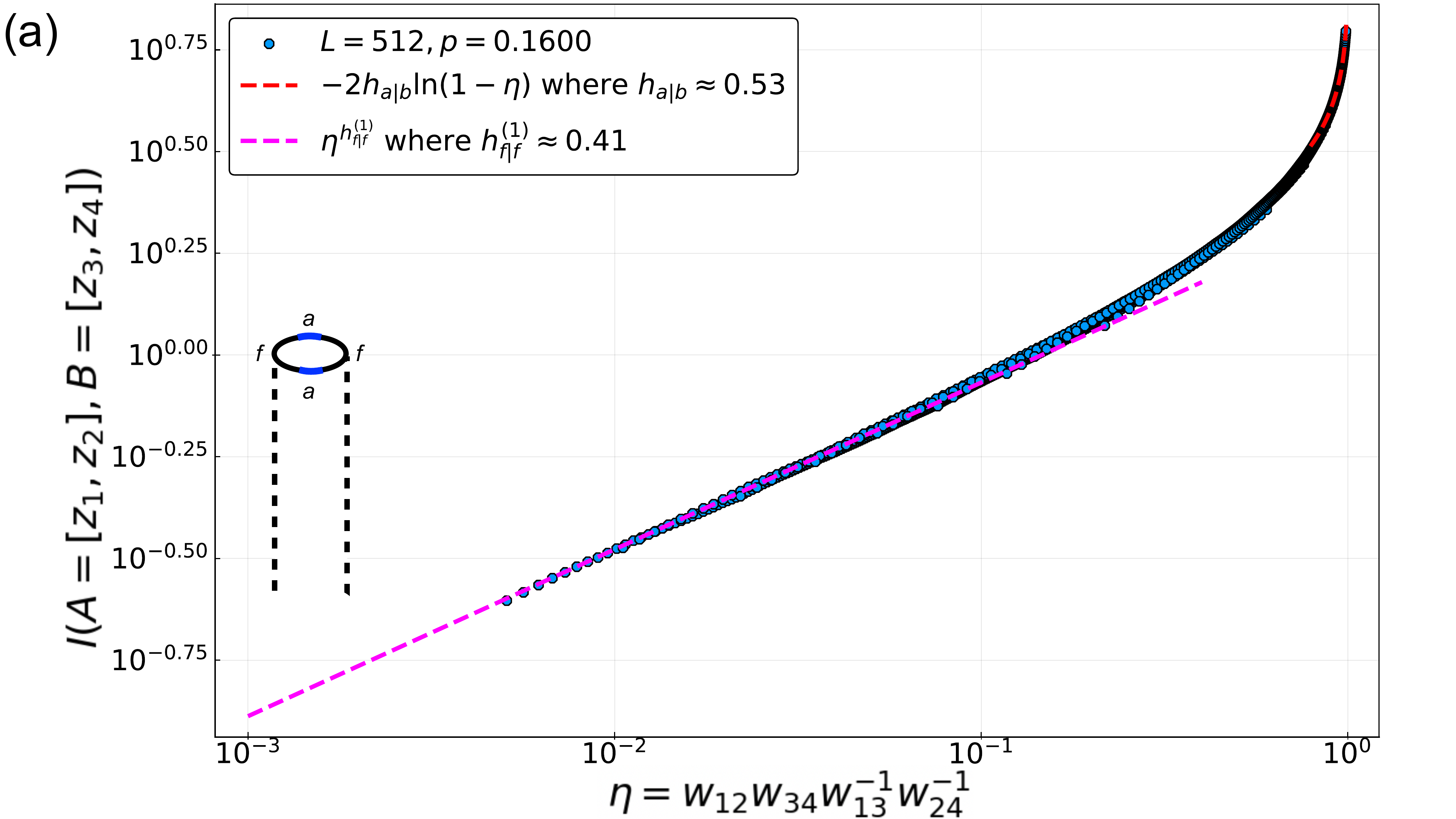}
    \includegraphics[width=.5\textwidth]{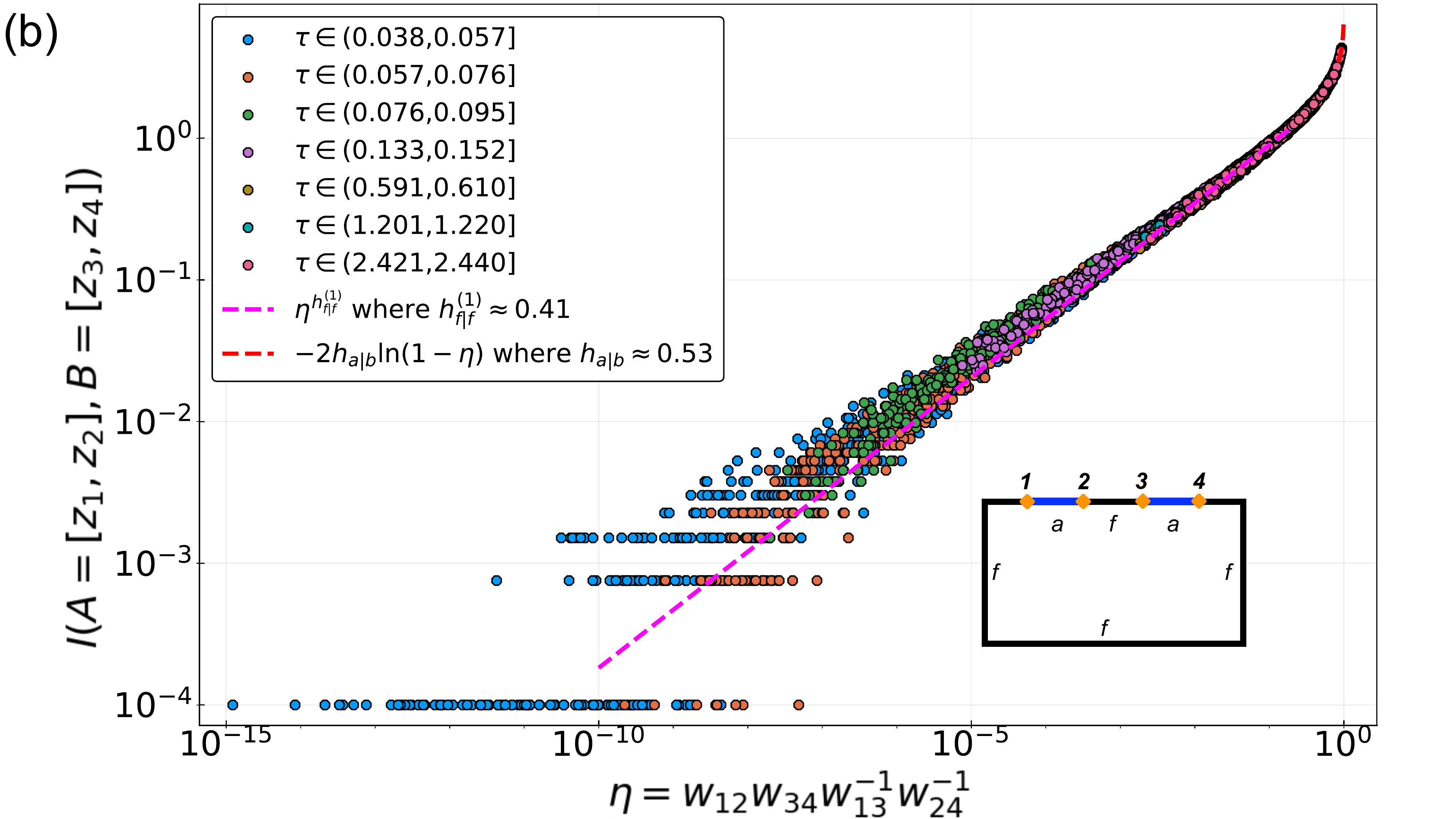}
    \caption{The mutual information between $A = [z_1, z_2]$ and $B = [z_3, z_4]$, after all qubits outside $A \cup B$ are projected out in the final state of the circuit.
    For both (a) periodic and (b) open boundary conditions, we find $I([z_1, z_2], [z_3, z_4]) \propto \eta^{h_\mathsfit{f|f}^{(1)}}$ as $\eta \to 0$, where the value of $h_\mathsfit{f|f}^{(1)} \approx 0.41$ is in excellent agreement with previous results (see Figs.~(\ref{fig:res_fafb}, \ref{fig:refQ}, \ref{fig:refQ_power_law})) and is markedly different from percolation (see Table~\ref{table:compare_clifford_percolation}).
    }
    \label{fig:localizable_MI}
\end{figure}

In this Appendix, we present another method for extracting the scaling dimension $h_\mathsfit{f|f}^{(1)}$, using a quantity similar to the so-called ``localizable entanglement''~\cite{Cirac2004LocalizableEntanglement}.
In this set up, the circuit initial state is taken to be a product state, corresponding to the boundary condition $\mathsfit{f}$ (see e.g. Fig.~\ref{fig:upc_rect}(a)).
In the final state of the circuit, we choose two disjoint subregions $A = [z_1, z_2]$ and $B = [z_3, z_4]$, and perform a projective measurement on every qubit outside $A \cup B$.
The projective measurements create a product state in $A \cup B$, thus also correspond to the boundary condition $\mathsfit{f}$, as we posit.\footnote{In the random Haar circuit, it can be shown that projective measurements do indeed create the free boundary condition, following the mapping developed in Refs.~\cite{andreas2019hybrid, choi2019spin}.}
The boundary conditions are shown in the insets of Fig.~\ref{fig:localizable_MI}(a,b).

For the circuit with periodic boundary condition, we focus on the steady state ($\tau = Y/L \gg 1$) and collapse the mutual information $I(A = [z_1, z_2], B = [z_3, z_4])$ against the cross ratio $\eta$, following the conformal mapping in Eq.~\eqref{eq:conformal_mapping_semi_infinite_cylinder} from the semi-infinite cylinder to the LHP.
The results are shown in Fig.~\ref{fig:localizable_MI}(a).
In particular, in the limit of small $\eta$, the OPE channel of two $\phi_{\mathsfit{f|a}}$ fields is relevant (see Eq.~\eqref{eq:ope_phi_phi_sub}), and we expect
\begin{align}
    I([z_1, z_2], [z_3, z_4]) \propto \eta^{h_\mathsfit{f|f}^{(1)}}, \quad \eta \to 0.
\end{align}
From Fig.~\ref{fig:localizable_MI}(a), we fit for $h_\mathsfit{f|f}^{(1)} \approx 0.41$, in excellent agreement with Figs.~(\ref{fig:res_fafb}, \ref{fig:refQ}, \ref{fig:refQ_power_law}).
Recall that in Figs.~(\ref{fig:res_fafb}, \ref{fig:refQ}) the estimate of $h_\mathsfit{f|f}^{(1)}$ relies on the fitting parameter $Y/T$, and in Fig.~\ref{fig:refQ_power_law} the estimate is restricted to an intermediate time scale.
The method here avoids both issues, and gives us an independent, consistent estimate of $h_\mathsfit{f|f}^{(1)}$, lending strong support that the Clifford CFT is distinct from percolation (see Table~\ref{table:compare_clifford_percolation}).

For the circuit with open boundary condition, the mutual information $I(A = [z_1, z_2], B = [z_3, z_4])$ is again given by the difference in free energy between
{boundary conditions of types}
$\bc{fafb}$ and $\bc{fafa}$, 
{completely analogous to the discussion in}
Sec.~\ref{sec:fafa}.
In particular, we find, using the OPE in Eq.~\eqref{eq:ope_phi_phi_sub}, that
\begin{align}
    I([z_1, z_2], [z_3, z_4]) \propto \eta^{h_\mathsfit{f|f}^{(1)}}, \quad \eta \to 0.
\end{align}
Fitting for $h_\mathsfit{f|f}^{(1)}$ in Fig.~\ref{fig:localizable_MI}(b), we again find $h_\mathsfit{f|f}^{(1)} \approx 0.41$, consistent with all previous results.


\section{Parallel results for the Hartley entropy in Haar circuits from minimal cuts in critical
first-passage percolation \label{sec:perc}}

\env{figure}{[b]
\includegraphics[width=.235\textwidth]{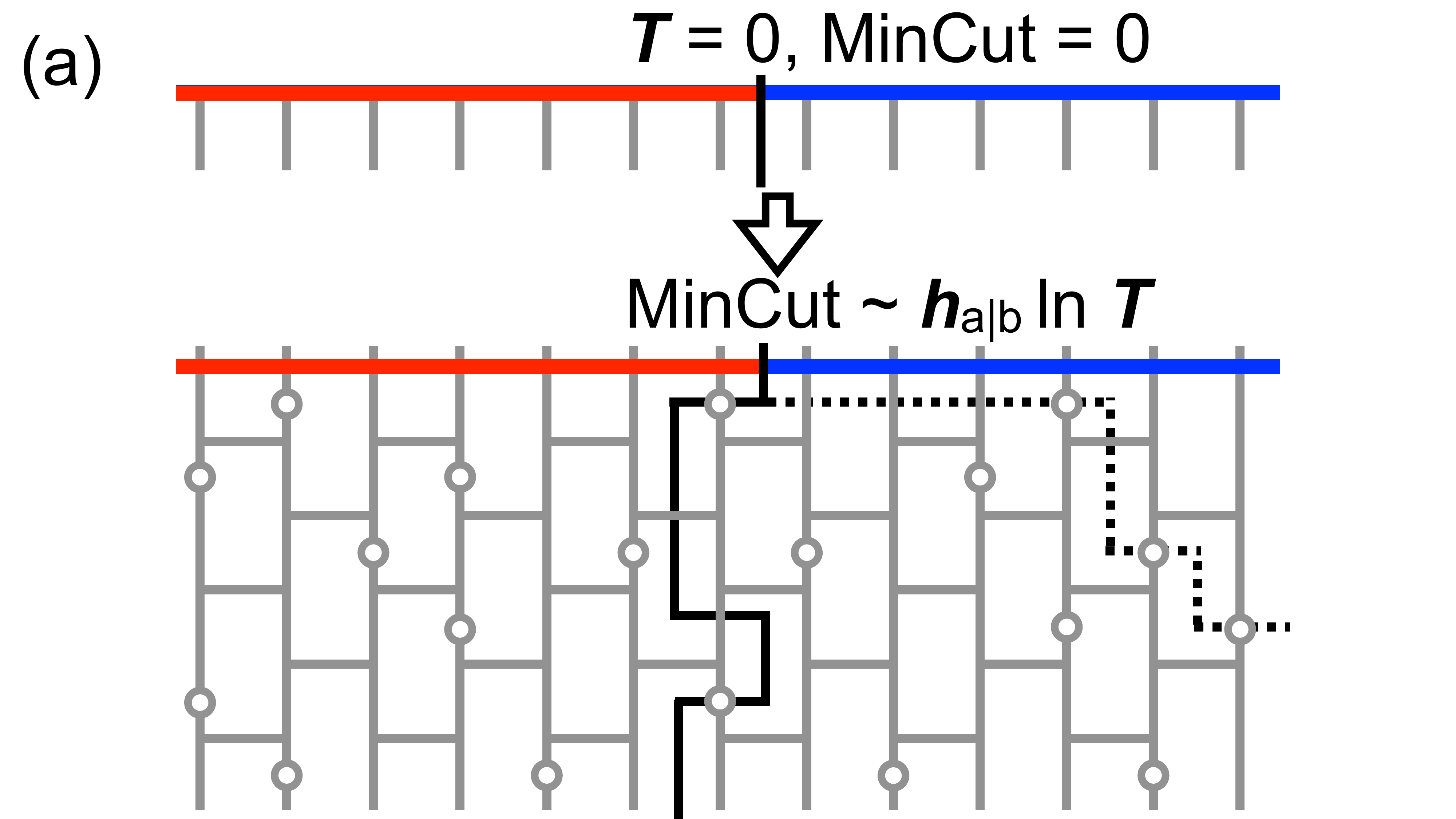}
\includegraphics[width=.235\textwidth]{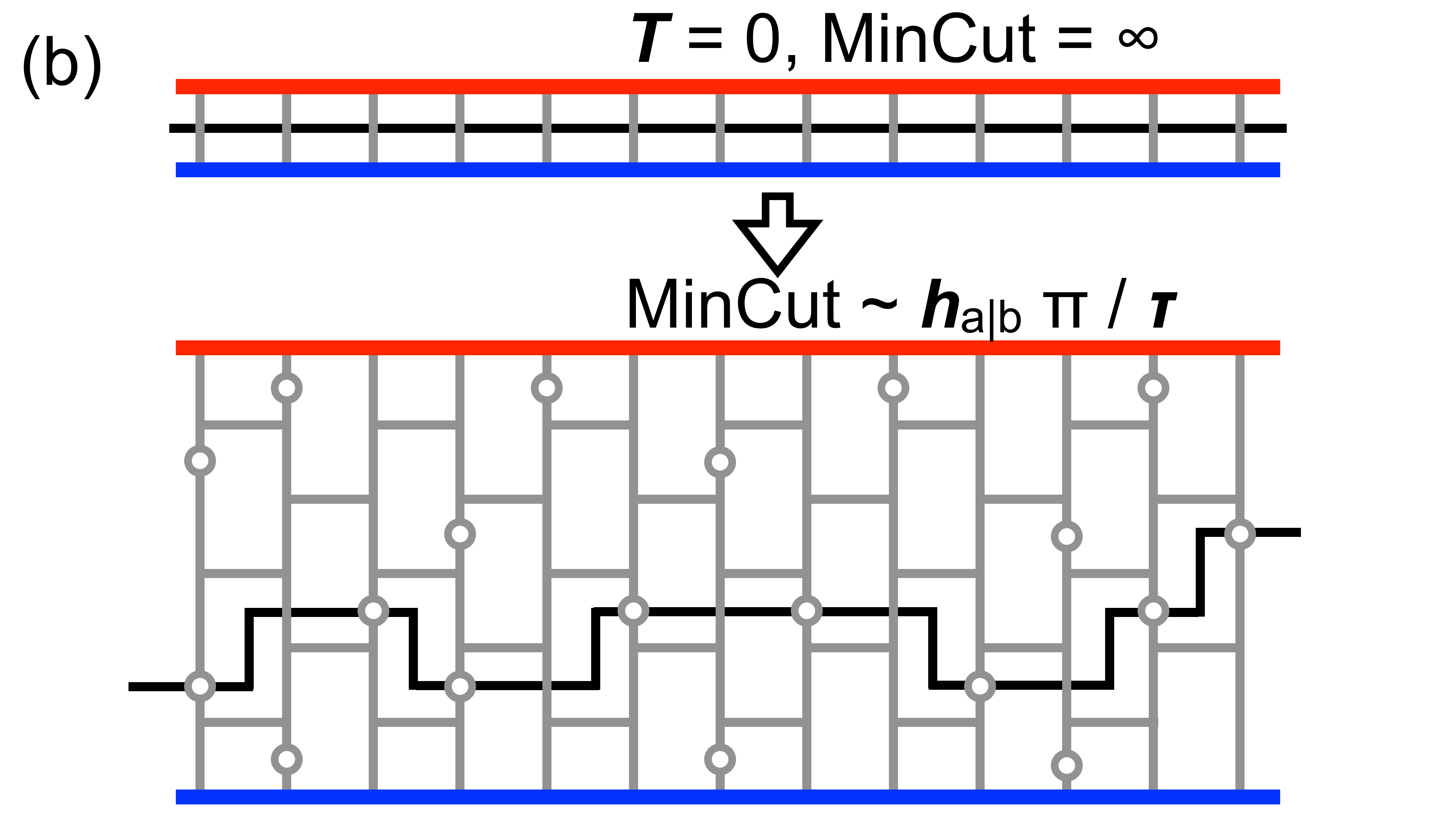}
\caption{
Minimal cuts for two sets of different boundary conditions. 
(a) should be compared with the $\bc{fffa}$ circuit in Fig.~\ref{fig:bc_fffa}, and (b) with the $\bc{fafa}$ circuit in Fig.~\ref{fig:bc_fafa}.
The finite time behavior follows from data collapse in Fig.~\ref{fig:res_perc_obc} and calculations in the main text.
}
\label{fig:perc_bc}
}

\env{figure}{[t]
\includegraphics[width=.5\textwidth]{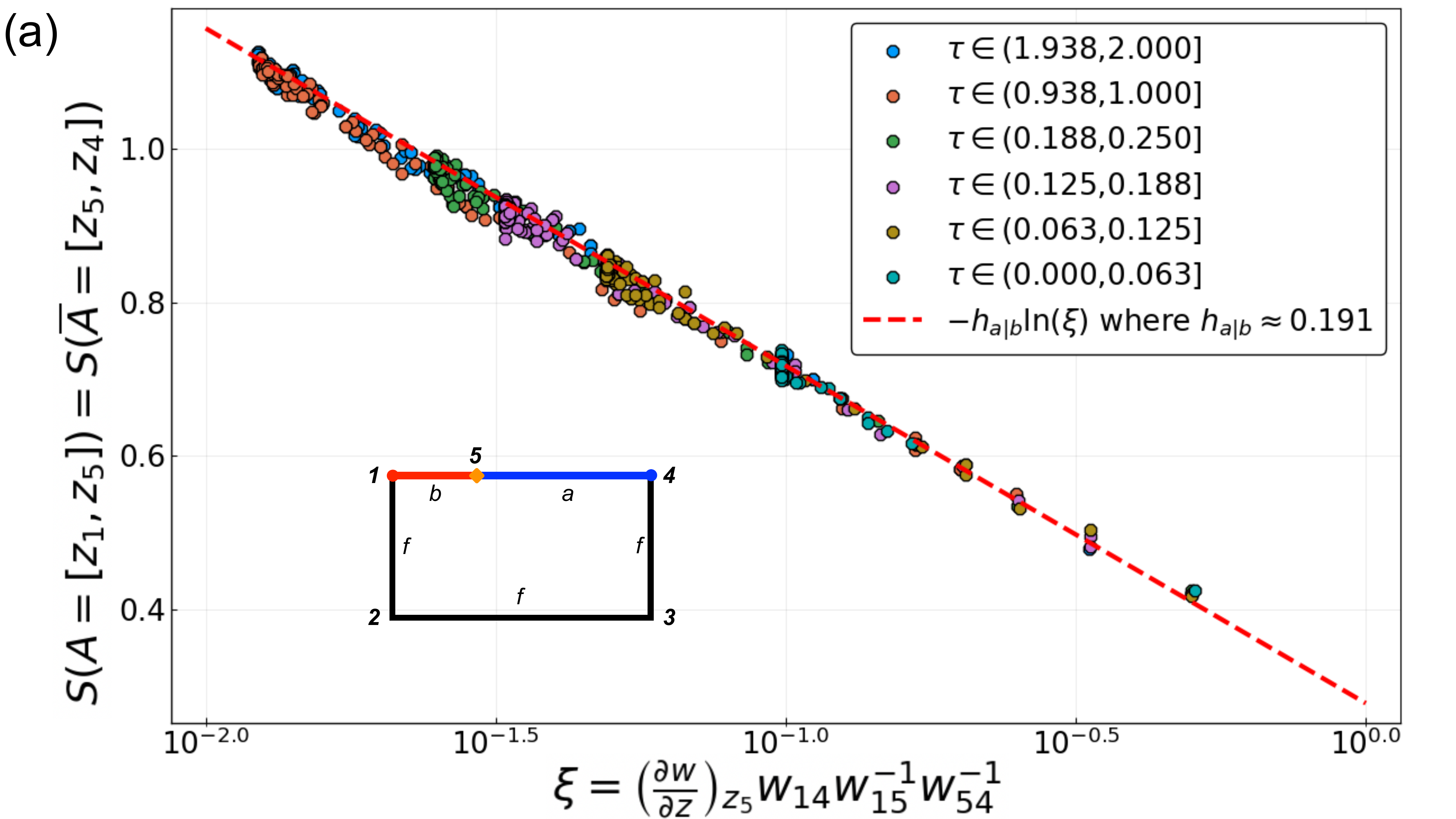}
\includegraphics[width=.5\textwidth]{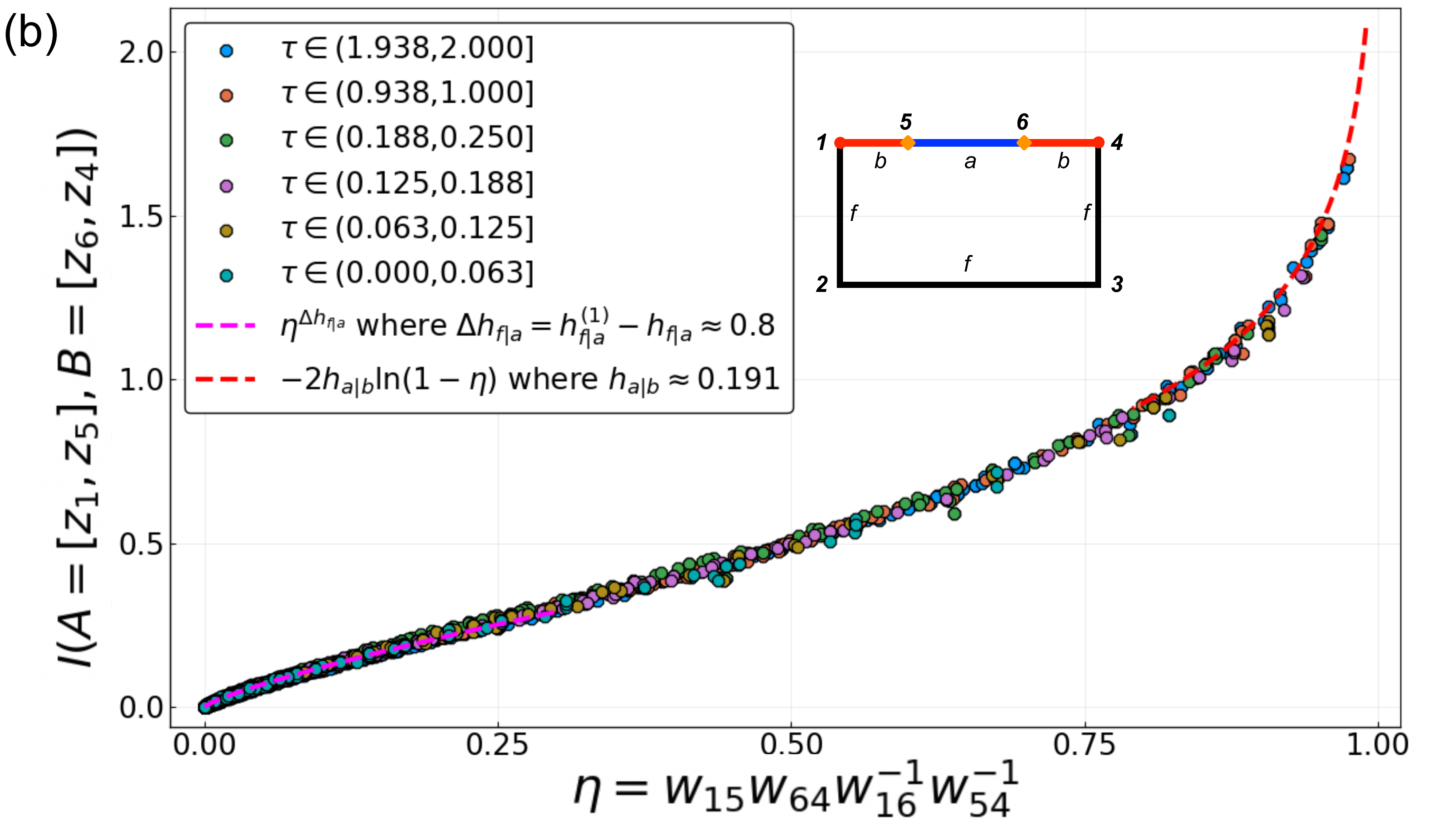}
\includegraphics[width=.5\textwidth]{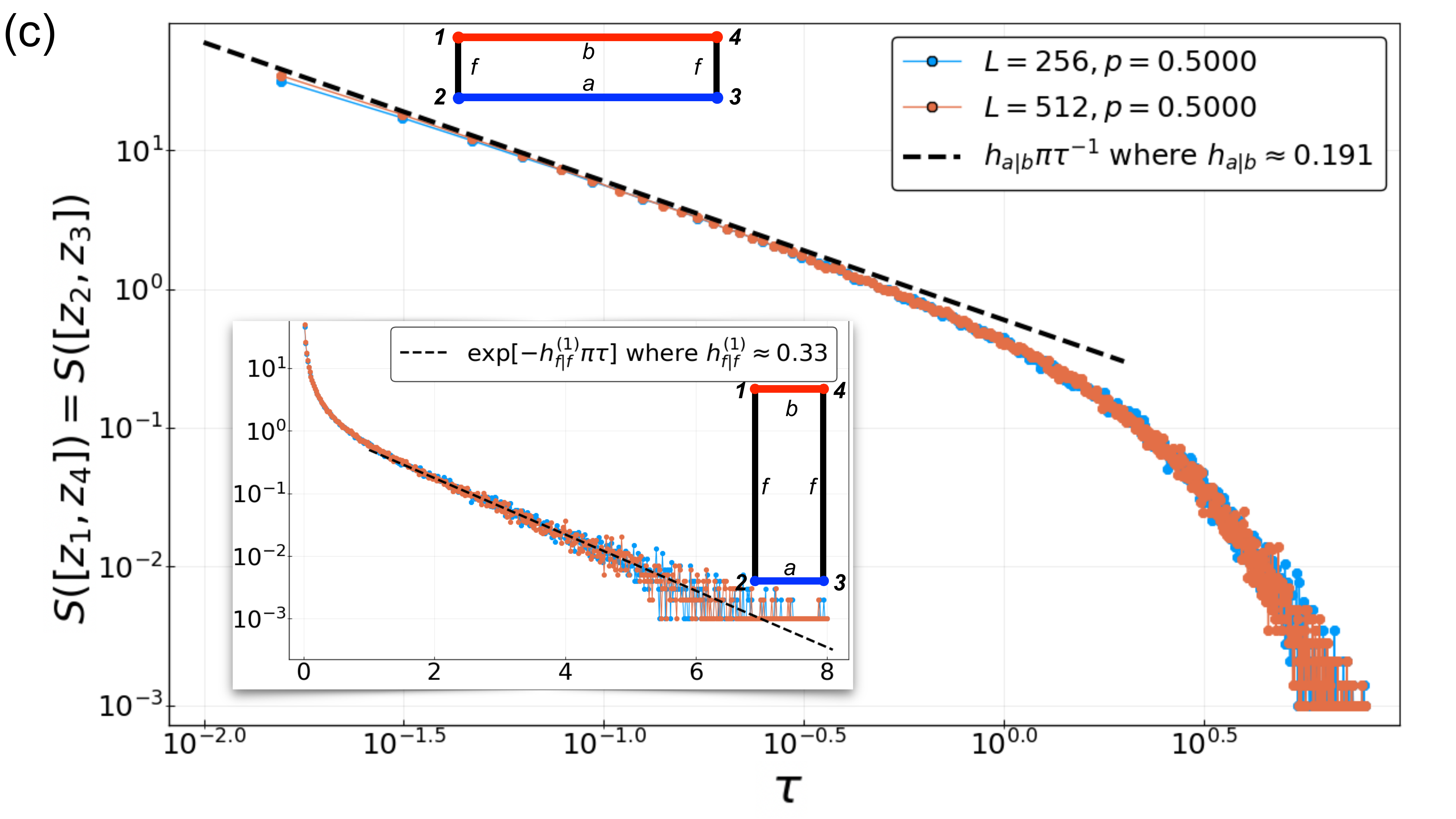}
\caption{
(a,b) Numerical results of minimal cuts as in the setup of Fig.~\ref{fig:perc_bc}(a), and should be compared with Fig.~\ref{fig:res_fffa}.
(c) Numerical results as in the setup of Fig.~\ref{fig:perc_bc}(b), and should be compared with Fig.~\ref{fig:res_fafb}.
The extracted scaling dimensions are summarized in Table~\ref{table:compare_clifford_percolation}.
}
\label{fig:res_perc_obc}
}

\env{figure}{[t]
\includegraphics[width=.5\textwidth]{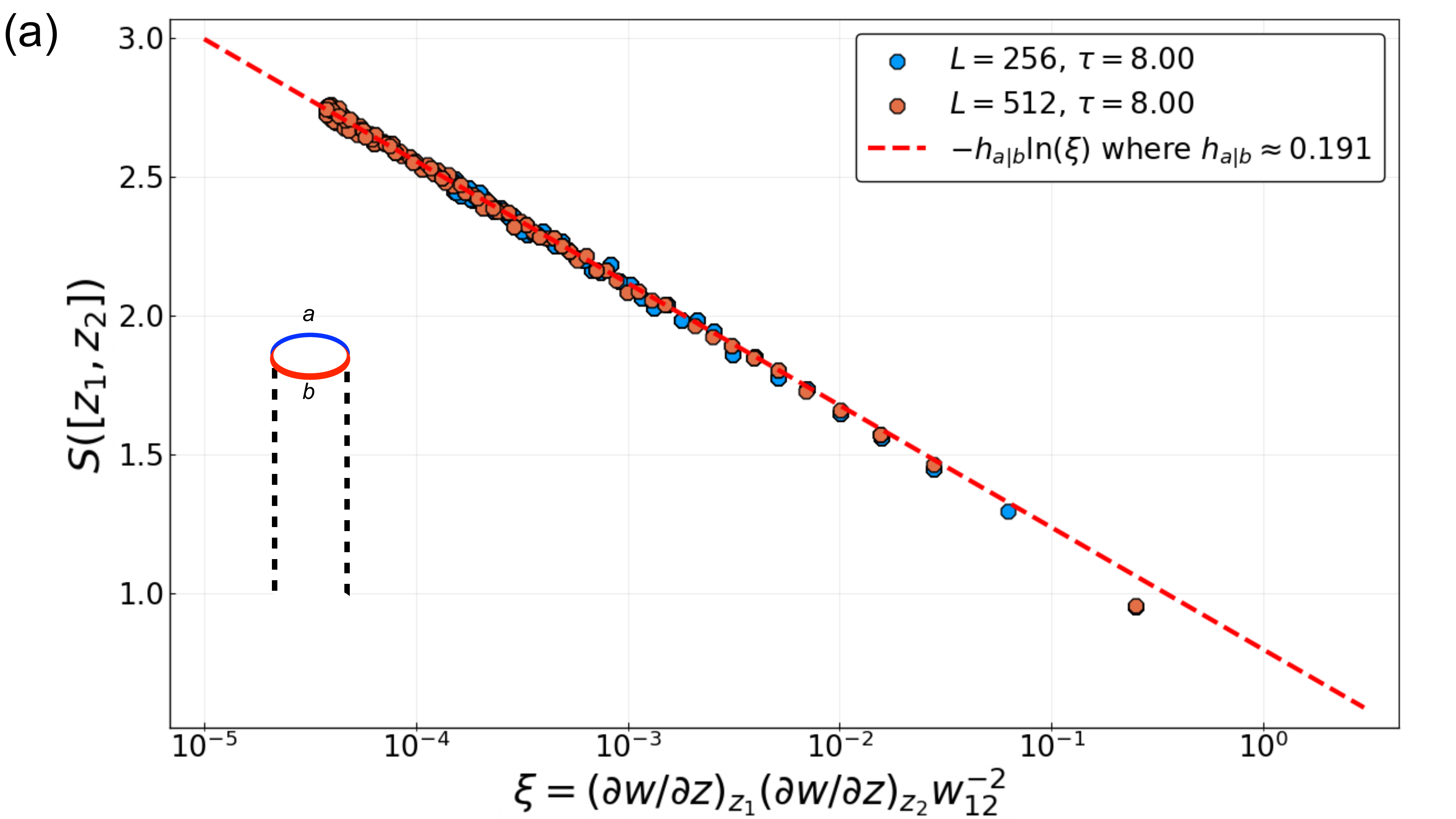}
\includegraphics[width=.5\textwidth]{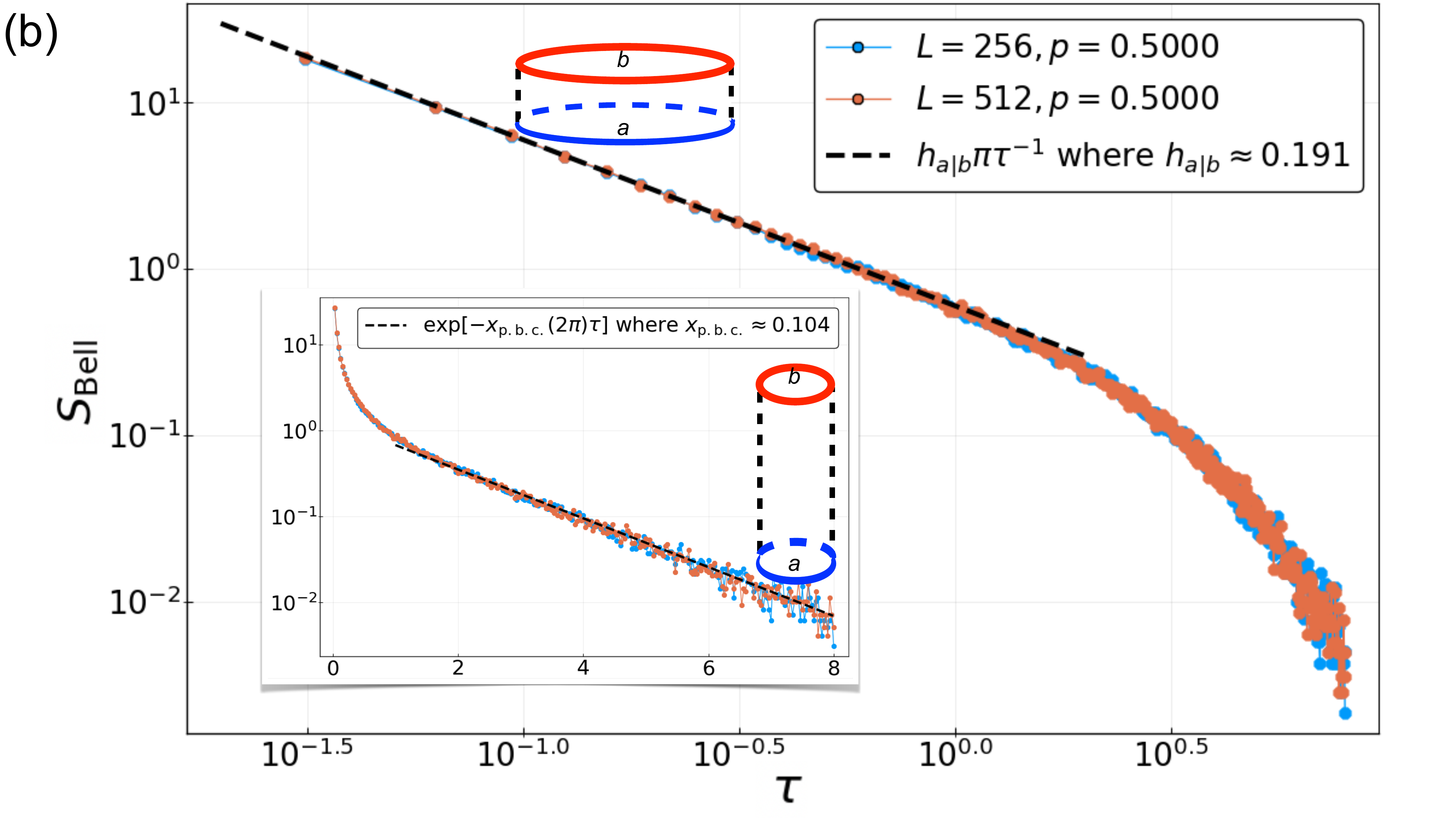}
\caption{
Numerical results for the
{Hartley ($0$-th R\'enyi) entropy in Haar random circuits with measurements, in a geometry of the type of}
Fig.~\ref{fig:perc_bc}(b), but with periodic spatial b.c.
This figure should be compared with Fig.~\ref{fig:res_pbc} in the main text.
}
\label{fig:res_perc_pbc}
}

In this Appendix we apply the same CFT formalism introduced in the main text to the analysis
{of the Hartley ($0$th R\'enyi) entropy in Haar random
unitary circuits with measurements, following Ref.~\cite{nahum2018hybrid}.} 

The goal of this Appendix is 
to further justify our conjectures 
presented in the main part of this paper for
the Clifford hybrid quantum circuits, by
analyzing corresponding setups for the Hartley entropy
in Haar random hybrid quantum circuits.
While the ability to describe the latter in terms  of ``minimal cuts'' in the theory of critical percolation has been established~\cite{nahum2018hybrid}, here we aim
at showing that various boundary condition setups discussed in the main text for Clifford circuits can be analyzed in 
a completely analogous way for the
Hartley entropy in the Haar circuits, and we obtain corresponding critical exponents for this case.

We consider two different 
{possibilities for performing the required ``minimal cuts'' on the underlying ``brickwall'' lattice 
as illustrated in Fig.~\ref{fig:perc_bc}.
In both cases, the lattice geometry
{is}
that of a rectangular hybrid circuit
{as}
in Fig.~\ref{fig:upc_rect}: the horizontal links are arranged in an even-odd fashion and represent two-qubit unitary gates, and the vertical links represent qubit propagation in time, which are interupted by hollow circles that break the link,
{representing}
the single-qubit measurements.
To make connections with the bond percolation problem on a square lattice, 
one can view the lattices in Fig.~\ref{fig:perc_bc} as obtained from a perfect square lattice, by breaking the vertical bonds at random (with probability $p$) and 
{by erasing  every other horizontal bond (i.e. in an alternating but regular fashion), which could be thought of as eliminating (or ``breaking'') with probability $=1/2$ exactly a fraction $1/2$ of all the horizontal bond (a regular version of the process that is implemented on the vertical bonds in a random fashion at criticality with the same probability $p=1/2$.)}\footnote{Notice that the microscopic details of this construction differs slightly from that in Ref.~\cite{nahum2018hybrid}, but can be exactly mapped to the latter by ``shrinking'' the two endpoints of each horizontal link (representing the unitary gate) to a single lattice site, thereby obtaining a square lattice rotated by 45$^\circ$.
Details of this construction should not affect the universal critical properties, as we have verified numerically, but chose not to display here.}
The ``minimal cut''
is defined to be the path that
{begins at the point on the boundary where the two differently colored (red and blue) boundary segments join (possibly at infinity), and
which crosses a minimal number of unbroken links in the bulk.}
{In other words, the ``minimal cut'' path is one which minimizes the ``cost'' defined to be number of unbroken links crosses by the path.
The ``cost'' of the ``minimal cut'' path is proportional to the Hartley ($0$-th R\'enyi) entropy~\cite{nahum2018hybrid}.
}
It is evident from this setup that the coloring pattern is a crucial input in defining the minimal cut.
\env{itemize}{
\item
In the first case (Fig.~\ref{fig:perc_bc}(a)), we label a segment
(left)
of the {upper edge} with red color, and the rest
(right)
of the {upper edge} blue, while the other three edges are uncolored  (denoting ``free'' b.c.s `$\bc{f}$').
In the figure, the minimal cut starts from the interface between red and blue segments, and can  terminate anywhere on the three uncolored ``free'' edges.
When the lattice 
{has}
zero depth, the minimal cut 
{has zero ``cost'',}
and
{its ``cost''}
grows as the lattice grows in depth.
\item
In the second case (Fig.~\ref{fig:perc_bc}(b)), we label the lower edge blue and the upper edge red, where a cut separating them must start from the left edge and terminate at the right edge, which are both ``free''.
Initially, the minimal cut must go through all vertical links, so is infinite in the thermodynamic limit; 
{however, as the circuit grows deeper, the minimal cut path can make use of broken links in the bulk (of which there will be more as the depth increases) to lower
its ``cost''.
Therefore the ``cost'' of the minimal cut path will {\it de}crease monotonically as the depth {\it in}creases, and so
will the Hartley entropy which this ``cost'' respresents.}
}


{Recall that the ``cost'' of the minimal cut path
in Fig.~\ref{fig:perc_bc}(a) exactly describes the {Hartley (zeroth Renyi)} entropy in a random Haar circuit
(Ref.~\cite{nahum2018hybrid}),}
where the initial state is a trivial product state (the situation is exactly like $\bc{fffa}$ in Fig.~\ref{fig:bc_fffa}(c));
while for Fig.~\ref{fig:perc_bc}(b), the minimal cut is exactly $S_{\sf Bell}$ for the $\bc{fafa}$ boundary condition (see Fig.~\ref{fig:bc_fafa}(b)).
The boundary conditions are also entirely similar: we follow the same coloring scheme, identifying ``blue'' with $\bc{a}$, ``red'' with $\bc{b}$, and ``uncolored'' with $\bc{f}$.

From Fig.~\ref{fig:perc_bc}(b), the symmetry between the upper and the lower edge is evident.
As emphasized in 
Sec.~\ref{sec:fafa},
the symmetry is only possible due to unitarity-breaking measurements that induce entanglement swapping (see a similar discussion in Ref.~\cite{nahum1911majorana}).

Recognizing that minimal cuts 
{have}
the meaning of
{(Hartley)}
entanglement entropy, we
{numerically}
compute ``entanglement entropies'' and ``mutual information'' at the critical point $p_c = 0.5$ as in Ref.~\cite{nahum2018hybrid}, making use of well-known 
algorithms for minimal cuts in graph theory~\cite{ford_fulkerson_1956, edmonds_karp_1972}.
The results are shown in Fig.~\ref{fig:res_perc_obc}.
We also consider similar setups with periodic boundary condition, where the results are shown in Fig.~\ref{fig:res_perc_pbc}.
In fitting the data, we have taken $Y = T$ for both open and periodic b.c. (see Sec.~\ref{sec:bcft} for definitions), due to the rotational symmetry of the percolation problem.

The extracted scaling dimensions are summarized in Table~\ref{table:compare_clifford_percolation}, and they match well with {those from the}
existing literature{, where available}.
This further supports our strategy of extracting scaling dimensions from the Clifford CFT. 
Comparing Clifford and percolation, we notice that the difference between the corresponding scaling dimensions are small but discernable under the present framework.
We also {observe} that the scaling dimensions in the Clifford CFT are consistently larger than or equal to their percolation counterparts.

\bibliography{refs}

\begin{thebibliography}{67}%
\makeatletter
\providecommand \@ifxundefined [1]{%
 \@ifx{#1\undefined}
}%
\providecommand \@ifnum [1]{%
 \ifnum #1\expandafter \@firstoftwo
 \else \expandafter \@secondoftwo
 \fi
}%
\providecommand \@ifx [1]{%
 \ifx #1\expandafter \@firstoftwo
 \else \expandafter \@secondoftwo
 \fi
}%
\providecommand \natexlab [1]{#1}%
\providecommand \enquote  [1]{``#1''}%
\providecommand \bibnamefont  [1]{#1}%
\providecommand \bibfnamefont [1]{#1}%
\providecommand \citenamefont [1]{#1}%
\providecommand \href@noop [0]{\@secondoftwo}%
\providecommand \href [0]{\begingroup \@sanitize@url \@href}%
\providecommand \@href[1]{\@@startlink{#1}\@@href}%
\providecommand \@@href[1]{\endgroup#1\@@endlink}%
\providecommand \@sanitize@url [0]{\catcode `\\12\catcode `\$12\catcode
  `\&12\catcode `\#12\catcode `\^12\catcode `\_12\catcode `\%12\relax}%
\providecommand \@@startlink[1]{}%
\providecommand \@@endlink[0]{}%
\providecommand \url  [0]{\begingroup\@sanitize@url \@url }%
\providecommand \@url [1]{\endgroup\@href {#1}{\urlprefix }}%
\providecommand \urlprefix  [0]{URL }%
\providecommand \Eprint [0]{\href }%
\providecommand \doibase [0]{https://doi.org/}%
\providecommand \selectlanguage [0]{\@gobble}%
\providecommand \bibinfo  [0]{\@secondoftwo}%
\providecommand \bibfield  [0]{\@secondoftwo}%
\providecommand \translation [1]{[#1]}%
\providecommand \BibitemOpen [0]{}%
\providecommand \bibitemStop [0]{}%
\providecommand \bibitemNoStop [0]{.\EOS\space}%
\providecommand \EOS [0]{\spacefactor3000\relax}%
\providecommand \BibitemShut  [1]{\csname bibitem#1\endcsname}%
\let\auto@bib@innerbib\@empty
\bibitem [{\citenamefont {Nielsen}\ and\ \citenamefont
  {Chuang}(2010)}]{nielsen2010qiqc}%
  \BibitemOpen
  \bibfield  {author} {\bibinfo {author} {\bibfnamefont {M.~A.}\ \bibnamefont
  {Nielsen}}\ and\ \bibinfo {author} {\bibfnamefont {I.~L.}\ \bibnamefont
  {Chuang}},\ }\href {https://doi.org/10.1017/CBO9780511976667} {\emph
  {\bibinfo {title} {Quantum Computation and Quantum Information}}}\ (\bibinfo
  {publisher} {Cambridge University Press},\ \bibinfo {year}
  {2010})\BibitemShut {NoStop}%
\bibitem [{\citenamefont {{Deutsch}}(1991)}]{Deutsch1991}%
  \BibitemOpen
  \bibfield  {author} {\bibinfo {author} {\bibfnamefont {J.~M.}\ \bibnamefont
  {{Deutsch}}},\ }\bibfield  {title} {\bibinfo {title} {{Quantum statistical
  mechanics in a closed system}},\ }\href
  {https://doi.org/10.1103/PhysRevA.43.2046} {\bibfield  {journal} {\bibinfo
  {journal} {\pra}\ }\textbf {\bibinfo {volume} {43}},\ \bibinfo {pages} {2046}
  (\bibinfo {year} {1991})}\BibitemShut {NoStop}%
\bibitem [{\citenamefont {{Srednicki}}(1994)}]{Srednicki1994}%
  \BibitemOpen
  \bibfield  {author} {\bibinfo {author} {\bibfnamefont {M.}~\bibnamefont
  {{Srednicki}}},\ }\bibfield  {title} {\bibinfo {title} {{Chaos and quantum
  thermalization}},\ }\href {https://doi.org/10.1103/PhysRevE.50.888}
  {\bibfield  {journal} {\bibinfo  {journal} {\pre}\ }\textbf {\bibinfo
  {volume} {50}},\ \bibinfo {pages} {888} (\bibinfo {year} {1994})},\ \Eprint
  {https://arxiv.org/abs/cond-mat/9403051} {arXiv:cond-mat/9403051 [cond-mat]}
  \BibitemShut {NoStop}%
\bibitem [{\citenamefont {{Calabrese}}\ and\ \citenamefont
  {{Cardy}}(2004)}]{Calabrese2004}%
  \BibitemOpen
  \bibfield  {author} {\bibinfo {author} {\bibfnamefont {P.}~\bibnamefont
  {{Calabrese}}}\ and\ \bibinfo {author} {\bibfnamefont {J.}~\bibnamefont
  {{Cardy}}},\ }\bibfield  {title} {\bibinfo {title} {{Entanglement entropy and
  quantum field theory}},\ }\href
  {https://doi.org/10.1088/1742-5468/2004/06/P06002} {\bibfield  {journal}
  {\bibinfo  {journal} {Journal of Statistical Mechanics: Theory and
  Experiment}\ }\textbf {\bibinfo {volume} {6}},\ \bibinfo {pages} {06002}
  (\bibinfo {year} {2004})},\ \Eprint {https://arxiv.org/abs/hep-th/0405152}
  {hep-th/0405152} \BibitemShut {NoStop}%
\bibitem [{\citenamefont {{Calabrese}}\ and\ \citenamefont
  {{Cardy}}(2009)}]{Calabrese_Cardy_2009}%
  \BibitemOpen
  \bibfield  {author} {\bibinfo {author} {\bibfnamefont {P.}~\bibnamefont
  {{Calabrese}}}\ and\ \bibinfo {author} {\bibfnamefont {J.}~\bibnamefont
  {{Cardy}}},\ }\bibfield  {title} {\bibinfo {title} {{Entanglement entropy and
  conformal field theory}},\ }\href
  {https://doi.org/10.1088/1751-8113/42/50/504005} {\bibfield  {journal}
  {\bibinfo  {journal} {Journal of Physics A Mathematical General}\ }\textbf
  {\bibinfo {volume} {42}},\ \bibinfo {eid} {504005} (\bibinfo {year}
  {2009})},\ \Eprint {https://arxiv.org/abs/0905.4013} {arXiv:0905.4013
  [cond-mat.stat-mech]} \BibitemShut {NoStop}%
\bibitem [{\citenamefont {{Levin}}\ and\ \citenamefont
  {{Wen}}(2006)}]{Levin2006}%
  \BibitemOpen
  \bibfield  {author} {\bibinfo {author} {\bibfnamefont {M.}~\bibnamefont
  {{Levin}}}\ and\ \bibinfo {author} {\bibfnamefont {X.-G.}\ \bibnamefont
  {{Wen}}},\ }\bibfield  {title} {\bibinfo {title} {{Detecting Topological
  Order in a Ground State Wave Function}},\ }\href
  {https://doi.org/10.1103/PhysRevLett.96.110405} {\bibfield  {journal}
  {\bibinfo  {journal} {Physical Review Letters}\ }\textbf {\bibinfo {volume}
  {96}},\ \bibinfo {eid} {110405} (\bibinfo {year} {2006})},\ \Eprint
  {https://arxiv.org/abs/cond-mat/0510613} {cond-mat/0510613} \BibitemShut
  {NoStop}%
\bibitem [{\citenamefont {{Kitaev}}\ and\ \citenamefont
  {{Preskill}}(2006)}]{Kitaev2006}%
  \BibitemOpen
  \bibfield  {author} {\bibinfo {author} {\bibfnamefont {A.}~\bibnamefont
  {{Kitaev}}}\ and\ \bibinfo {author} {\bibfnamefont {J.}~\bibnamefont
  {{Preskill}}},\ }\bibfield  {title} {\bibinfo {title} {{Topological
  Entanglement Entropy}},\ }\href
  {https://doi.org/10.1103/PhysRevLett.96.110404} {\bibfield  {journal}
  {\bibinfo  {journal} {Physical Review Letters}\ }\textbf {\bibinfo {volume}
  {96}},\ \bibinfo {eid} {110404} (\bibinfo {year} {2006})},\ \Eprint
  {https://arxiv.org/abs/hep-th/0510092} {hep-th/0510092} \BibitemShut
  {NoStop}%
\bibitem [{\citenamefont {{Ryu}}\ and\ \citenamefont
  {{Takayanagi}}(2006)}]{Ryu2006}%
  \BibitemOpen
  \bibfield  {author} {\bibinfo {author} {\bibfnamefont {S.}~\bibnamefont
  {{Ryu}}}\ and\ \bibinfo {author} {\bibfnamefont {T.}~\bibnamefont
  {{Takayanagi}}},\ }\bibfield  {title} {\bibinfo {title} {{Holographic
  Derivation of Entanglement Entropy from the anti de Sitter Space/Conformal
  Field Theory Correspondence}},\ }\href
  {https://doi.org/10.1103/PhysRevLett.96.181602} {\bibfield  {journal}
  {\bibinfo  {journal} {Physical Review Letters}\ }\textbf {\bibinfo {volume}
  {96}},\ \bibinfo {eid} {181602} (\bibinfo {year} {2006})},\ \Eprint
  {https://arxiv.org/abs/hep-th/0603001} {hep-th/0603001} \BibitemShut
  {NoStop}%
\bibitem [{\citenamefont {{Nandkishore}}\ and\ \citenamefont
  {{Huse}}(2015)}]{Nandkishore2015}%
  \BibitemOpen
  \bibfield  {author} {\bibinfo {author} {\bibfnamefont {R.}~\bibnamefont
  {{Nandkishore}}}\ and\ \bibinfo {author} {\bibfnamefont {D.~A.}\ \bibnamefont
  {{Huse}}},\ }\bibfield  {title} {\bibinfo {title} {{Many-Body Localization
  and Thermalization in Quantum Statistical Mechanics}},\ }\href
  {https://doi.org/10.1146/annurev-conmatphys-031214-014726} {\bibfield
  {journal} {\bibinfo  {journal} {Annual Review of Condensed Matter Physics}\
  }\textbf {\bibinfo {volume} {6}},\ \bibinfo {pages} {15} (\bibinfo {year}
  {2015})},\ \Eprint {https://arxiv.org/abs/1404.0686} {arXiv:1404.0686
  [cond-mat.stat-mech]} \BibitemShut {NoStop}%
\bibitem [{\citenamefont {{Abanin}}\ \emph {et~al.}(2018)\citenamefont
  {{Abanin}}, \citenamefont {{Altman}}, \citenamefont {{Bloch}},\ and\
  \citenamefont {{Serbyn}}}]{abanin2018rmp}%
  \BibitemOpen
  \bibfield  {author} {\bibinfo {author} {\bibfnamefont {D.~A.}\ \bibnamefont
  {{Abanin}}}, \bibinfo {author} {\bibfnamefont {E.}~\bibnamefont {{Altman}}},
  \bibinfo {author} {\bibfnamefont {I.}~\bibnamefont {{Bloch}}},\ and\ \bibinfo
  {author} {\bibfnamefont {M.}~\bibnamefont {{Serbyn}}},\ }\bibfield  {title}
  {\bibinfo {title} {{Ergodicity, Entanglement and Many-Body Localization}},\
  }\href@noop {} {\bibfield  {journal} {\bibinfo  {journal} {arXiv e-prints}\
  ,\ \bibinfo {eid} {arXiv:1804.11065}} (\bibinfo {year} {2018})},\ \Eprint
  {https://arxiv.org/abs/1804.11065} {arXiv:1804.11065 [cond-mat.dis-nn]}
  \BibitemShut {NoStop}%
\bibitem [{\citenamefont {Einstein}\ \emph {et~al.}(1935)\citenamefont
  {Einstein}, \citenamefont {Podolsky},\ and\ \citenamefont
  {Rosen}}]{einstein1935can}%
  \BibitemOpen
  \bibfield  {author} {\bibinfo {author} {\bibfnamefont {A.}~\bibnamefont
  {Einstein}}, \bibinfo {author} {\bibfnamefont {B.}~\bibnamefont {Podolsky}},\
  and\ \bibinfo {author} {\bibfnamefont {N.}~\bibnamefont {Rosen}},\ }\bibfield
   {title} {\bibinfo {title} {{Can Quantum-Mechanical Description of Physical
  Reality Be Considered Complete?}},\ }\href
  {https://doi.org/10.1103/PhysRev.47.777} {\bibfield  {journal} {\bibinfo
  {journal} {Phys. Rev.}\ }\textbf {\bibinfo {volume} {47}},\ \bibinfo {pages}
  {777} (\bibinfo {year} {1935})}\BibitemShut {NoStop}%
\bibitem [{\citenamefont {Bell}(1964)}]{bell1964on}%
  \BibitemOpen
  \bibfield  {author} {\bibinfo {author} {\bibfnamefont {J.~S.}\ \bibnamefont
  {Bell}},\ }\bibfield  {title} {\bibinfo {title} {{On the Einstein Podolsky
  Rosen paradox}},\ }\href
  {https://doi.org/10.1103/PhysicsPhysiqueFizika.1.195} {\bibfield  {journal}
  {\bibinfo  {journal} {Physics Physique Fizika}\ }\textbf {\bibinfo {volume}
  {1}},\ \bibinfo {pages} {195} (\bibinfo {year} {1964})}\BibitemShut {NoStop}%
\bibitem [{\citenamefont {Bennett}\ \emph {et~al.}(1993)\citenamefont
  {Bennett}, \citenamefont {Brassard}, \citenamefont {Cr\'epeau}, \citenamefont
  {Jozsa}, \citenamefont {Peres},\ and\ \citenamefont
  {Wootters}}]{bennett1993teleporting}%
  \BibitemOpen
  \bibfield  {author} {\bibinfo {author} {\bibfnamefont {C.~H.}\ \bibnamefont
  {Bennett}}, \bibinfo {author} {\bibfnamefont {G.}~\bibnamefont {Brassard}},
  \bibinfo {author} {\bibfnamefont {C.}~\bibnamefont {Cr\'epeau}}, \bibinfo
  {author} {\bibfnamefont {R.}~\bibnamefont {Jozsa}}, \bibinfo {author}
  {\bibfnamefont {A.}~\bibnamefont {Peres}},\ and\ \bibinfo {author}
  {\bibfnamefont {W.~K.}\ \bibnamefont {Wootters}},\ }\bibfield  {title}
  {\bibinfo {title} {{Teleporting an unknown quantum state via dual classical
  and Einstein-Podolsky-Rosen channels}},\ }\href
  {https://doi.org/10.1103/PhysRevLett.70.1895} {\bibfield  {journal} {\bibinfo
   {journal} {Phys. Rev. Lett.}\ }\textbf {\bibinfo {volume} {70}},\ \bibinfo
  {pages} {1895} (\bibinfo {year} {1993})}\BibitemShut {NoStop}%
\bibitem [{\citenamefont {{Bouwmeester}}\ \emph {et~al.}(1997)\citenamefont
  {{Bouwmeester}}, \citenamefont {{Pan}}, \citenamefont {{Mattle}},
  \citenamefont {{Eibl}}, \citenamefont {{Weinfurter}},\ and\ \citenamefont
  {{Zeilinger}}}]{Bouwmeester1997teleportation}%
  \BibitemOpen
  \bibfield  {author} {\bibinfo {author} {\bibfnamefont {{\color{gr}
  Dirk}.}~\bibnamefont {{Bouwmeester}}}, \bibinfo {author} {\bibfnamefont
  {J.-W.}\ \bibnamefont {{Pan}}}, \bibinfo {author} {\bibfnamefont
  {K.}~\bibnamefont {{Mattle}}}, \bibinfo {author} {\bibfnamefont
  {M.}~\bibnamefont {{Eibl}}}, \bibinfo {author} {\bibfnamefont
  {H.}~\bibnamefont {{Weinfurter}}},\ and\ \bibinfo {author} {\bibfnamefont
  {A.}~\bibnamefont {{Zeilinger}}},\ }\bibfield  {title} {\bibinfo {title}
  {{Experimental quantum teleportation}},\ }\href
  {https://doi.org/10.1038/37539} {\bibfield  {journal} {\bibinfo  {journal}
  {\nat}\ }\textbf {\bibinfo {volume} {390}},\ \bibinfo {pages} {575} (\bibinfo
  {year} {1997})},\ \Eprint {https://arxiv.org/abs/1901.11004}
  {arXiv:1901.11004 [quant-ph]} \BibitemShut {NoStop}%
\bibitem [{\citenamefont {\ifmmode~\dot{Z}\else \.{Z}\fi{}ukowski}\ \emph
  {et~al.}(1993)\citenamefont {\ifmmode~\dot{Z}\else \.{Z}\fi{}ukowski},
  \citenamefont {Zeilinger}, \citenamefont {Horne},\ and\ \citenamefont
  {Ekert}}]{zukowski1993event}%
  \BibitemOpen
  \bibfield  {author} {\bibinfo {author} {\bibfnamefont {M.}~\bibnamefont
  {\ifmmode~\dot{Z}\else \.{Z}\fi{}ukowski}}, \bibinfo {author} {\bibfnamefont
  {A.}~\bibnamefont {Zeilinger}}, \bibinfo {author} {\bibfnamefont {M.~A.}\
  \bibnamefont {Horne}},\ and\ \bibinfo {author} {\bibfnamefont {A.~K.}\
  \bibnamefont {Ekert}},\ }\bibfield  {title} {\bibinfo {title}
  {{``Event-ready-detectors'' Bell experiment via entanglement swapping}},\
  }\href {https://doi.org/10.1103/PhysRevLett.71.4287} {\bibfield  {journal}
  {\bibinfo  {journal} {Phys. Rev. Lett.}\ }\textbf {\bibinfo {volume} {71}},\
  \bibinfo {pages} {4287} (\bibinfo {year} {1993})}\BibitemShut {NoStop}%
\bibitem [{\citenamefont {Pan}\ \emph {et~al.}(1998)\citenamefont {Pan},
  \citenamefont {Bouwmeester}, \citenamefont {Weinfurter},\ and\ \citenamefont
  {Zeilinger}}]{pan1998experimental}%
  \BibitemOpen
  \bibfield  {author} {\bibinfo {author} {\bibfnamefont {J.-W.}\ \bibnamefont
  {Pan}}, \bibinfo {author} {\bibfnamefont {{\color{gr} Dirk}.}~\bibnamefont
  {Bouwmeester}}, \bibinfo {author} {\bibfnamefont {H.}~\bibnamefont
  {Weinfurter}},\ and\ \bibinfo {author} {\bibfnamefont {A.}~\bibnamefont
  {Zeilinger}},\ }\bibfield  {title} {\bibinfo {title} {{Experimental
  Entanglement Swapping: Entangling Photons That Never Interacted}},\ }\href
  {https://doi.org/10.1103/PhysRevLett.80.3891} {\bibfield  {journal} {\bibinfo
   {journal} {Phys. Rev. Lett.}\ }\textbf {\bibinfo {volume} {80}},\ \bibinfo
  {pages} {3891} (\bibinfo {year} {1998})}\BibitemShut {NoStop}%
\bibitem [{\citenamefont {Jennewein}\ \emph {et~al.}(2001)\citenamefont
  {Jennewein}, \citenamefont {Weihs}, \citenamefont {Pan},\ and\ \citenamefont
  {Zeilinger}}]{jennewein2001experimental}%
  \BibitemOpen
  \bibfield  {author} {\bibinfo {author} {\bibfnamefont {T.}~\bibnamefont
  {Jennewein}}, \bibinfo {author} {\bibfnamefont {G.}~\bibnamefont {Weihs}},
  \bibinfo {author} {\bibfnamefont {J.-W.}\ \bibnamefont {Pan}},\ and\ \bibinfo
  {author} {\bibfnamefont {A.}~\bibnamefont {Zeilinger}},\ }\bibfield  {title}
  {\bibinfo {title} {{Experimental Nonlocality Proof of Quantum Teleportation
  and Entanglement Swapping}},\ }\href
  {https://doi.org/10.1103/PhysRevLett.88.017903} {\bibfield  {journal}
  {\bibinfo  {journal} {Phys. Rev. Lett.}\ }\textbf {\bibinfo {volume} {88}},\
  \bibinfo {pages} {017903} (\bibinfo {year} {2001})}\BibitemShut {NoStop}%
\bibitem [{\citenamefont {Lieb}\ and\ \citenamefont
  {Robinson}(1972)}]{lieb1972finite}%
  \BibitemOpen
  \bibfield  {author} {\bibinfo {author} {\bibfnamefont {E.~H.}\ \bibnamefont
  {Lieb}}\ and\ \bibinfo {author} {\bibfnamefont {D.~W.}\ \bibnamefont
  {Robinson}},\ }\bibfield  {title} {\bibinfo {title} {The finite group
  velocity of quantum spin systems},\ }\href
  {https://projecteuclid.org:443/euclid.cmp/1103858407} {\bibfield  {journal}
  {\bibinfo  {journal} {Comm. Math. Phys.}\ }\textbf {\bibinfo {volume} {28}},\
  \bibinfo {pages} {251} (\bibinfo {year} {1972})}\BibitemShut {NoStop}%
\bibitem [{\citenamefont {{Hastings}}(2010)}]{hastings1008locality}%
  \BibitemOpen
  \bibfield  {author} {\bibinfo {author} {\bibfnamefont {M.~B.}\ \bibnamefont
  {{Hastings}}},\ }\bibfield  {title} {\bibinfo {title} {{Locality in Quantum
  Systems}},\ }\href@noop {} {\bibfield  {journal} {\bibinfo  {journal} {arXiv
  e-prints}\ ,\ \bibinfo {eid} {arXiv:1008.5137}} (\bibinfo {year} {2010})},\
  \Eprint {https://arxiv.org/abs/1008.5137} {arXiv:1008.5137 [math-ph]}
  \BibitemShut {NoStop}%
\bibitem [{\citenamefont {{Cao}}\ \emph {et~al.}(2019)\citenamefont {{Cao}},
  \citenamefont {{Tilloy}},\ and\ \citenamefont {{De
  Luca}}}]{cao2018monitoring}%
  \BibitemOpen
  \bibfield  {author} {\bibinfo {author} {\bibfnamefont {X.}~\bibnamefont
  {{Cao}}}, \bibinfo {author} {\bibfnamefont {A.}~\bibnamefont {{Tilloy}}},\
  and\ \bibinfo {author} {\bibfnamefont {A.}~\bibnamefont {{De Luca}}},\
  }\bibfield  {title} {\bibinfo {title} {{Entanglement in a fermion chain under
  continuous monitoring}},\ }\href
  {https://doi.org/10.21468/SciPostPhys.7.2.024} {\bibfield  {journal}
  {\bibinfo  {journal} {SciPost Physics}\ }\textbf {\bibinfo {volume} {7}},\
  \bibinfo {eid} {024} (\bibinfo {year} {2019})},\ \Eprint
  {https://arxiv.org/abs/1804.04638} {arXiv:1804.04638 [cond-mat.stat-mech]}
  \BibitemShut {NoStop}%
\bibitem [{\citenamefont {{Chan}}\ \emph {et~al.}(2019)\citenamefont {{Chan}},
  \citenamefont {{Nandkishore}}, \citenamefont {{Pretko}},\ and\ \citenamefont
  {{Smith}}}]{nandkishore2018hybrid}%
  \BibitemOpen
  \bibfield  {author} {\bibinfo {author} {\bibfnamefont {A.}~\bibnamefont
  {{Chan}}}, \bibinfo {author} {\bibfnamefont {R.~M.}\ \bibnamefont
  {{Nandkishore}}}, \bibinfo {author} {\bibfnamefont {M.}~\bibnamefont
  {{Pretko}}},\ and\ \bibinfo {author} {\bibfnamefont {G.}~\bibnamefont
  {{Smith}}},\ }\bibfield  {title} {\bibinfo {title} {{Unitary-projective
  entanglement dynamics}},\ }\href {https://doi.org/10.1103/PhysRevB.99.224307}
  {\bibfield  {journal} {\bibinfo  {journal} {\prb}\ }\textbf {\bibinfo
  {volume} {99}},\ \bibinfo {eid} {224307} (\bibinfo {year} {2019})},\ \Eprint
  {https://arxiv.org/abs/1808.05949} {arXiv:1808.05949 [cond-mat.stat-mech]}
  \BibitemShut {NoStop}%
\bibitem [{\citenamefont {{Skinner}}\ \emph {et~al.}(2019)\citenamefont
  {{Skinner}}, \citenamefont {{Ruhman}},\ and\ \citenamefont
  {{Nahum}}}]{nahum2018hybrid}%
  \BibitemOpen
  \bibfield  {author} {\bibinfo {author} {\bibfnamefont {B.}~\bibnamefont
  {{Skinner}}}, \bibinfo {author} {\bibfnamefont {J.}~\bibnamefont
  {{Ruhman}}},\ and\ \bibinfo {author} {\bibfnamefont {A.}~\bibnamefont
  {{Nahum}}},\ }\bibfield  {title} {\bibinfo {title} {{Measurement-Induced
  Phase Transitions in the Dynamics of Entanglement}},\ }\href
  {https://doi.org/10.1103/PhysRevX.9.031009} {\bibfield  {journal} {\bibinfo
  {journal} {Physical Review X}\ }\textbf {\bibinfo {volume} {9}},\ \bibinfo
  {eid} {031009} (\bibinfo {year} {2019})},\ \Eprint
  {https://arxiv.org/abs/1808.05953} {arXiv:1808.05953 [cond-mat.stat-mech]}
  \BibitemShut {NoStop}%
\bibitem [{\citenamefont {{Li}}\ \emph {et~al.}(2018)\citenamefont {{Li}},
  \citenamefont {{Chen}},\ and\ \citenamefont {{Fisher}}}]{li1808hybrid}%
  \BibitemOpen
  \bibfield  {author} {\bibinfo {author} {\bibfnamefont {Y.}~\bibnamefont
  {{Li}}}, \bibinfo {author} {\bibfnamefont {X.}~\bibnamefont {{Chen}}},\ and\
  \bibinfo {author} {\bibfnamefont {M.~P.~A.}\ \bibnamefont {{Fisher}}},\
  }\bibfield  {title} {\bibinfo {title} {{Quantum Zeno effect and the many-body
  entanglement transition}},\ }\href
  {https://doi.org/10.1103/PhysRevB.98.205136} {\bibfield  {journal} {\bibinfo
  {journal} {Physical Review B}\ }\textbf {\bibinfo {volume} {98}},\ \bibinfo
  {eid} {205136} (\bibinfo {year} {2018})},\ \Eprint
  {https://arxiv.org/abs/1808.06134} {arXiv:1808.06134 [quant-ph]} \BibitemShut
  {NoStop}%
\bibitem [{\citenamefont {{Li}}\ \emph {et~al.}(2019)\citenamefont {{Li}},
  \citenamefont {{Chen}},\ and\ \citenamefont {{Fisher}}}]{li1901hybrid}%
  \BibitemOpen
  \bibfield  {author} {\bibinfo {author} {\bibfnamefont {Y.}~\bibnamefont
  {{Li}}}, \bibinfo {author} {\bibfnamefont {X.}~\bibnamefont {{Chen}}},\ and\
  \bibinfo {author} {\bibfnamefont {M.~P.~A.}\ \bibnamefont {{Fisher}}},\
  }\bibfield  {title} {\bibinfo {title} {{Measurement-driven entanglement
  transition in hybrid quantum circuits}},\ }\href
  {https://doi.org/10.1103/PhysRevB.100.134306} {\bibfield  {journal} {\bibinfo
   {journal} {Physical Review B}\ }\textbf {\bibinfo {volume} {100}},\ \bibinfo
  {eid} {134306} (\bibinfo {year} {2019})},\ \Eprint
  {https://arxiv.org/abs/1901.08092} {arXiv:1901.08092 [cond-mat.stat-mech]}
  \BibitemShut {NoStop}%
\bibitem [{\citenamefont {{Choi}}\ \emph {et~al.}(2019)\citenamefont {{Choi}},
  \citenamefont {{Bao}}, \citenamefont {{Qi}},\ and\ \citenamefont
  {{Altman}}}]{choi2019qec}%
  \BibitemOpen
  \bibfield  {author} {\bibinfo {author} {\bibfnamefont {S.}~\bibnamefont
  {{Choi}}}, \bibinfo {author} {\bibfnamefont {Y.}~\bibnamefont {{Bao}}},
  \bibinfo {author} {\bibfnamefont {X.-L.}\ \bibnamefont {{Qi}}},\ and\
  \bibinfo {author} {\bibfnamefont {E.}~\bibnamefont {{Altman}}},\ }\bibfield
  {title} {\bibinfo {title} {{Quantum error correction and entanglement phase
  transition in random unitary circuits with projective measurements}},\
  }\href@noop {} {\bibfield  {journal} {\bibinfo  {journal} {arXiv e-prints}\
  ,\ \bibinfo {eid} {arXiv:1903.05124}} (\bibinfo {year} {2019})},\ \Eprint
  {https://arxiv.org/abs/1903.05124} {arXiv:1903.05124 [quant-ph]} \BibitemShut
  {NoStop}%
\bibitem [{\citenamefont {{Szyniszewski}}\ \emph {et~al.}(2019)\citenamefont
  {{Szyniszewski}}, \citenamefont {{Romito}},\ and\ \citenamefont
  {{Schomerus}}}]{szyniszewski1903measurement}%
  \BibitemOpen
  \bibfield  {author} {\bibinfo {author} {\bibfnamefont {M.}~\bibnamefont
  {{Szyniszewski}}}, \bibinfo {author} {\bibfnamefont {A.}~\bibnamefont
  {{Romito}}},\ and\ \bibinfo {author} {\bibfnamefont {H.}~\bibnamefont
  {{Schomerus}}},\ }\bibfield  {title} {\bibinfo {title} {{Entanglement
  transition from variable-strength weak measurements}},\ }\href
  {https://doi.org/10.1103/PhysRevB.100.064204} {\bibfield  {journal} {\bibinfo
   {journal} {\prb}\ }\textbf {\bibinfo {volume} {100}},\ \bibinfo {eid}
  {064204} (\bibinfo {year} {2019})},\ \Eprint
  {https://arxiv.org/abs/1903.05452} {arXiv:1903.05452 [cond-mat.stat-mech]}
  \BibitemShut {NoStop}%
\bibitem [{\citenamefont {{Gullans}}\ and\ \citenamefont
  {{Huse}}(2019{\natexlab{a}})}]{gullans1905purification}%
  \BibitemOpen
  \bibfield  {author} {\bibinfo {author} {\bibfnamefont {M.~J.}\ \bibnamefont
  {{Gullans}}}\ and\ \bibinfo {author} {\bibfnamefont {D.~A.}\ \bibnamefont
  {{Huse}}},\ }\bibfield  {title} {\bibinfo {title} {{Dynamical purification
  phase transition induced by quantum measurements}},\ }\href@noop {}
  {\bibfield  {journal} {\bibinfo  {journal} {arXiv e-prints}\ } (\bibinfo
  {year} {2019}{\natexlab{a}})},\ \Eprint {https://arxiv.org/abs/1905.05195}
  {arXiv:1905.05195 [quant-ph]} \BibitemShut {NoStop}%
\bibitem [{\citenamefont {{Bao}}\ \emph {et~al.}(2020)\citenamefont {{Bao}},
  \citenamefont {{Choi}},\ and\ \citenamefont {{Altman}}}]{choi2019spin}%
  \BibitemOpen
  \bibfield  {author} {\bibinfo {author} {\bibfnamefont {Y.}~\bibnamefont
  {{Bao}}}, \bibinfo {author} {\bibfnamefont {S.}~\bibnamefont {{Choi}}},\ and\
  \bibinfo {author} {\bibfnamefont {E.}~\bibnamefont {{Altman}}},\ }\bibfield
  {title} {\bibinfo {title} {{Theory of the phase transition in random unitary
  circuits with measurements}},\ }\href
  {https://doi.org/10.1103/PhysRevB.101.104301} {\bibfield  {journal} {\bibinfo
   {journal} {\prb}\ }\textbf {\bibinfo {volume} {101}},\ \bibinfo {eid}
  {104301} (\bibinfo {year} {2020})},\ \Eprint
  {https://arxiv.org/abs/1908.04305} {arXiv:1908.04305 [cond-mat.stat-mech]}
  \BibitemShut {NoStop}%
\bibitem [{\citenamefont {{Jian}}\ \emph {et~al.}(2020)\citenamefont {{Jian}},
  \citenamefont {{You}}, \citenamefont {{Vasseur}},\ and\ \citenamefont
  {{Ludwig}}}]{andreas2019hybrid}%
  \BibitemOpen
  \bibfield  {author} {\bibinfo {author} {\bibfnamefont {C.-M.}\ \bibnamefont
  {{Jian}}}, \bibinfo {author} {\bibfnamefont {Y.-Z.}\ \bibnamefont {{You}}},
  \bibinfo {author} {\bibfnamefont {R.}~\bibnamefont {{Vasseur}}},\ and\
  \bibinfo {author} {\bibfnamefont {A.~W.~W.}\ \bibnamefont {{Ludwig}}},\
  }\bibfield  {title} {\bibinfo {title} {{Measurement-induced criticality in
  random quantum circuits}},\ }\href
  {https://doi.org/10.1103/PhysRevB.101.104302} {\bibfield  {journal} {\bibinfo
   {journal} {\prb}\ }\textbf {\bibinfo {volume} {101}},\ \bibinfo {eid}
  {104302} (\bibinfo {year} {2020})},\ \Eprint
  {https://arxiv.org/abs/1908.08051} {arXiv:1908.08051 [cond-mat.stat-mech]}
  \BibitemShut {NoStop}%
\bibitem [{\citenamefont {{Tang}}\ and\ \citenamefont
  {{Zhu}}(2020)}]{Tang2019}%
  \BibitemOpen
  \bibfield  {author} {\bibinfo {author} {\bibfnamefont {Q.}~\bibnamefont
  {{Tang}}}\ and\ \bibinfo {author} {\bibfnamefont {W.}~\bibnamefont {{Zhu}}},\
  }\bibfield  {title} {\bibinfo {title} {{Measurement-induced phase transition:
  A case study in the nonintegrable model by density-matrix renormalization
  group calculations}},\ }\href
  {https://doi.org/10.1103/PhysRevResearch.2.013022} {\bibfield  {journal}
  {\bibinfo  {journal} {Physical Review Research}\ }\textbf {\bibinfo {volume}
  {2}},\ \bibinfo {eid} {013022} (\bibinfo {year} {2020})},\ \Eprint
  {https://arxiv.org/abs/1908.11253} {arXiv:1908.11253 [cond-mat.stat-mech]}
  \BibitemShut {NoStop}%
\bibitem [{\citenamefont {{Gullans}}\ and\ \citenamefont
  {{Huse}}(2019{\natexlab{b}})}]{gullans1910scalable}%
  \BibitemOpen
  \bibfield  {author} {\bibinfo {author} {\bibfnamefont {M.~J.}\ \bibnamefont
  {{Gullans}}}\ and\ \bibinfo {author} {\bibfnamefont {D.~A.}\ \bibnamefont
  {{Huse}}},\ }\bibfield  {title} {\bibinfo {title} {{Scalable probes of
  measurement-induced criticality}},\ }\href@noop {} {\bibfield  {journal}
  {\bibinfo  {journal} {arXiv e-prints}\ ,\ \bibinfo {eid} {arXiv:1910.00020}}
  (\bibinfo {year} {2019}{\natexlab{b}})},\ \Eprint
  {https://arxiv.org/abs/1910.00020} {arXiv:1910.00020 [cond-mat.stat-mech]}
  \BibitemShut {NoStop}%
\bibitem [{\citenamefont {{Zabalo}}\ \emph {et~al.}(2020)\citenamefont
  {{Zabalo}}, \citenamefont {{Gullans}}, \citenamefont {{Wilson}},
  \citenamefont {{Gopalakrishnan}}, \citenamefont {{Huse}},\ and\ \citenamefont
  {{Pixley}}}]{huse1911tripartite}%
  \BibitemOpen
  \bibfield  {author} {\bibinfo {author} {\bibfnamefont {A.}~\bibnamefont
  {{Zabalo}}}, \bibinfo {author} {\bibfnamefont {M.~J.}\ \bibnamefont
  {{Gullans}}}, \bibinfo {author} {\bibfnamefont {J.~H.}\ \bibnamefont
  {{Wilson}}}, \bibinfo {author} {\bibfnamefont {S.}~\bibnamefont
  {{Gopalakrishnan}}}, \bibinfo {author} {\bibfnamefont {D.~A.}\ \bibnamefont
  {{Huse}}},\ and\ \bibinfo {author} {\bibfnamefont {J.~H.}\ \bibnamefont
  {{Pixley}}},\ }\bibfield  {title} {\bibinfo {title} {{Critical properties of
  the measurement-induced transition in random quantum circuits}},\ }\href
  {https://doi.org/10.1103/PhysRevB.101.060301} {\bibfield  {journal} {\bibinfo
   {journal} {\prb}\ }\textbf {\bibinfo {volume} {101}},\ \bibinfo {eid}
  {060301} (\bibinfo {year} {2020})},\ \Eprint
  {https://arxiv.org/abs/1911.00008} {arXiv:1911.00008 [cond-mat.dis-nn]}
  \BibitemShut {NoStop}%
\bibitem [{\citenamefont {Zhang}\ \emph {et~al.}(2020)\citenamefont {Zhang},
  \citenamefont {Reyes}, \citenamefont {Kourtis}, \citenamefont {Chamon},
  \citenamefont {Mucciolo},\ and\ \citenamefont
  {Ruckenstein}}]{zhang2020nonuniversal}%
  \BibitemOpen
  \bibfield  {author} {\bibinfo {author} {\bibfnamefont {L.}~\bibnamefont
  {Zhang}}, \bibinfo {author} {\bibfnamefont {J.~A.}\ \bibnamefont {Reyes}},
  \bibinfo {author} {\bibfnamefont {S.}~\bibnamefont {Kourtis}}, \bibinfo
  {author} {\bibfnamefont {C.}~\bibnamefont {Chamon}}, \bibinfo {author}
  {\bibfnamefont {E.~R.}\ \bibnamefont {Mucciolo}},\ and\ \bibinfo {author}
  {\bibfnamefont {A.~E.}\ \bibnamefont {Ruckenstein}},\ }\href@noop {}
  {\bibinfo {title} {Nonuniversal entanglement level statistics in
  projection-driven quantum circuits}} (\bibinfo {year} {2020}),\ \Eprint
  {https://arxiv.org/abs/2001.11428} {arXiv:2001.11428 [cond-mat.stat-mech]}
  \BibitemShut {NoStop}%
\bibitem [{\citenamefont {Fan}\ \emph {et~al.}(2020)\citenamefont {Fan},
  \citenamefont {Vijay}, \citenamefont {Vishwanath},\ and\ \citenamefont
  {You}}]{fan2020selforganized}%
  \BibitemOpen
  \bibfield  {author} {\bibinfo {author} {\bibfnamefont {R.}~\bibnamefont
  {Fan}}, \bibinfo {author} {\bibfnamefont {S.}~\bibnamefont {Vijay}}, \bibinfo
  {author} {\bibfnamefont {A.}~\bibnamefont {Vishwanath}},\ and\ \bibinfo
  {author} {\bibfnamefont {Y.-Z.}\ \bibnamefont {You}},\ }\href@noop {}
  {\bibinfo {title} {Self-organized error correction in random unitary circuits
  with measurement}} (\bibinfo {year} {2020}),\ \Eprint
  {https://arxiv.org/abs/2002.12385} {arXiv:2002.12385 [cond-mat.stat-mech]}
  \BibitemShut {NoStop}%
\bibitem [{\citenamefont {{Yunger Halpern}}\ and\ \citenamefont
  {{Crosson}}(2019)}]{halpern1711qiqcog}%
  \BibitemOpen
  \bibfield  {author} {\bibinfo {author} {\bibfnamefont {N.}~\bibnamefont
  {{Yunger Halpern}}}\ and\ \bibinfo {author} {\bibfnamefont {E.}~\bibnamefont
  {{Crosson}}},\ }\bibfield  {title} {\bibinfo {title} {{Quantum information in
  quantum cognition}},\ }\href@noop {} {\bibfield  {journal} {\bibinfo
  {journal} {Annals of Physics}\ }\textbf {\bibinfo {volume} {407}},\ \bibinfo
  {pages} {92} (\bibinfo {year} {2019})},\ \Eprint
  {https://arxiv.org/abs/1711.04801} {arXiv:1711.04801 [quant-ph]} \BibitemShut
  {NoStop}%
\bibitem [{\citenamefont {{Nahum}}\ \emph {et~al.}(2017)\citenamefont
  {{Nahum}}, \citenamefont {{Ruhman}}, \citenamefont {{Vijay}},\ and\
  \citenamefont {{Haah}}}]{nahum2017KPZ}%
  \BibitemOpen
  \bibfield  {author} {\bibinfo {author} {\bibfnamefont {A.}~\bibnamefont
  {{Nahum}}}, \bibinfo {author} {\bibfnamefont {J.}~\bibnamefont {{Ruhman}}},
  \bibinfo {author} {\bibfnamefont {S.}~\bibnamefont {{Vijay}}},\ and\ \bibinfo
  {author} {\bibfnamefont {J.}~\bibnamefont {{Haah}}},\ }\bibfield  {title}
  {\bibinfo {title} {{Quantum Entanglement Growth under Random Unitary
  Dynamics}},\ }\href {https://doi.org/10.1103/PhysRevX.7.031016} {\bibfield
  {journal} {\bibinfo  {journal} {Physical Review X}\ }\textbf {\bibinfo
  {volume} {7}},\ \bibinfo {eid} {031016} (\bibinfo {year} {2017})},\ \Eprint
  {https://arxiv.org/abs/1608.06950} {arXiv:1608.06950 [cond-mat.stat-mech]}
  \BibitemShut {NoStop}%
\bibitem [{\citenamefont {{Nahum}}\ \emph {et~al.}(2018)\citenamefont
  {{Nahum}}, \citenamefont {{Vijay}},\ and\ \citenamefont
  {{Haah}}}]{nahum2018operator}%
  \BibitemOpen
  \bibfield  {author} {\bibinfo {author} {\bibfnamefont {A.}~\bibnamefont
  {{Nahum}}}, \bibinfo {author} {\bibfnamefont {S.}~\bibnamefont {{Vijay}}},\
  and\ \bibinfo {author} {\bibfnamefont {J.}~\bibnamefont {{Haah}}},\
  }\bibfield  {title} {\bibinfo {title} {{Operator Spreading in Random Unitary
  Circuits}},\ }\href {https://doi.org/10.1103/PhysRevX.8.021014} {\bibfield
  {journal} {\bibinfo  {journal} {Physical Review X}\ }\textbf {\bibinfo
  {volume} {8}},\ \bibinfo {eid} {021014} (\bibinfo {year} {2018})},\ \Eprint
  {https://arxiv.org/abs/1705.08975} {arXiv:1705.08975 [cond-mat.str-el]}
  \BibitemShut {NoStop}%
\bibitem [{\citenamefont {{von Keyserlingk}}\ \emph {et~al.}(2018)\citenamefont
  {{von Keyserlingk}}, \citenamefont {{Rakovszky}}, \citenamefont
  {{Pollmann}},\ and\ \citenamefont {{Sondhi}}}]{keyserlingk2018operator}%
  \BibitemOpen
  \bibfield  {author} {\bibinfo {author} {\bibfnamefont {C.~W.}\ \bibnamefont
  {{von Keyserlingk}}}, \bibinfo {author} {\bibfnamefont {T.}~\bibnamefont
  {{Rakovszky}}}, \bibinfo {author} {\bibfnamefont {F.}~\bibnamefont
  {{Pollmann}}},\ and\ \bibinfo {author} {\bibfnamefont {S.~L.}\ \bibnamefont
  {{Sondhi}}},\ }\bibfield  {title} {\bibinfo {title} {{Operator Hydrodynamics,
  OTOCs, and Entanglement Growth in Systems without Conservation Laws}},\
  }\href {https://doi.org/10.1103/PhysRevX.8.021013} {\bibfield  {journal}
  {\bibinfo  {journal} {Physical Review X}\ }\textbf {\bibinfo {volume} {8}},\
  \bibinfo {eid} {021013} (\bibinfo {year} {2018})},\ \Eprint
  {https://arxiv.org/abs/1705.08910} {arXiv:1705.08910 [cond-mat.str-el]}
  \BibitemShut {NoStop}%
\bibitem [{\citenamefont {{Napp}}\ \emph {et~al.}(2019)\citenamefont {{Napp}},
  \citenamefont {{La Placa}}, \citenamefont {{Dalzell}}, \citenamefont
  {{Brandao}},\ and\ \citenamefont {{Harrow}}}]{harrow2001efficient}%
  \BibitemOpen
  \bibfield  {author} {\bibinfo {author} {\bibfnamefont {J.}~\bibnamefont
  {{Napp}}}, \bibinfo {author} {\bibfnamefont {R.~L.}\ \bibnamefont {{La
  Placa}}}, \bibinfo {author} {\bibfnamefont {A.~M.}\ \bibnamefont
  {{Dalzell}}}, \bibinfo {author} {\bibfnamefont {F.~G.~S.~L.}\ \bibnamefont
  {{Brandao}}},\ and\ \bibinfo {author} {\bibfnamefont {A.~W.}\ \bibnamefont
  {{Harrow}}},\ }\bibfield  {title} {\bibinfo {title} {{Efficient classical
  simulation of random shallow 2D quantum circuits}},\ }\href@noop {}
  {\bibfield  {journal} {\bibinfo  {journal} {arXiv e-prints}\ ,\ \bibinfo
  {eid} {arXiv:2001.00021}} (\bibinfo {year} {2019})},\ \Eprint
  {https://arxiv.org/abs/2001.00021} {arXiv:2001.00021 [quant-ph]} \BibitemShut
  {NoStop}%
\bibitem [{\citenamefont {{Vasseur}}\ \emph {et~al.}(2019)\citenamefont
  {{Vasseur}}, \citenamefont {{Potter}}, \citenamefont {{You}},\ and\
  \citenamefont {{Ludwig}}}]{vasseur2018rtn}%
  \BibitemOpen
  \bibfield  {author} {\bibinfo {author} {\bibfnamefont {R.}~\bibnamefont
  {{Vasseur}}}, \bibinfo {author} {\bibfnamefont {A.~C.}\ \bibnamefont
  {{Potter}}}, \bibinfo {author} {\bibfnamefont {Y.-Z.}\ \bibnamefont
  {{You}}},\ and\ \bibinfo {author} {\bibfnamefont {A.~W.~W.}\ \bibnamefont
  {{Ludwig}}},\ }\bibfield  {title} {\bibinfo {title} {{Entanglement
  transitions from holographic random tensor networks}},\ }\href
  {https://doi.org/10.1103/PhysRevB.100.134203} {\bibfield  {journal} {\bibinfo
   {journal} {\prb}\ }\textbf {\bibinfo {volume} {100}},\ \bibinfo {eid}
  {134203} (\bibinfo {year} {2019})},\ \Eprint
  {https://arxiv.org/abs/1807.07082} {arXiv:1807.07082 [cond-mat.stat-mech]}
  \BibitemShut {NoStop}%
\bibitem [{\citenamefont {{Gottesman}}(1996)}]{gottesman9604hamming}%
  \BibitemOpen
  \bibfield  {author} {\bibinfo {author} {\bibfnamefont {D.}~\bibnamefont
  {{Gottesman}}},\ }\bibfield  {title} {\bibinfo {title} {{Class of quantum
  error-correcting codes saturating the quantum Hamming bound}},\ }\href
  {https://doi.org/10.1103/PhysRevA.54.1862} {\bibfield  {journal} {\bibinfo
  {journal} {Physical Review A}\ }\textbf {\bibinfo {volume} {54}},\ \bibinfo
  {pages} {1862} (\bibinfo {year} {1996})},\ \Eprint
  {https://arxiv.org/abs/quant-ph/9604038} {arXiv:quant-ph/9604038 [quant-ph]}
  \BibitemShut {NoStop}%
\bibitem [{\citenamefont {{Gottesman}}(1998)}]{gottesman9807heisenberg}%
  \BibitemOpen
  \bibfield  {author} {\bibinfo {author} {\bibfnamefont {D.}~\bibnamefont
  {{Gottesman}}},\ }\bibfield  {title} {\bibinfo {title} {{The Heisenberg
  Representation of Quantum Computers}},\ }\href@noop {} {\bibfield  {journal}
  {\bibinfo  {journal} {arXiv e-prints}\ ,\ \bibinfo {eid} {quant-ph/9807006}}
  (\bibinfo {year} {1998})},\ \Eprint {https://arxiv.org/abs/quant-ph/9807006}
  {arXiv:quant-ph/9807006 [quant-ph]} \BibitemShut {NoStop}%
\bibitem [{\citenamefont {{Aaronson}}\ and\ \citenamefont
  {{Gottesman}}(2004)}]{aaronson0406chp}%
  \BibitemOpen
  \bibfield  {author} {\bibinfo {author} {\bibfnamefont {S.}~\bibnamefont
  {{Aaronson}}}\ and\ \bibinfo {author} {\bibfnamefont {D.}~\bibnamefont
  {{Gottesman}}},\ }\bibfield  {title} {\bibinfo {title} {{Improved simulation
  of stabilizer circuits}},\ }\href
  {https://doi.org/10.1103/PhysRevA.70.052328} {\bibfield  {journal} {\bibinfo
  {journal} {Physical Review A}\ }\textbf {\bibinfo {volume} {70}},\ \bibinfo
  {eid} {052328} (\bibinfo {year} {2004})},\ \Eprint
  {https://arxiv.org/abs/quant-ph/0406196} {arXiv:quant-ph/0406196 [quant-ph]}
  \BibitemShut {NoStop}%
\bibitem [{\citenamefont {{Belavin}}\ \emph {et~al.}(1984)\citenamefont
  {{Belavin}}, \citenamefont {{Polyakov}},\ and\ \citenamefont
  {{Zamolodchikov}}}]{BPZ1984}%
  \BibitemOpen
  \bibfield  {author} {\bibinfo {author} {\bibfnamefont {A.~A.}\ \bibnamefont
  {{Belavin}}}, \bibinfo {author} {\bibfnamefont {A.~M.}\ \bibnamefont
  {{Polyakov}}},\ and\ \bibinfo {author} {\bibfnamefont {A.~B.}\ \bibnamefont
  {{Zamolodchikov}}},\ }\bibfield  {title} {\bibinfo {title} {{Infinite
  conformal symmetry in two-dimensional quantum field theory}},\ }\href
  {https://doi.org/10.1016/0550-3213(84)90052-X} {\bibfield  {journal}
  {\bibinfo  {journal} {Nuclear Physics B}\ }\textbf {\bibinfo {volume}
  {241}},\ \bibinfo {pages} {333} (\bibinfo {year} {1984})}\BibitemShut
  {NoStop}%
\bibitem [{\citenamefont {{Cardy}}(2004)}]{cardy0411bcft}%
  \BibitemOpen
  \bibfield  {author} {\bibinfo {author} {\bibfnamefont {J.}~\bibnamefont
  {{Cardy}}},\ }\bibfield  {title} {\bibinfo {title} {{Boundary Conformal Field
  Theory}},\ }\href@noop {} {\bibfield  {journal} {\bibinfo  {journal} {arXiv
  e-prints}\ ,\ \bibinfo {eid} {hep-th/0411189}} (\bibinfo {year} {2004})},\
  \Eprint {https://arxiv.org/abs/hep-th/0411189} {arXiv:hep-th/0411189
  [hep-th]} \BibitemShut {NoStop}%
\bibitem [{\citenamefont {{Hayden}}\ \emph {et~al.}(2016)\citenamefont
  {{Hayden}}, \citenamefont {{Nezami}}, \citenamefont {{Qi}}, \citenamefont
  {{Thomas}}, \citenamefont {{Walter}},\ and\ \citenamefont
  {{Yang}}}]{Hayden2016}%
  \BibitemOpen
  \bibfield  {author} {\bibinfo {author} {\bibfnamefont {P.}~\bibnamefont
  {{Hayden}}}, \bibinfo {author} {\bibfnamefont {S.}~\bibnamefont {{Nezami}}},
  \bibinfo {author} {\bibfnamefont {X.-L.}\ \bibnamefont {{Qi}}}, \bibinfo
  {author} {\bibfnamefont {N.}~\bibnamefont {{Thomas}}}, \bibinfo {author}
  {\bibfnamefont {M.}~\bibnamefont {{Walter}}},\ and\ \bibinfo {author}
  {\bibfnamefont {Z.}~\bibnamefont {{Yang}}},\ }\bibfield  {title} {\bibinfo
  {title} {{Holographic duality from random tensor networks}},\ }\href
  {https://doi.org/10.1007/JHEP11(2016)009} {\bibfield  {journal} {\bibinfo
  {journal} {Journal of High Energy Physics}\ }\textbf {\bibinfo {volume}
  {2016}},\ \bibinfo {eid} {9} (\bibinfo {year} {2016})},\ \Eprint
  {https://arxiv.org/abs/1601.01694} {arXiv:1601.01694 [hep-th]} \BibitemShut
  {NoStop}%
\bibitem [{\citenamefont {Driscoll}\ and\ \citenamefont
  {Trefethen}(2002)}]{scmappingbook}%
  \BibitemOpen
  \bibfield  {author} {\bibinfo {author} {\bibfnamefont {T.~A.}\ \bibnamefont
  {Driscoll}}\ and\ \bibinfo {author} {\bibfnamefont {L.~N.}\ \bibnamefont
  {Trefethen}},\ }\href {https://doi.org/10.1017/CBO9780511546808} {\emph
  {\bibinfo {title} {Schwarz-Christoffel Mapping}}},\ Cambridge Monographs on
  Applied and Computational Mathematics\ (\bibinfo  {publisher} {Cambridge
  University Press},\ \bibinfo {year} {2002})\BibitemShut {NoStop}%
\bibitem [{\citenamefont {{Cardy}}(1992)}]{cardy9111finite}%
  \BibitemOpen
  \bibfield  {author} {\bibinfo {author} {\bibfnamefont {J.~L.}\ \bibnamefont
  {{Cardy}}},\ }\bibfield  {title} {\bibinfo {title} {{Critical percolation in
  finite geometries}},\ }\href {https://doi.org/10.1088/0305-4470/25/4/009}
  {\bibfield  {journal} {\bibinfo  {journal} {Journal of Physics A Mathematical
  General}\ }\textbf {\bibinfo {volume} {25}},\ \bibinfo {pages} {L201}
  (\bibinfo {year} {1992})},\ \Eprint {https://arxiv.org/abs/hep-th/9111026}
  {arXiv:hep-th/9111026 [hep-th]} \BibitemShut {NoStop}%
\bibitem [{\citenamefont {{Fattal}}\ \emph {et~al.}(2004)\citenamefont
  {{Fattal}}, \citenamefont {{Cubitt}}, \citenamefont {{Yamamoto}},
  \citenamefont {{Bravyi}},\ and\ \citenamefont
  {{Chuang}}}]{Fattal2004stabilizer}%
  \BibitemOpen
  \bibfield  {author} {\bibinfo {author} {\bibfnamefont {D.}~\bibnamefont
  {{Fattal}}}, \bibinfo {author} {\bibfnamefont {T.~S.}\ \bibnamefont
  {{Cubitt}}}, \bibinfo {author} {\bibfnamefont {Y.}~\bibnamefont
  {{Yamamoto}}}, \bibinfo {author} {\bibfnamefont {S.}~\bibnamefont
  {{Bravyi}}},\ and\ \bibinfo {author} {\bibfnamefont {I.~L.}\ \bibnamefont
  {{Chuang}}},\ }\bibfield  {title} {\bibinfo {title} {{Entanglement in the
  stabilizer formalism}},\ }\href@noop {} {\bibfield  {journal} {\bibinfo
  {journal} {arXiv e-prints}\ ,\ \bibinfo {eid} {quant-ph/0406168}} (\bibinfo
  {year} {2004})},\ \Eprint {https://arxiv.org/abs/quant-ph/0406168}
  {arXiv:quant-ph/0406168 [quant-ph]} \BibitemShut {NoStop}%
\bibitem [{\citenamefont {{Hamma}}\ \emph
  {et~al.}(2005{\natexlab{a}})\citenamefont {{Hamma}}, \citenamefont
  {{Ionicioiu}},\ and\ \citenamefont {{Zanardi}}}]{hamma0406ground}%
  \BibitemOpen
  \bibfield  {author} {\bibinfo {author} {\bibfnamefont {A.}~\bibnamefont
  {{Hamma}}}, \bibinfo {author} {\bibfnamefont {R.}~\bibnamefont
  {{Ionicioiu}}},\ and\ \bibinfo {author} {\bibfnamefont {P.}~\bibnamefont
  {{Zanardi}}},\ }\bibfield  {title} {\bibinfo {title} {{Ground state
  entanglement and geometric entropy in the Kitaev model}},\ }\href
  {https://doi.org/10.1016/j.physleta.2005.01.060} {\bibfield  {journal}
  {\bibinfo  {journal} {Physics Letters A}\ }\textbf {\bibinfo {volume}
  {337}},\ \bibinfo {pages} {22} (\bibinfo {year} {2005}{\natexlab{a}})},\
  \Eprint {https://arxiv.org/abs/quant-ph/0406202} {arXiv:quant-ph/0406202
  [quant-ph]} \BibitemShut {NoStop}%
\bibitem [{\citenamefont {{Hamma}}\ \emph
  {et~al.}(2005{\natexlab{b}})\citenamefont {{Hamma}}, \citenamefont
  {{Ionicioiu}},\ and\ \citenamefont {{Zanardi}}}]{hamma0409bipartite}%
  \BibitemOpen
  \bibfield  {author} {\bibinfo {author} {\bibfnamefont {A.}~\bibnamefont
  {{Hamma}}}, \bibinfo {author} {\bibfnamefont {R.}~\bibnamefont
  {{Ionicioiu}}},\ and\ \bibinfo {author} {\bibfnamefont {P.}~\bibnamefont
  {{Zanardi}}},\ }\bibfield  {title} {\bibinfo {title} {{Bipartite entanglement
  and entropic boundary law in lattice spin systems}},\ }\href
  {https://doi.org/10.1103/PhysRevA.71.022315} {\bibfield  {journal} {\bibinfo
  {journal} {\pra}\ }\textbf {\bibinfo {volume} {71}},\ \bibinfo {eid} {022315}
  (\bibinfo {year} {2005}{\natexlab{b}})},\ \Eprint
  {https://arxiv.org/abs/quant-ph/0409073} {arXiv:quant-ph/0409073 [quant-ph]}
  \BibitemShut {NoStop}%
\bibitem [{\citenamefont {{Klappenecker}}\ and\ \citenamefont
  {{Roetteler}}(2000)}]{Klappenecker2002stabilizer}%
  \BibitemOpen
  \bibfield  {author} {\bibinfo {author} {\bibfnamefont {A.}~\bibnamefont
  {{Klappenecker}}}\ and\ \bibinfo {author} {\bibfnamefont {M.}~\bibnamefont
  {{Roetteler}}},\ }\bibfield  {title} {\bibinfo {title} {{Beyond Stabilizer
  Codes II: Clifford Codes}},\ }\href@noop {} {\bibfield  {journal} {\bibinfo
  {journal} {arXiv e-prints}\ ,\ \bibinfo {eid} {quant-ph/0010076}} (\bibinfo
  {year} {2000})},\ \Eprint {https://arxiv.org/abs/quant-ph/0010076}
  {arXiv:quant-ph/0010076 [quant-ph]} \BibitemShut {NoStop}%
\bibitem [{\citenamefont {{Linden}}\ \emph {et~al.}(2013)\citenamefont
  {{Linden}}, \citenamefont {{Mat{\'u}{\v{s}}}}, \citenamefont {{Ruskai}},\
  and\ \citenamefont {{Winter}}}]{Linden2013stabilizer}%
  \BibitemOpen
  \bibfield  {author} {\bibinfo {author} {\bibfnamefont {N.}~\bibnamefont
  {{Linden}}}, \bibinfo {author} {\bibfnamefont {F.}~\bibnamefont
  {{Mat{\'u}{\v{s}}}}}, \bibinfo {author} {\bibfnamefont {M.~B.}\ \bibnamefont
  {{Ruskai}}},\ and\ \bibinfo {author} {\bibfnamefont {A.}~\bibnamefont
  {{Winter}}},\ }\bibfield  {title} {\bibinfo {title} {{The Quantum Entropy
  Cone of Stabiliser States}},\ }\href@noop {} {\bibfield  {journal} {\bibinfo
  {journal} {arXiv e-prints}\ ,\ \bibinfo {eid} {arXiv:1302.5453}} (\bibinfo
  {year} {2013})},\ \Eprint {https://arxiv.org/abs/1302.5453} {arXiv:1302.5453
  [quant-ph]} \BibitemShut {NoStop}%
\bibitem [{\citenamefont {{Gurarie}}\ and\ \citenamefont
  {{Ludwig}}(2005)}]{GurarieLudwigCirc2004}%
  \BibitemOpen
  \bibfield  {author} {\bibinfo {author} {\bibfnamefont {V.}~\bibnamefont
  {{Gurarie}}}\ and\ \bibinfo {author} {\bibfnamefont {A.~W.~W.}\ \bibnamefont
  {{Ludwig}}},\ }\bibfield  {title} {\bibinfo {title} {{Conformal Field Theory
  at Central charge c=0 and Two-Dimensional Critical Systems with Quenched
  Disorder}},\ }\href {https://doi.org/10.1142/9789812775344_0032} {\bibfield
  {journal} {\bibinfo  {journal} {In ``From Fields to Strings: Circumnavigating
  Theoretical Physics'', Editors: M. Shifman, A. Vainshtein, J. Wheater (World
  Scientific)}\ ,\ \bibinfo {pages} {1384}} (\bibinfo {year} {2005})},\ \Eprint
  {https://arxiv.org/abs/0409105} {arXiv:0409105 [hep-th]} \BibitemShut
  {NoStop}%
\bibitem [{\citenamefont {{Cardy}}(1986)}]{CARDY1986}%
  \BibitemOpen
  \bibfield  {author} {\bibinfo {author} {\bibfnamefont {J.~L.}\ \bibnamefont
  {{Cardy}}},\ }\bibfield  {title} {\bibinfo {title} {{Effect of boundary
  conditions on the operator content of two-dimensional conformally invariant
  theories}},\ }\href {https://doi.org/10.1016/0550-3213(86)90596-1} {\bibfield
   {journal} {\bibinfo  {journal} {Nuclear Physics B}\ }\textbf {\bibinfo
  {volume} {275}},\ \bibinfo {pages} {200} (\bibinfo {year}
  {1986})}\BibitemShut {NoStop}%
\bibitem [{\citenamefont {{Jiang}}\ and\ \citenamefont
  {{Yao}}(2016)}]{FirstPassageMath}%
  \BibitemOpen
  \bibfield  {author} {\bibinfo {author} {\bibfnamefont {J.}~\bibnamefont
  {{Jiang}}}\ and\ \bibinfo {author} {\bibfnamefont {C.-L.}\ \bibnamefont
  {{Yao}}},\ }\bibfield  {title} {\bibinfo {title} {{Critical first-passage
  percolation starting on the boundary}},\ }\href@noop {} {\bibfield  {journal}
  {\bibinfo  {journal} {arXiv e-prints}\ ,\ \bibinfo {eid} {arXiv:1612.01803}}
  (\bibinfo {year} {2016})},\ \Eprint {https://arxiv.org/abs/1612.01803}
  {arXiv:1612.01803 [math.PR]} \BibitemShut {NoStop}%
\bibitem [{\citenamefont {Stauffer}\ and\ \citenamefont
  {Aharony}(1992)}]{intro_perc_theory}%
  \BibitemOpen
  \bibfield  {author} {\bibinfo {author} {\bibfnamefont {D.}~\bibnamefont
  {Stauffer}}\ and\ \bibinfo {author} {\bibfnamefont {A.}~\bibnamefont
  {Aharony}},\ }\href {https://doi.org/doi.org/10.1201/9781315274386} {\emph
  {\bibinfo {title} {Introduction To Percolation Theory, Second Edition}}},\
  \bibinfo {edition} {2nd}\ ed.\ (\bibinfo  {publisher} {Taylor \& Francis},\
  \bibinfo {year} {1992})\BibitemShut {NoStop}%
\bibitem [{\citenamefont {{Page}}(1993)}]{Page1993}%
  \BibitemOpen
  \bibfield  {author} {\bibinfo {author} {\bibfnamefont {D.~N.}\ \bibnamefont
  {{Page}}},\ }\bibfield  {title} {\bibinfo {title} {{Average entropy of a
  subsystem}},\ }\href {https://doi.org/10.1103/PhysRevLett.71.1291} {\bibfield
   {journal} {\bibinfo  {journal} {\prl}\ }\textbf {\bibinfo {volume} {71}},\
  \bibinfo {pages} {1291} (\bibinfo {year} {1993})},\ \Eprint
  {https://arxiv.org/abs/gr-qc/9305007} {arXiv:gr-qc/9305007 [gr-qc]}
  \BibitemShut {NoStop}%
\bibitem [{\citenamefont {{Ashida}}\ and\ \citenamefont
  {{Ueda}}(2017)}]{ueda1709nonlocal}%
  \BibitemOpen
  \bibfield  {author} {\bibinfo {author} {\bibfnamefont {Y.}~\bibnamefont
  {{Ashida}}}\ and\ \bibinfo {author} {\bibfnamefont {M.}~\bibnamefont
  {{Ueda}}},\ }\bibfield  {title} {\bibinfo {title} {{Full-Counting
  Many-Particle Dynamics: Nonlocal and Chiral Propagation of Correlations}},\
  }\href@noop {} {\bibfield  {journal} {\bibinfo  {journal} {arXiv e-prints}\
  ,\ \bibinfo {eid} {arXiv:1709.03704}} (\bibinfo {year} {2017})},\ \Eprint
  {https://arxiv.org/abs/1709.03704} {arXiv:1709.03704 [cond-mat.quant-gas]}
  \BibitemShut {NoStop}%
\bibitem [{\citenamefont {{Negnevitsky}}\ \emph {et~al.}(2018)\citenamefont
  {{Negnevitsky}}, \citenamefont {{Marinelli}}, \citenamefont {{Mehta}},
  \citenamefont {{Lo}}, \citenamefont {{Fl{\"u}hmann}},\ and\ \citenamefont
  {{Home}}}]{negnevitsky1804repeated}%
  \BibitemOpen
  \bibfield  {author} {\bibinfo {author} {\bibfnamefont {V.}~\bibnamefont
  {{Negnevitsky}}}, \bibinfo {author} {\bibfnamefont {M.}~\bibnamefont
  {{Marinelli}}}, \bibinfo {author} {\bibfnamefont {K.~K.}\ \bibnamefont
  {{Mehta}}}, \bibinfo {author} {\bibfnamefont {H.~Y.}\ \bibnamefont {{Lo}}},
  \bibinfo {author} {\bibfnamefont {C.}~\bibnamefont {{Fl{\"u}hmann}}},\ and\
  \bibinfo {author} {\bibfnamefont {J.~P.}\ \bibnamefont {{Home}}},\ }\bibfield
   {title} {\bibinfo {title} {{Repeated multi-qubit readout and feedback with a
  mixed-species trapped-ion register}},\ }\href
  {https://doi.org/10.1038/s41586-018-0668-z} {\bibfield  {journal} {\bibinfo
  {journal} {\nat}\ }\textbf {\bibinfo {volume} {563}},\ \bibinfo {pages} {527}
  (\bibinfo {year} {2018})},\ \Eprint {https://arxiv.org/abs/1804.09703}
  {arXiv:1804.09703 [quant-ph]} \BibitemShut {NoStop}%
\bibitem [{\citenamefont {{Minev}}\ \emph {et~al.}(2019)\citenamefont
  {{Minev}}, \citenamefont {{Mundhada}}, \citenamefont {{Shankar}},
  \citenamefont {{Reinhold}}, \citenamefont {{Guti{\'e}rrez-J{\'a}uregui}},
  \citenamefont {{Schoelkopf}}, \citenamefont {{Mirrahimi}}, \citenamefont
  {{Carmichael}},\ and\ \citenamefont {{Devoret}}}]{minev1803catch}%
  \BibitemOpen
  \bibfield  {author} {\bibinfo {author} {\bibfnamefont {Z.~K.}\ \bibnamefont
  {{Minev}}}, \bibinfo {author} {\bibfnamefont {S.~O.}\ \bibnamefont
  {{Mundhada}}}, \bibinfo {author} {\bibfnamefont {S.}~\bibnamefont
  {{Shankar}}}, \bibinfo {author} {\bibfnamefont {P.}~\bibnamefont
  {{Reinhold}}}, \bibinfo {author} {\bibfnamefont {R.}~\bibnamefont
  {{Guti{\'e}rrez-J{\'a}uregui}}}, \bibinfo {author} {\bibfnamefont {R.~J.}\
  \bibnamefont {{Schoelkopf}}}, \bibinfo {author} {\bibfnamefont
  {M.}~\bibnamefont {{Mirrahimi}}}, \bibinfo {author} {\bibfnamefont {H.~J.}\
  \bibnamefont {{Carmichael}}},\ and\ \bibinfo {author} {\bibfnamefont {M.~H.}\
  \bibnamefont {{Devoret}}},\ }\bibfield  {title} {\bibinfo {title} {{To catch
  and reverse a quantum jump mid-flight}},\ }\href
  {https://doi.org/10.1038/s41586-019-1287-z} {\bibfield  {journal} {\bibinfo
  {journal} {\nat}\ }\textbf {\bibinfo {volume} {570}},\ \bibinfo {pages} {200}
  (\bibinfo {year} {2019})},\ \Eprint {https://arxiv.org/abs/1803.00545}
  {arXiv:1803.00545 [quant-ph]} \BibitemShut {NoStop}%
\bibitem [{\citenamefont {et. al.}(2019)}]{google2019supremacy}%
  \BibitemOpen
  \bibfield  {author} {\bibinfo {author} {\bibfnamefont {F.~A.}\ \bibnamefont
  {et. al.}},\ }\bibfield  {title} {\bibinfo {title} {{Quantum supremacy using
  a programmable superconducting processor}},\ }\href
  {https://doi.org/10.1038/s41586-019-1666-5} {\bibfield  {journal} {\bibinfo
  {journal} {\nat}\ }\textbf {\bibinfo {volume} {574}},\ \bibinfo {pages} {505}
  (\bibinfo {year} {2019})},\ \Eprint {https://arxiv.org/abs/1910.11333}
  {arXiv:1910.11333 [quant-ph]} \BibitemShut {NoStop}%
\bibitem [{\citenamefont {Polyakov}(1970)}]{Polyakov:1970xd}%
  \BibitemOpen
  \bibfield  {author} {\bibinfo {author} {\bibfnamefont {A.~M.}\ \bibnamefont
  {Polyakov}},\ }\bibfield  {title} {\bibinfo {title} {{Conformal symmetry of
  critical fluctuations}},\ }\href@noop {} {\bibfield  {journal} {\bibinfo
  {journal} {JETP Lett.}\ }\textbf {\bibinfo {volume} {12}},\ \bibinfo {pages}
  {381} (\bibinfo {year} {1970})},\ \bibinfo {note} {[Pisma Zh. Eksp. Teor.
  Fiz.12,538(1970)]}\BibitemShut {NoStop}%
\bibitem [{\citenamefont {{Popp}}\ \emph {et~al.}(2005)\citenamefont {{Popp}},
  \citenamefont {{Verstraete}}, \citenamefont {{Mart{\'\i}n-Delgado}},\ and\
  \citenamefont {{Cirac}}}]{Cirac2004LocalizableEntanglement}%
  \BibitemOpen
  \bibfield  {author} {\bibinfo {author} {\bibfnamefont {M.}~\bibnamefont
  {{Popp}}}, \bibinfo {author} {\bibfnamefont {F.}~\bibnamefont
  {{Verstraete}}}, \bibinfo {author} {\bibfnamefont {M.~A.}\ \bibnamefont
  {{Mart{\'\i}n-Delgado}}},\ and\ \bibinfo {author} {\bibfnamefont {J.~I.}\
  \bibnamefont {{Cirac}}},\ }\bibfield  {title} {\bibinfo {title} {{Localizable
  entanglement}},\ }\href {https://doi.org/10.1103/PhysRevA.71.042306}
  {\bibfield  {journal} {\bibinfo  {journal} {\pra}\ }\textbf {\bibinfo
  {volume} {71}},\ \bibinfo {eid} {042306} (\bibinfo {year} {2005})},\ \Eprint
  {https://arxiv.org/abs/quant-ph/0411123} {arXiv:quant-ph/0411123 [quant-ph]}
  \BibitemShut {NoStop}%
\bibitem [{\citenamefont {{Nahum}}\ and\ \citenamefont
  {{Skinner}}(2019)}]{nahum1911majorana}%
  \BibitemOpen
  \bibfield  {author} {\bibinfo {author} {\bibfnamefont {A.}~\bibnamefont
  {{Nahum}}}\ and\ \bibinfo {author} {\bibfnamefont {B.}~\bibnamefont
  {{Skinner}}},\ }\bibfield  {title} {\bibinfo {title} {{Entanglement and
  dynamics of diffusion-annihilation processes with Majorana defects}},\
  }\href@noop {} {\bibfield  {journal} {\bibinfo  {journal} {arXiv e-prints}\
  ,\ \bibinfo {eid} {arXiv:1911.11169}} (\bibinfo {year} {2019})},\ \Eprint
  {https://arxiv.org/abs/1911.11169} {arXiv:1911.11169 [cond-mat.stat-mech]}
  \BibitemShut {NoStop}%
\bibitem [{\citenamefont {Ford}\ and\ \citenamefont
  {Fulkerson}(1956)}]{ford_fulkerson_1956}%
  \BibitemOpen
  \bibfield  {author} {\bibinfo {author} {\bibfnamefont {L.~R.}\ \bibnamefont
  {Ford}}\ and\ \bibinfo {author} {\bibfnamefont {D.~R.}\ \bibnamefont
  {Fulkerson}},\ }\bibfield  {title} {\bibinfo {title} {Maximal flow through a
  network},\ }\href {https://doi.org/10.4153/CJM-1956-045-5} {\bibfield
  {journal} {\bibinfo  {journal} {Canadian Journal of Mathematics}\ }\textbf
  {\bibinfo {volume} {8}},\ \bibinfo {pages} {399–404} (\bibinfo {year}
  {1956})}\BibitemShut {NoStop}%
\bibitem [{\citenamefont {Edmonds}\ and\ \citenamefont
  {Karp}(1972)}]{edmonds_karp_1972}%
  \BibitemOpen
  \bibfield  {author} {\bibinfo {author} {\bibfnamefont {J.}~\bibnamefont
  {Edmonds}}\ and\ \bibinfo {author} {\bibfnamefont {R.~M.}\ \bibnamefont
  {Karp}},\ }\bibfield  {title} {\bibinfo {title} {Theoretical improvements in
  algorithmic efficiency for network flow problems},\ }\href
  {https://doi.org/10.1145/321694.321699} {\bibfield  {journal} {\bibinfo
  {journal} {J. ACM}\ }\textbf {\bibinfo {volume} {19}},\ \bibinfo {pages}
  {248–264} (\bibinfo {year} {1972})}\BibitemShut {NoStop}%
\end{thebibliography}%
\end{document}